\documentclass[apj,numberedappendix]{emulateapj}

\newcommand{\MEarth}{\mbox{$M_{\mathrm{E}}$}}        % Mearth
\newcommand{\MJup}{\mbox{$M_{\mathrm{J}}$}}          % Mjup
\newcommand{\Mp}{\mbox{$M_{p}$}}                     % Mp
\newcommand{\Ms}{\mbox{$M_{s}$}}                     % Ms
\newcommand{\dMp}{\dot{M}_{p}}                       % dot{Mp}
\newcommand{\Rhill}{\mbox{$R_\mathrm{H}$}}           % Rhill
\newcommand{\Rbondi}{\mbox{$R_\mathrm{B}$}}          % Rbondi
\newcommand{\Rcapt}{S_\mathrm{c}}                    % Rcapt
\newcommand{\densu}{\mbox{$M_{s}\,a^{-2}_{0}$}}      % Sigma units
\newcommand{\viscu}{\mbox{$a^{2}_{0}\,\Omega_{0}$}}  % Viscosity units
\newcommand{\AU}{\mbox{AU}}                          % AU
\usepackage%
[ps2pdf,     % or dvips, dvipsone
pagebackref, % or backref
pdfpagescrop={53 436 389 704}, %dvipsone; 19 lines
pdfpagescrop={53 486 389 754}, %dvips; 19 lines
colorlinks=true,
linkcolor=red,
filecolor=red,
citecolor=blue,
urlcolor=magenta,
pdftitle={Evolution of Migrating Planets Undergoing Gas Accretion},
pdfauthor={D'Angelo & Lubow},
pdfsubject={Migrating Planets Undergoing Gas Accretion},
pdfkeywords={planetary systems: protoplanetary disks},
bookmarks=true,
breaklinks=true,
hypertexnames=false,
pdfstartview=FitV,
hyperfootnotes=false,
bookmarksopen=false,
pdfpagemode=None]{hyperref}
\shorttitle{\textsc{Migrating Planets Undergoing Gas Accretion}}
\shortauthors{\textsc{D'Angelo \& Lubow}}
\begin{document}

\title{Evolution of Migrating Planets Undergoing Gas Accretion\altaffilmark{\dag}}
\author{Gennaro D'Angelo\altaffilmark{1}}
\affil{NASA Ames Research Center, Space Science and Astrobiology Division,
       MS 245-3, Moffett Field, CA 94035}
\email{gennaro.dangelo@nasa.gov}
\and
\author{Stephen H. Lubow}
\affil{Space Telescope Science Institute, 3700 San Martin
       Drive, Baltimore, MD 21218}
\email{lubow@stsci.edu}
\slugcomment{\today}
\altaffiltext{1}{NASA Postdoctoral Fellow.}
\altaffiltext{$^\dag$}{%
                      To appear in % 
                      \textsc{The Astrophysical Journal} %
                      (v684 n1 September 20, 2008 issue).
                      Also available as ApJ preprint doi: %
                      \texttt{10.1086/590904}.
%                     and as \textsc{arXiv} preprint:
%                     \texttt{arXiv:0801.0002}.
                                        }
\begin{abstract}
We analyze the orbital and mass evolution of planets that undergo 
run-away gas accretion by means of two- and three-dimensional 
hydrodynamic simulations.
The disk torque distribution per unit disk mass as a function
of radius provides an important diagnostic for the nature of the 
disk-planet interactions. 
We first consider torque distributions for nonmigrating planets
of fixed mass and show that there is general agreement with the 
expectations of resonance theory. We then present results of
simulations for mass-gaining, migrating planets.
For planets with an initial mass of $5$ Earth masses ($\MEarth$), 
which are embedded in disks with standard parameters and which undergo 
run-away gas accretion to one Jupiter mass ($\MJup$), the torque
distributions per unit disk mass are largely unaffected by
migration and accretion for a given planet mass. 
The migration rates for these planets are in agreement with the
predictions of the standard theory for planet migration (Type~I 
and Type~II migration). 
The planet mass growth occurs through gas capture within the
planet's Bondi radius at lower planet masses, the Hill radius at 
intermediate planet masses, and through reduced accretion at 
higher planet masses due to gap formation.
During run-away mass growth, a planet migrates inwards by only 
about $20$\% in radius before achieving a mass of $\sim 1\,\MJup$. 
For the above models, we find no evidence of fast migration
driven by coorbital torques, known as Type~III migration.  
We do find evidence of Type~III migration for a fixed mass
planet of Saturn's mass that is immersed in a cold and massive
disk. In this case the planet migration is assumed to begin 
before gap formation completes.
The migration is understood through a model in which the torque
is due to an asymmetry in density between trapped gas on the
leading side of the planet and ambient gas on the trailing side
of the planet. 
\end{abstract}
\keywords{%
accretion, accretion disks --- hydrodynamics --- 
methods: numerical ---
planetary systems: formation --- 
planetary systems: protoplanetary disks ---
solar system: formation}

\section{Introduction}

In the core accretion picture of planet formation 
\citep[and references therein]{bodenheimer1986,wuchterl1991b,%
pollack1996,hubickyj2005},
a small mass solid core initially rapidly accretes solid material, 
followed by a slow evolution phase of gas and solid accretion.
During this slow evolution phase, the planet is limited in its ability
to accrete gas by the thermal heating caused by the impacting solids.
Once the planet's gas mass is greater than its solid mass,
typically at several Earth masses, the planet undergoes ``run-away''
gas accretion, in which it can accrete whatever mass is provided to it.
These processes have been treated by one-dimensional, spherically
symmetric structure calculations in the above papers.

On the other hand, multi-dimensional hydrodynamical calculations of a 
protostellar disk interacting with the planet has revealed various flow 
properties of the gas, including the gap opening by tidal effects, 
previously anticipated by one-dimensional disk models \citep{lin1986}. 
In addition, planet migration that results from disk-planet interactions 
has been analyzed by means of such simulations.
Good agreement is often, but not always, found between the simulations 
and the expectations of theory 
\citep{rnelson2000,bate2003,gennaro2003b,anelson2003b,li2005,gennaro2006}.
These calculations typically do not include the mass
evolution of the planet. Usually they apply accretion boundary 
conditions onto the planet as a means of modelling the run-away gas 
accretion process. One aim of this paper is to analyze the effects
of planet mass growth on migration.

Several controversies remain on the effects of gas. The role of 
coorbital torques on planet migration, in the subgiant mass range,
is not well understood. \citet[hereafter MP03]{masset2003}
suggested on the basis of a model and simulations that a fast mode 
of migration (sometimes called Type III migration) can occur due 
to strong coorbital torques. \citet[hereafter OL06]{ogilvie2006} 
found support for the concept of coorbital dominated migration under 
certain conditions.
At higher grid resolution under the conditions specified by MP03, 
simulations by \citet[hereafter DBL05]{gennaro2005} found that the
migration rate was much slower.

Another subject of interest is how planet masses may be limited by 
a reduction in the gas accretion rate. 
\citet{lin1986} proposed such a reduction by tidal torques that open 
a gap about the orbit of the planet. The value of the highest planet 
mass achieved in the presence of gap opening is somewhat 
controversial. Some studies 
\citep{lubow1999,bate2003,gennaro2003b} have suggested that the 
maximum planet mass is about $6$--$10\,\MJup$, corresponding to 
the upper limit of the observed range of extrasolar planets 
\citep{marcy2005,butler2006}. 
This limit suggests that some other process, such as disk 
dispersal or other self-limiting feedback on planetary accretion, 
is responsible for the lower masses ($\sim 1\,\MJup$) typically 
found observationally.
Other studies suggest that the tidal limit is $\sim 1\,\MJup$ and 
therefore no additional process is required to explain the typical 
masses \citep[e.g.,][]{dobbs-dixon2007}. 

We will address these and other issues in this paper by analyzing
the orbital evolution of a mass-gaining planet embedded in a gas 
disk.
In section~\ref{sec:TorqueDistributions} we analyze the torque 
distributions for planets of constant mass on fixed circular 
orbits. In section~\ref{sec:GrowingMigratingPlanets} we analyze 
the orbital and mass evolution of migrating planets that undergo 
run-away mass accretion. Section~\ref{sec:typeiii} describes a 
model that appears to exhibit migration that is dominated by 
coorbital torques, i.e., Type~III migration.
Section~\ref{sec:summary} contains the summary and discussion.

\section{Torque Distribution for a Non-Migrating Planet}
\label{sec:TorqueDistributions}

Disk-planet gravitational torques result in planet migration 
\citep{gt1980,lin1993,ward1997}. 
The distribution of torque with disk radius provides a means of 
connecting the theory with simulations. 
In this section, we model the disk as a three-dimensional 
system and consider fixed mass planets on fixed circular orbits. 
The torque per unit radius for a planet embedded in a disk was 
previously considered in \citet{bate2003}.
Here we reconsider the analysis with higher resolution, 
especially in the coorbital region, and apply the torque 
distribution per unit disk mass.

\subsection{Numerical Procedure}
\label{sec:NumericalProcedure}

In this section, we describe the torques exerted by a disk 
on an embedded planet 
with mass, \Mp, equal to $1\,\MEarth$, $10\,\MEarth$, 
$0.3\,\MJup$, and $1\,\MJup$. For the two smallest mass planets 
we consider, the planet's Hill radius is smaller than the 
vertical disk thickness of several percent of the distance 
to the star. For the two largest mass planets, the Hill radius
is comparable or larger than the disk thickness.

\subsubsection{Disk Model}
\label{sec:DiskModel}

We use spherical polar coordinates $\{R, \theta, \phi\}$, with 
the origin located at the star-planet center of mass.
The reference frame corotates with the star-planet system. 
The planet's orbit lies in the plane $\theta=\pi/2$. 
The disk is assumed to be symmetric with respect to this plane, hence
only the disk's northern hemisphere (i.e., the volume $\theta\le\pi/2$) 
is simulated. 

We assume that the material in the disk is locally isothermal and
that the pressure $p$ is given by
\begin{equation}
  p(R, \theta, \phi) = \rho(R, \theta, \phi) c^2_{s}(r),
\end{equation}
where $ \rho(R, \theta, \phi)$ is the mass density.
Quantity $ c_{s}(r)$ is the gas sound speed, which is taken
to be a function of cylindrical radius $r= R \sin{\theta}$.
The aspect ratio of the disk, $H/r$, is taken to be constant
and equal to $0.05$. Therefore, the temperature distribution in the
disk is only a function of the distance from the disk's rotation axis, 
$r$, and decreases as $c^2_{s}\propto 1/r$.
Viscous forces are calculated by adopting the stress tensor for a 
Newtonian fluid \citep{M&M} with constant kinematic 
viscosity, $\nu$ and zero bulk viscosity. Disk self-gravity is ignored.
In Appendix~\ref{sec:diskgrav}, we discuss some effects of disk 
self-gravity and of the axisymmetric component of disk gravity on
the migration rates.

\subsubsection{Disk and Planet Parameters}
\label{sec:DiskPlanetParameters}

We adopt the stellar mass $\Ms$ as unit of mass, the
orbital radius $a$ as unit of length, and 
$\Omega^{-1}_{p}=\left[G\,(\Ms+\Mp)/a^3\right]^{-1/2}$ as 
unit of time.  In converting to dimensional units
we consider $a= 5.2\,\AU$ and $\Ms = 1\, M_{\odot}$.

The disk extends from $0$ to $2 \pi$ in azimuth around the star and,
in radius, from $0.4$ to either $4.0$ (Jupiter-mass case) or $2.5$
(lower mass cases). In the $\theta$-direction, the disk domain 
extends above the midplane ($\theta=\pi/2$) for $10$ degrees, 
comprising $3.5$ pressure scale heights, $H$. The initial mass density 
distribution is independent of $\phi$, has a Gaussian profile in the 
$\theta$-direction, and has a radial profile proportional to 
$R^{-3/2}$, so that the initial (unperturbed) surface density varies 
as $R^{-1/2}$.
We adopt a constant dimensionless kinematic viscosity $\nu$ equal to 
$10^{-5}$, corresponding to a turbulent viscosity parameter 
$\alpha=0.004$ at the cylindrical radius $r=1$ ($5.2\,\AU$).

As mentioned above, we perform calculations for four planet masses: 
$\Mp=3\times10^{-6}$, $3\times10^{-5}$, $3\times10^{-4}$, and 
$1\times10^{-3}$, which correspond, respectively, to $1\,\MEarth$, 
$10\,\MEarth$, $0.3\,\MJup$, and $1\,\MJup$.
The gravitational potential, $\Phi_{p}$, of the planet is smoothed 
over a length $\epsilon$ equal to $0.1\,\Rhill$ and is given by
\begin{equation}
\Phi_{p}=-\frac{G\Mp}{\sqrt{S^2+\epsilon^2}}, 
\label{eq:Phip}
\end{equation} 
where $S$ is the distance from the planet and $\Rhill$ is the Hill
radius of the planet.

\subsubsection{Numerical Method}
\label{sec:NumericalMethod}

The mass and momentum equations that describe the evolution of the
disk (e.g., DBL05) are solved numerically by means of 
a finite-difference scheme that applies an operator splitting 
procedure to perform the spatial integration 
of advection and source terms \citep{ziegler1997}.
The algorithm is second-order accurate in space and semi-second-order 
in time. 
The equations are discretized over a mesh with constant grid spacing 
in each coordinate direction.
Nested grids are used to enhance the numerical resolution in 
(arbitrarily large) regions around the planet 
\citep{gennaro2002,gennaro2003b}. This strategy allows the volume 
resolution to be increased by a factor $2^3$ for each added grid level.
These calculations are executed with grid systems involving $5$
levels of grid nesting. The linear base resolution is 
$\Delta R=a\,\Delta\theta=a\,\Delta\phi=0.014\,a$.
The linear resolution achieved in the coorbital region around 
the planet is approximately $9\times10^{-4}\,a$, which corresponds to 
$\sim0.01\,\Rhill$ and $\sim0.1\,\Rhill$ in the Jupiter-mass and 
Earth-mass cases, respectively.
To quantify resolution effects in the Earth-mass case, we also
applied a linear resolution twice as high throughout the entire 
grid system (base resolution of $7\times10^{-3}\,a$ and 
resolution in the coorbital region around the planet of 
$4\times10^{-4}\,a$). 
The torques at the two resolutions, integrated over the disk domain, 
differ by about $5$\%.

The boundary condition near the planet involves removing gas from 
$\sim 0.1\,\Rhill$ of the planet at each timestep. The procedure 
for mass removal is described in more detail in 
section~\ref{sec:GasAccretion}.
In the calculations reported in section~\ref{sec:numerical_results},
the removed mass is not added to the planet's mass in order to keep
it fixed. In sections~\ref{sec:GrowingMigratingPlanets} and
\ref{sec:typeiii} (as well as in 
Appendix~\ref{sec:NumericalSensitivityStudy} and \ref{sec:diskgrav}), 
we will 
present cases in which the planet's mass is augmented by the mass of 
the gas removed from the disk.

The outer boundary of the disk domain is closed to both inflow and
outflow, whereas the inner boundary allows outflow (material can flow
out of the grid domain) but not inflow. 
Reflective and symmetry boundary conditions are applied at colatitude
$\theta=\theta_{\mathrm{min}}$ and at the disk mid-plane 
($\theta=\pi/2$), respectively.

Simulations are run for about $100$ orbital periods. In models with 
$0.3\,\MJup$ and $1\,\MJup$ mass planets, the initial density 
distribution includes a gap along the planet's orbit to account for 
an approximate balance between viscous and tidal torques, which
reduces the relaxation time towards steady state. In all calculations
discussed here, the flow achieves a fairly steady state within 
$\sim 100$ orbits.

\subsection{Theoretical Considerations}

\subsubsection{Torque Density}

Consider a cylindrical coordinate system $\{r, \phi, z\}$ centered
on the star-planet center of mass.
The disk torque along the rotation axis per unit radius exerted on 
the planet is given by
\begin{equation}
\frac{d T}{ d r} (r,t) = \left \langle r\!\!% 
                       \int_0^{2\pi}\!\!\!\! d \phi\! 
                       \int_{-\infty}^{\infty}\!\!\!\! dz \, 
                       \rho({\mathbf{r}},t) \, 
                       \partial_{\phi} \Phi_p({\mathbf{r}},t) 
                         \right \rangle,
\label{eq:dTdr}
\end{equation}
where $\langle X(t) \rangle$ denotes the time-average of $X$ over 
an orbit period centered about time $t$, $\rho$ is the gas 
density, and $\Phi_p$ is the potential due to the planet 
(eq.~\ref{eq:Phip}).

\subsubsection{Radial Overlap Regions}
\label{sec:radoverlap}

The linear theory of Lindblad resonances for disk-planet interactions
demonstrates that the strongest contributing resonances have
azimuthal wavenumbers
$m \sim r/H$. This estimate comes from considering the so-called
torque cutoff effect that arises from Lindblad resonances that lie 
close to the planet \citep{gt1980,ward1986,pawel1993a}.
As a consequence of the resonance condition, we expect the peak torque 
density to be at a distance of roughly $H$ from the planet. The torque 
cutoff is not sharp and there are torque contributions from resonances 
that lie closer than distance $\sim H$ from the planet, although at 
a decreasing level as they get closer to the planet.
As we will see, the numerical results show the torque density
peak to be close to distance $H$ from the planet.
However, the torque cutoff calculations assume that the orbits are such 
that the gas azimuthally passes by the planet, i.e, lies on circulating 
orbits. 
On the other hand, close to the planet's orbit, this assumption breaks 
down and the gas flows on librating streamlines of the horseshoe orbit 
region.
This region \textit{generally} extends in the radial direction to 
a distance of about $3\,\Rhill$ from the planet's orbital radius, 
where $\Rhill$ is the planet's Hill radius. 
But, close to the planet, the region becomes less extended 
radially, spanning only to approximately $\Rhill$. 
That is, the noncoorbital (circulating) streamlines pass closest 
to the planet at a distance about equal to $\Rhill$
(see streamline \textit{a} in \citealp{lubow1999} and 
Figure~5 in \citealp{bate2003}).
In the horseshoe orbit region, the corotational resonance can play 
a role.

These two regions, the coorbital region (extending up to about 
$3\,\Rhill$ from the planet's orbital radius) and Lindblad torque 
region (extending beyond about distance $H$ from the planet's 
orbital radius), overlap in a one-dimensional radial sense for 
planet-to-star mass ratios 
\begin{equation}
q \gtrsim \frac{1}{9} \left(\frac{H}{r} \right)^3.
\end{equation}
This condition does not necessarily imply a physical
overlap in two or three dimensions. But it does affect our 
interpretation of the torque density reduced to one dimension, 
$dT(r)/dr$. The reason is that for a given radius $r$ such that 
$\Rhill < |r-a|  <  3\,\Rhill$,
the gas lies in either the coorbital (librating) or noncoorbital 
(circulating) region, depending on the azimuth.

For the disk parameters considered in this section,
the one-dimensional overlap occurs for planet masses greater than 
about $4.6\,\MEarth$, which covers all, but one, of the planet 
masses considered.  
For a $1\,\MJup$ planet, this overlap occurs out to a radius of 
about $1.2\,a$ or a radial distance of about $4\,H$ from  planet. 

The two regions physically overlap in a two- or three-dimensional 
sense, when the closest approach of all noncoorbital (circulating) 
streamlines, which occurs at a distance $\sim \Rhill$ from the planet,
is greater than the distance where there are maximum Lindblad torques
($\sim H$). 
This occurs when 
\begin{equation}
q \gtrsim 3 \left(\frac{H}{r} \right)^3.
\end{equation}
In this case, the usual torque cutoff condition for Lindblad 
resonances is questionable. This argument suggests that the torque 
density maximum for Lindblad resonances should occur at a radial 
distance from the planet
\begin{equation}
|r-a| \simeq \max{(\Rhill,H)}.
\label{eq:rTm}
\end{equation}
When this condition is satisfied, the overall torque on the planet 
will be reduced, even if $\Rhill \lesssim H$, since resonances that 
lie closer than distance $H$ from the planet are suppressed\footnote{%
They may still partially contribute, due to their finite widths.}.
For the disk parameters considered in this section, this condition 
is satisfied for $\Mp \ga 4\times 10^{-4}\,\Ms$ (or $0.4\,\MJup$). 

\subsubsection{Saturation Effects of Coorbital Torques}
\label{sec:sat}

The flow in the coorbital region is trapped in horseshoe
orbits. For a time-reversible system (e.g., no dissipation or 
migration), the streamlines are exactly periodic and no net 
torque occurs on the planet due to the disk (i.e., the torque 
saturates), except for possible initial transients due to 
initial conditions.  However, turbulent viscosity introduces 
irreversibility that can lead to a net torque.
The condition for saturation within the framework of the 
$\alpha$-disk model is that the libration timescale of the fluid 
in the coorbital region is shorter than the viscous radial diffusion 
timescale across this region.
Based on scaling arguments, the saturation condition is given by
\citep{ward1992}
\begin{equation}
\alpha \lesssim q^{3/2} \left( \frac{r}{H} \right)^{7/2}.
\end{equation}
For the parameters in this section, this constraint implies that
for planets of order $10\,\MEarth$ or greater, the corotation
torques should be saturated (small). Saturation effects should be 
important for the larger planet masses we consider. 

\subsection{Numerical Results}
\label{sec:numerical_results}

The torque per unit disk mass is defined by
\begin{equation}
\frac{d T}{d M} (r,t) =\left \langle \frac{1}{2 \pi \Sigma(r,t)}\!
                       \int_0^{2\pi}\!\!\!\! d \phi\! 
                       \int_{-\infty}^{\infty}\!\!\!\! dz \,
                       \rho({\mathbf{r}},t) \, 
                       \partial_{\phi} \Phi_p({\mathbf{r}},t) 
                       \right \rangle,
\label{eq:dTdM}
\end{equation}
where $\Sigma(r,t)$ is the axisymmetric disk density (i.e.,
the surface density averaged over the azimuth $\phi$)
and notation $\langle X(t) \rangle$ is defined below 
equation~(\ref{eq:dTdr}).

Numerically, the torque distribution per unit disk mass is determined 
by dividing the (three-dimensional) disk into a series of concentric 
shells, of radius $R$ and thickness $\Delta R$, centered at the origin
and calculating the torque exerted by the shell and the mass of the 
shell. The torque per unit disk mass is obtain from the ratio
of these two quantities\footnote{%
There is a slight error of order $(H/r)^2$ in this procedure due 
to the difference between the spherical coordinate system used
in the calculations and the cylindrical coordinates that apply 
to the definition of the torque in equation~(\ref{eq:dTdr}).}, 
averaged over an orbit period.
We use the radial grid spacing on the base grid for the value of 
$\Delta R$. 
The torques arising from within the Hill sphere of the planet are 
ignored in this section, but are included in later sections of this 
paper.
We ignore such considerations here in order to compare results
with the standard theory of coorbital and Lindblad torques, which
does not include such contributions \citep{tanaka2002}.

\begin{figure}
\centering%
\resizebox{\linewidth}{!}{%
\includegraphics{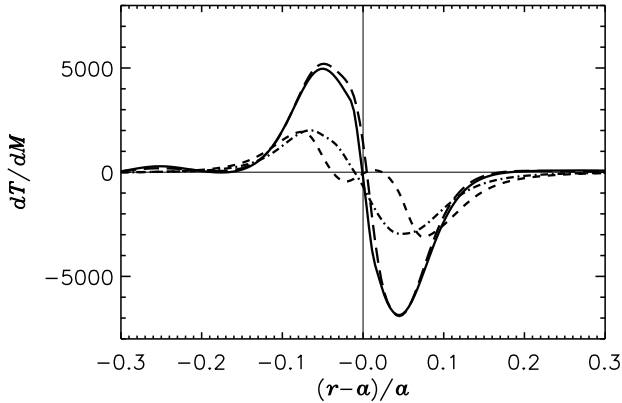}}
\caption{Torque per unit disk mass on the planet as a function of 
         radius in units of the planet's semi-major axis, $a$. The 
         vertical scale is in units of $G\Ms (\Mp/\Ms)^2/a$. 
         The solid, long-dashed, dot-dashed, and short-dashed curves
         are for 
         $1\,\MEarth$, $10\,\MEarth$, $0.3\,\MJup$, and $1\,\MJup$ 
         mass planets, respectively. 
         The disk is modeled as a three-dimensional system.
         The vertical disk thickness is $H/r=0.05$ for all the cases.
         Torque distributions are averaged over one orbital period.
         }
\label{fig:dTdM}
\end{figure}
The torque per unit disk mass for four planet mass cases is shown
in Figure~\ref{fig:dTdM}.  
The plots are normalized such that the torque densities in the four 
cases would be the same, according to linear theory, if the 
axisymmetric disk density gradients and gas properties (sound speeds 
and viscosities) were the same. 
That is, the torque density per unit disk mass is scaled by the square 
of the star-to-planet mass ratio.
The $1\,\MEarth$ (\textit{solid line}) and $10\,\MEarth$ 
(\textit{long-dashed line}) cases nearly exactly overlap as 
predicted, while the $0.3\,\MJup$ (\textit{dot-dashed line}) and
$1\,\MJup$ (\textit{short-dashed line}) cases have a smaller scaled 
torque density. 
The scaling in the plot masks the fact that the results span a large 
range of parameter space. In going from $1\,\MEarth$ to $1\,\MJup$
there is a change in torque density by a large factor, $10^5$, while 
the discrepancy is about a factor of 2.5.
 
The deviations in the $0.3\,\MJup$ and $1\,\MJup$ cases could be due 
to the modified torque cutoff, pressure gradients, and nonlinearities.
Since $\Rhill \ga H$ in these cases, Lindblad resonance contributions 
are weakened by the modified torque cutoff, as discussed in 
Section~\ref{sec:radoverlap}.
Pressure gradients cause shifts in the resonance locations.  
For mild pressure gradients that change sign across the orbit
of the planet (as would occur for a mild gap), the resonances shift 
away from the orbit of the planet \citep[see eq.~26 of][]{ward1986}.
The shift would then cause the torques per unit disk mass to be weaker, 
as seen in the figure. The situation is more complicated in the case
of stronger pressure gradients, as may occur for deep gaps, and the 
sign of the effect on the torque depends on the detailed shape
of the density profile.
Nonlinearities may play a role in the $1\,\MJup$ case, since there 
are shocks in the disk in that case, due to the strong forcing. 
But the total torque is not expected to be substantially effected 
by nonlinearity.
For a fixed smooth background disk density distribution, resonant 
torques are quite insensitive to the level of nonlinearity 
\citep{yuan1994}.
For a $1\,\MJup$ planet and a resonance with azimuthal wavenumber 
$m=20=H/a$, the nonlinearity is mild with nonlinearity parameter 
$f=0.6$, as defined by \citet{yuan1994}.
Some broadening of the torque density profile is predicted, while 
the total torque is reduced by only about $1$\%. For much stronger 
nonlinearity, $f=3$, the torque reduction is only $5$\%.
This estimate is based on considering only a single resonance. 
Many resonances overlap, increasing the level of nonlinearity.
However, the theory does not describe overlapping resonances.
So, although we cannot be definite about the importance of 
nonlinearities, indications for a single resonance suggest that 
they are not important.

The torque density per unit disk mass for the $1\,\MJup$ planet 
in Figure\,\ref{fig:dTdM} (\textit{short-dashed line}) shows 
indications of saturation for $|r-a| < \Rhill$. 
As discussed above, this effect is suggested by theoretical 
considerations. The torque density peak for the $1\,\MJup$ case 
is slightly displaced away from the planet relative to the smaller 
mass cases and lies close to a distance $\Rhill \simeq 0.07\,a$ 
from the planet.
This result is consistent with equation~(\ref{eq:rTm}) in the 
$1\,\MJup$ case, $|r-a| \simeq 0.07\,a =1.4\,H$.

\begin{figure}
\centering%
\resizebox{\linewidth}{!}{%
\includegraphics{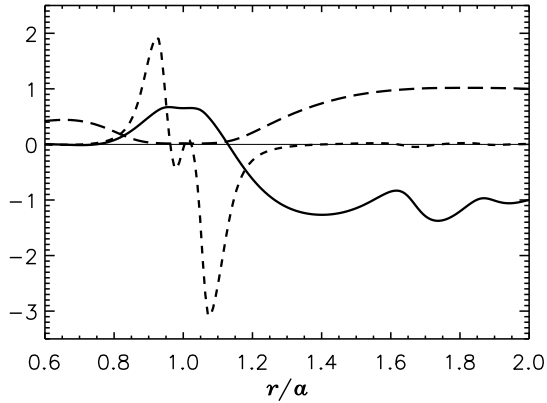}}
\caption{Azimuthally averaged surface density 
         (\textit{long-dashed curve}), 
         disk torque per unit disk radius exerted on the planet 
         (\textit{short-dashed curve}), and cumulative torque 
         (\textit{solid curve}),
         i.e., torque per unit radius integrated outward, as a function 
         of radius for a $1\,\MJup$ planet on a fixed circular orbit.
         The disk is modeled as a three-dimensional system.
         The unit of radius is the
         planet's orbital radius $a$. The surface density and 
         cumulative torque are normalized by their absolute values 
         at $r=2 $. The disk torque per unit disk mass is normalized
         by  $10^{3}\,G\Ms (\Mp/\Ms)^2/a$. The plotted values are 
         averaged over one orbital period.
         }
\label{fig:dTdr}
\end{figure}
Figure~\ref{fig:dTdr} shows that the torque in the $1\,\MJup$ case
is acquired close to the planet, well within the gap region.
Most of the torque is accumulated by material with intermediate/low 
density interacting with an intermediate magnitude torque
per unit disk mass. About $80$\% of the torque is due to material
within a radial distance of $0.25 \,a$ from the planet.

\section{Migrating and Growing Planets}
\label{sec:GrowingMigratingPlanets}

We investigate the orbital migration of a planet that is undergoing
run-away gas accretion. 
We consider several disk configurations, by changing
the initial surface density, the pressure scale height, and the 
kinematic viscosity. 
We use disk models and numerical procedures similar to those 
introduced in section~\ref{sec:NumericalProcedure}.
Throughout this section, the disk is modeled as a 
three-dimensional system.
The origin of the coordinate system is taken to be the star.
The coordinate system rotates about the origin at a rate
equal to the rotation rate of the planet around the star.
We integrate the equations
of motion of the planet, under the action of disk torques and 
apparent forces arising from the rotation of the reference frame, 
as described in DBL05. The unit of length is the
initial star-planet separation $a_0$ (or $5.2\,\AU$ when 
converting into physical units). The unit of
time is the inverse of $\Omega_{0}$, the initial angular
speed of the planet. The unit of mass is the stellar mass $\Ms$
($1\,M_{\odot}$).

The grid system achieves a linear base resolution of 
$\Delta R=a_{0}\,\Delta\theta=a_{0}\,\Delta\phi=0.014\,a_{0}$. 
In the coorbital region around the planet, the linear resolution is 
about $9\times10^{-4}\, a_0$. Nested grid levels cover extended
radial regions of the disk so that the planet remains within the 
domain covered by the most refined grid level over the entire orbital 
evolution.
Convergence tests were carried out with a grid system that used a 
volume resolution $(3/2)^3$ times as high throughout the whole disk 
domain and on all grid levels. No significant differences are
observed (see Appendix~\ref{sec:ResolutionTest}).
To avoid depletion of the disk interior of the planet's orbit, we
apply nonreflecting boundary conditions to the inner grid (radial) 
border. 
We test our results against possible boundary condition effects
in Appendix~\ref{sec:BoundaryConditionEffects} by applying outflow
boundary conditions and moving radial disk boundaries farther away 
from the planet's orbit in both directions. No important effects are
observed. Near the planet we apply accreting boundary
conditions on the gas, as described in section~\ref{sec:GasAccretion}.
We consider planetary mass increases that extend over more than two 
orders of magnitude and a range of disk surface densities.

To avoid possible spurious torques exerted by material
gravitationally bound to the planet, contributions from within 
$\Rhill/2$ of the planet are not taken into account. We report in 
Appendix~\ref{sec:ExcludedTorquesEffects} on the sensitivity of 
the results to the radius of the excluded region by considering 
a smaller radius. We find that the changes are not significant.

We generally initiate the calculations with a planet mass 
$\Mp=1.5\times10^{-5}\,\Ms$, or $5\,\MEarth$. 
However, in some applications discussed in section~\ref{sec:typeiii},
we use an initial mass $\Mp=3\times10^{-4}\,\Ms$ (about $0.3\,\MJup$)
in order to study the effects on migration of releasing a more massive
planet in an unperturbed disk.

\subsection{Planet Mass Growth}
\label{sec:PlanetMassGrowth}

\subsubsection{Gas Accretion}
\label{sec:GasAccretion}

In the core accretion scenario of giant planet formation, prior 
to the phase of run-away gas accretion, the rate at which gas is 
accreted is largely determined by the ability of a planetary 
core's envelope to radiate away the energy delivered by gas and 
solids \citep[phase of slow gas accretion, see e.g.,][]{hubickyj2005}. 
During the initial stages of planet growth, the accretion of 
solids dominates, and the dissipation of the kinetic energy of 
the impacting solids provides an important heat source for the 
accreted gaseous envelope.
Models of \citet{hubickyj2005}, which ignore the effects of 
planet migration, experience a depletion of solid disk material 
in the vicinity of the planet and consequently a reduction in the 
envelope heating rate.
When the mass of the gas (in the envelope) is comparable to 
the mass of solids (in the core), the pressure gradient cannot 
prevent the gravitational collapse of the envelope.
This situation results in a sudden increase of the gas accretion 
rate and a rapid growth of the planet's mass, the so-called 
run-away gas accretion phase \citep[e.g.,][]{wuchterl1993,pollack1996}. 

The models presented here assume run-away gas accretion.
They do not account for the thermal structure
and detailed microphysics of a planet's envelope. 
Therefore, we do not determine self-consistent gas accretion rates,
prior to the phase of run-away gas accretion 
($\Mp\lesssim 10\,\MEarth$). The models also ignore the effects
of heating by impacting solids that act to slow the gas accretion, 
as the planet migrates out of the region of depleted solids.
During the run-away gas accretion phase, the 
accretion rate onto the planet is only limited by the amount of
gas that the disk is able to supply. The calculations 
described here provide estimates of such limiting gas accretion rates
during the run-away gas accretion phase.

In these models, we adopt a prescription that gas within a distance of 
$R_{\mathrm{acc}}=0.1\,\Rhill$ from the planet can accrete onto it. 
Accreted gas is removed from the disk and its mass is added to the 
planet mass. For the models we consider, this distance
is safely smaller than the possible characteristic accretion
radii: the Hill radius, $\Rhill$, and the Bondi radius, $\Rbondi$ 
(distance beyond which the thermal energy of the gas is larger than 
the gravitational energy that binds the gas to the planet).
The distance $R_{\mathrm{acc}}$ is at least a factor of $3$ smaller 
than $\Rbondi$.
Therefore, this mass removal prescription should not determine the 
accretion rate for the case of run-away gas accretion
\citep[see also][]{tanigawa2002}.
The amount of material accreted per time-step $\Delta t$ is given by 
$(\Delta t/\tau_{\mathrm{acc}})\int\! \rho\,dV$, 
where $dV$ is the volume element and
$\tau_{\mathrm{acc}}$ is a removal timescale. 
The integral is performed over the sphere of radius $0.1\,\Rhill$ 
centered on the planet.
Here we set
$\tau_{\mathrm{acc}}=0.1\,\Omega^{-1}_{0}$ within the sphere of 
radius $0.05\,\Rhill$ and
$\tau_{\mathrm{acc}}=0.3\,\Omega^{-1}_{0}$ for 
$0.05\,\Rhill <S< 0.1 \,\Rhill$ ($S$ is the distance from the
planet).

\subsubsection{Mass Evolution}
\label{sec:MassEvolution}

\begin{figure}
\centering%
\resizebox{\linewidth}{!}{%
\includegraphics{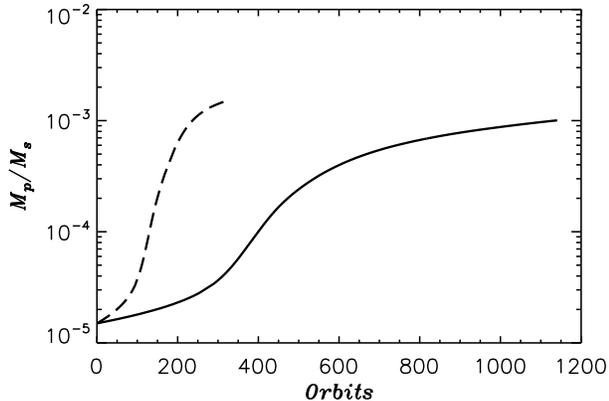}}
\caption{Mass evolution of a protoplanet having initial planet 
         mass $5\,\MEarth$ undergoing run-away gas accretion
         in a three-dimensional disk with initial surface density 
         $\Sigma_{p}=3\times10^{-4}\,\densu$ or about 
         $100\,\mathrm{g}\,\mathrm{cm}^{-2}$ at the planet's 
         initial orbital radius of $5.2\,\AU$ 
         (\textit{solid line}) and
         $\Sigma_{p}=9\times10^{-4}\densu$ or about 
         $300\,\mathrm{g}\,\mathrm{cm}^{-2}$ 
         (\textit{dashed line}).
         In both cases, the disk thickness is $H/r=0.05$ and
         the turbulent viscosity parameter is 
         $\nu=1\times10^{-5}\,\viscu$ 
         ($\alpha=0.004$ at $5.2\,\AU$). 
         The time refers to orbits at $a_{0}=5.2\,\AU$ or about
         $12$ years.
         }
\label{fig:mp_sig}
\end{figure}
In this section we describe the accretion rates
of migrating, mass-gaining planets.
Figure~\ref{fig:mp_sig} shows the planet mass as a function of 
time, $\Mp=\Mp(t)$, for a model with initial
(unperturbed) surface density at the initial orbital radius of
the planet 
$\Sigma_{p}=3\times10^{-4}\,\densu$ (\textit{solid line}).
For a planet orbiting a Solar mass star at $5.2\,\AU$, this density
is about $100\,\mathrm{g}\,\mathrm{cm}^{-2}$, roughly corresponding 
to the minimum mass solar nebula.

\begin{figure}
\centering%
\resizebox{\linewidth}{!}{%
\includegraphics{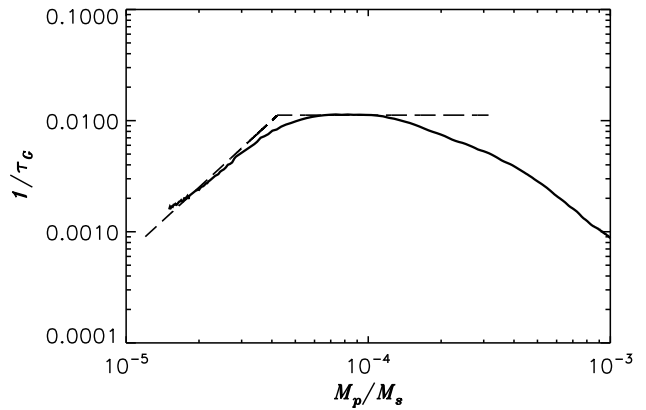}}
\caption{Mass growth rate $1/\tau_{G}=\dMp/\Mp$ in units of 
         inverse orbital periods at the initial radius of the 
         planet, $\Omega_0/(2\pi)$, plotted against $\Mp/\Ms$ 
         for the solid curve case in Figure~\ref{fig:mp_sig}.
         The dashed line plots the growth rate according to 
         equation~(\ref{eq:taubh}). The slopes of the two 
         dashed line segments are predicted by the model. 
         The two free parameters, $C_{\mathrm{B}}$ and 
         $C_{\mathrm{H}}$, are dimensionless constants of order
         unity that control the intercepts and are fit to the
         solid curve. The slanted portion of dashed line
         corresponds to accretion within the Bondi radius,
         given by $1/\tau_{\mathrm{B}}$ in 
         equation~(\ref{eq:taubh}) with $C_{\mathrm{B}}=2.6$.
         The horizontal portion of the dashed line corresponds
         to accretion within the Hill radius for a disk with no gap,
         given by $1/\tau_{\mathrm{H}}$ in equation~(\ref{eq:taubh})
         with $C_{\mathrm{H}}=0.89$.
         At higher planet masses, the growth rates drop due to 
         the presence of the tidally produced gap.
         }
\label{fig:tauG_sn_fr}
\end{figure}
The mass evolution can be understood in terms of Bondi and Hill
accretion. Consider a simple model in which gas is captured 
within some radius, $\Rcapt$, of a planet and assume $\Rcapt<H$. 
Mass is accreted with some velocity relative to the planet of 
order $\Omega\,\Rcapt$, and so the mass accretion rate in a 
three-dimensional disk (where $\rho\approx\Sigma/H$) is 
estimated as
\begin{equation}
\dMp \sim \frac{\Sigma}{H} \, \Omega\,\Rcapt^3,
\label{eq:Mpscale}
\end{equation}
where we take $\Rcapt$ as either the Bondi or Hill radius,
with the Bondi radius given by $\Rbondi=G\,\Mp/c^2_{s}$
and the Hill radius given by $\Rhill = a\,[\Mp/(3\,\Ms)]^{1/3}$.

In the case that gas pressure prevents the gas from being
bound to the planet within the Hill sphere (or, equivalently, that
pressure forces dominate over gravitational three-body forces), 
we expect the Bondi description to be appropriate. This condition 
is that
\begin{equation}
c^2_{s}  \gtrsim \frac{G \, \Mp}{\Rhill}
\end{equation}
or
\begin{equation}
\Rbondi \lesssim \Rhill.
\end{equation}
Therefore in the general case we take
\begin{equation}
\Rcapt = \min{(\Rbondi , \Rhill)}.
\end{equation}

It then follows that the Bondi and Hill mass growth rates, 
$\dMp/\Mp$, of the planet are given by
\begin{eqnarray}
1/\tau_{\mathrm{B}} &=& C_{\mathrm{B}}\,\Omega\,\frac{\Sigma\,a^2}{\Ms}%
                        \left(\frac{a}{H}\right)^7%
                        \left(\frac{\Mp}{\Ms}\right)^2, \\ \label{eq:tauBinv}
1/\tau_{\mathrm{H}} &=& \frac{1}{3}\,C_{\mathrm{H}}\,\Omega\,%
                        \frac{\Sigma\,a^2}{M_s}\,%
                        \left(\frac{a}{H}\right),          \label{eq:tauHinv}
\end{eqnarray}
where $C_{\mathrm{B}}$ and $C_{\mathrm{H}}$ are dimensionless
coefficients of order unity.
The overall mass growth rate is given by
\begin{equation}
1/\tau_{G} = \left \{\!%
             \begin{array}{ll}
             1/\tau_{\mathrm{B}} & \mbox{for $\Mp <   M_{\mathrm{t}}$}\\
             1/\tau_{\mathrm{H}} & \mbox{for $\Mp \ge M_{\mathrm{t}}$}%
             \end{array}
             \right.
\label{eq:taubh}
\end{equation}
where
\begin{equation}
M_{\mathrm{t}} = \frac{\Ms}{\sqrt{3}}\,%
                 \sqrt{\frac{C_{\mathrm{H}}}{C_{\mathrm{B}}}}\,%
                 \left(\frac{H}{a}\right)^3
\end{equation}
is the transition planet mass where 
$\tau_{\mathrm{H}}=\tau_{\mathrm{B}}$.

In Figure~\ref{fig:tauG_sn_fr}, we plot the mass growth rate,
$1/\tau_{G}$, for the solid curve case in Figure~\ref{fig:mp_sig}.
We applied equation~(\ref{eq:taubh}) and adopted constant 
values of $\Sigma=\Sigma(a_{0})$, at time $t=0$, and 
$\Omega=\Omega_{0}$.
The figure shows that the Bondi and Hill accretion rates
in equation~(\ref{eq:taubh}) agree with the simulation results
for values of $C_{\mathrm{B}}=2.6$ and $C_{\mathrm{H}}= 0.89$.
The transition mass in this case evaluates to 
$M_{\mathrm{t}} = 4.2 \times 10^{-5}\,\Ms$.
It lies between the Bondi and Hill accretion regimes in the figure,
at the intersection between the two dashed line segments.
For larger values of planet mass, 
$\Mp \ga 2 \times 10^{-4}\,\Ms\approx 4.8\,M_{\mathrm{t}}$,
this simple estimate of the mass growth rate breaks down because 
the density is depleted near the planet due to the onset of gap 
formation.
The density near the planet is reduced by about $40$\% when
$\Mp\approx 2 \times 10^{-4}\,\Ms$ 
(see Fig.~\ref{fig:a_sn}, \textit{right panel}).
In addition, the Hill radius becomes comparable to $H$, since
$\Rhill = H = 0.05\,a$ for $\Mp = 3.75 \times 10^{-4}\,\Ms$.

Simulations carried out in two dimensions would have different
scaling behavior, since the right-hand side of 
equation~(\ref{eq:Mpscale}) would be $\Sigma\,\Omega\,\Rcapt^2$.
The dependence of the mass accretion rate on planet mass and 
disk sound speed then artificially deviates from the 
three-dimensional case. In two dimensions we have that
$1/\tau_{\mathrm{B}} \propto (\Mp/\Ms)\,(a/H)^4$ and 
$1/\tau_{\mathrm{H}} \propto (\Ms/\Mp)^{1/3}$.

The maximum of the accretion rate for the solid curve case of 
Figure~\ref{fig:mp_sig} is 
$\dMp\sim 5\times10^{-3}\,\Sigma_{p}\,a^2\simeq 1.5\times10^{-3}\,\MJup$ 
per orbit and occurs when  $\Mp \approx 0.3\,\MJup$. 
This result is consistent with the previous findings of 
\citet{gennaro2003b} and \citet{bate2003}, who considered
planets on fixed orbits.
Also displayed in Figure~\ref{fig:mp_sig} is the planet's mass 
evolution in a disk with initial $\Sigma_{p}=9\times10^{-4}\,\densu$ 
(\textit{dashed line}) or about 
$300\,\mathrm{g}\,\mathrm{cm}^{-2}$ at $5.2\,\AU$. 
For $\Mp/\Ms\lesssim 10^{-4}$, the accretion rate is a factor of $3$
larger than that of the lower density disk case (\textit{solid line}). 
Hence, equation~(\ref{eq:taubh}) applies to the growth rate with the 
same coefficients $C_{\mathrm{B}}$ and $C_{\mathrm{H}}$ as those 
given above.
For larger planet masses, the accretion rate keeps increasing 
until $\Mp\approx 0.7\,\MJup$, at which point $\dMp$ starts to 
decline very rapidly as $\Mp$ grows further. 
This is because effects due to gap formation are delayed.
The timescale required to form a gap of half-width $\xi\Rhill$ 
is $\tau_{\mathrm{gap}}\sim \xi^5\,q^{-1/3}\,\Omega^{-1}$
\citep[see, e.g.,][]{bryden1999}, where $\xi\approx 2$ 
(see \textit{long-dashed line} in Fig.~\ref{fig:dTdr}). 
In the lower density disk model (\textit{solid curve} in 
Fig.~\ref{fig:mp_sig}), 
$\tau_{\mathrm{gap}}<\tau_{G}$ for $\Mp/\Ms\gtrsim 10^{-4}$.
In the higher density disk model (\textit{dashed curve}), 
$\tau_{\mathrm{gap}}$ becomes shorter than $\tau_{G}$ 
only when $\Mp\gtrsim 0.7\,\MJup$.

\begin{figure}
\centering%
\resizebox{\linewidth}{!}{%
\includegraphics{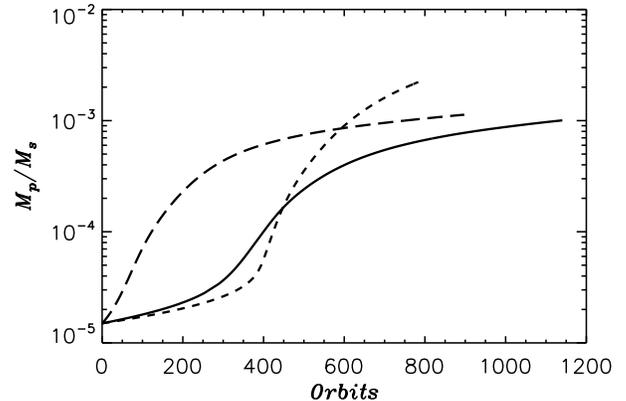}}
\caption{Mass evolution of a protoplanet having initial planet 
         mass $5\,\MEarth$ and undergoing run-away gas accretion in 
         a three-dimensional disk with initial surface density 
         $\Sigma_{p}=3\times10^{-4}\,\densu\approx 100\,\mathrm{g}\,\mathrm{cm}^{-2}$
         at the planet's initial orbital radius $a_0=5.2\,\AU$. 
         The solid line represents a case with $H/r=0.05$ and 
         turbulent viscosity $\nu=1\times10^{-5}\,\viscu$ 
         ($\alpha=0.004$ at $5.2\,\AU$), the long-dashed
         line refers to a variant model with $H/r=0.04$ and 
         the same $\nu$ value 
         ($\alpha=0.006$ at $5.2\,\AU$), and the 
         short-dashed line represents a variant model with 
         $\nu=1\times10^{-4}\,\viscu$ ($\alpha=0.04$ at $5.2\,\AU$).
         The time refers to orbits at $a_{0}=5.2\,\AU$ or about 
         $12$ years.
         }
\label{fig:mp_hnu}
\end{figure}
In Figure~\ref{fig:mp_hnu}, the mass evolution is shown for cases 
in which 
$\Sigma_{p}=3\times10^{-4}\,\densu\approx 100\,\mathrm{g}\,\mathrm{cm}^{-2}$, 
but with different scale heights, $H$, and kinematic 
viscosities, $\nu$. 
Near $\Mp=1\,\MJup$, the accretion rates of the two models 
with different $H/r$ (\textit{solid} and \textit{long-dashed lines}), 
but the same $\Sigma_{p}$ and $\nu$, are nearly equal, with
$\dMp \approx 3\times10^{-3}\,\Sigma_{p}\,a^2\simeq 9\times10^{-4}\MJup$ 
per orbit.
At larger planet masses, $\dMp$ is smaller in the case of a colder 
disk (\textit{long-dashed line}) because of the stronger tidal torques 
exerted by the planet on the disk material that produce a wider 
gap. 
When $\Mp\approx 1\,\MJup$, the simulation with $10$ times larger 
viscosity (\textit{short-dashed line}) yields an accretion rate
that is a factor of nearly $8$ larger.
This result is consistent with previous two-dimensional studies 
of planets on fixed orbits that do not gain mass.
For $\Mp\approx 1\,\MJup$ these studies showed that $\dMp$ scales 
approximately linearly with $\nu \Sigma$, the overall disk accretion 
rate evaluated just outside the gap \citep{kley1999,lubow2006}.

\subsubsection{Mass Within the Hill Sphere}
\label{sec:HillSphereMass}

We discuss here the relevance of torques exerted on a 
planet and originating within the planet's Hill sphere. 
We may expect that material gravitationally bound to the
planet should not be capable of exerting significantly strong
torques, if resolution is appropriate (DBL05).
In some situations, if the local density is large, any torque 
imbalance can be easily amplified by lack of numerical resolution
(because torques depend on $1/S^2$, where $S$ is the distance to
the planet). Artificial effects may arise 
when the mass within $\sim\Rhill$ of the planet is larger than 
the planet's mass.
However, not all this material is necessarily bound to the planet.
Because of the nonspherical nature of the Roche lobe,
the Hill radius represents an overestimate for the size of the 
region where gas is bound to the planet 
\citep{paczynski1971,eggleton1983}.
We have found that accumulated gas may be bound to the planet 
within distances shorter than $\Rhill/2$ from the planet
(see Appendix~\ref{sec:BoundGas}).

In all the cases discussed in this section, the amount of material
that lies within $\Rhill/2$ of the planet is smaller than $\Mp$,
throughout the evolution, by several orders of magnitude. 
For models in Figure~\ref{fig:mp_sig}, as well as for those in 
Figure~\ref{fig:mp_hnu}, the ratio of these two masses ranges 
from less than $\sim 10^{-3}$ to $\sim 10^{-2}$, depending mainly 
on the planet's mass. We also consider models with initial 
densities larger than those discussed here (described in 
section~\ref{sec:typeiii}). 
However, this mass ratio remains on the order of $10^{-2}$ or 
smaller.
Therefore, due to the accretion boundary condition employed here
at the planet location, these models do not experience a build-up 
of mass near the planet (with possible effects on planet migration).
The accreted mass is accounted for by the increase in the planet 
mass. 

\subsection{Planet Migration}
\label{sec:PlanetMigration}

\subsubsection{Theoretical Regimes of Migration}
\label{sec:RegimesofMigration}

A planet that grows in mass from a few Earth-masses to a few 
Jupiter-masses is susceptible to two ``classical'' regimes of
migration. The Type~I regime is expected when the planet
causes small, linear disk density perturbations 
\citep[e.g.,][]{ward1997,tanaka2002}. In the opposite
limit, Type~II occurs when the planet mass is large enough
to cause nonlinear density perturbations that result in a 
density gap along its orbit \citep{lin1986}.

For the parameters we adopt (pressure scale height $H/r\sim 0.05$,
kinematic viscosity of disk $\nu\ge 1\times10^{-5}\,\viscu$,
and initial planet mass $\Mp/\Ms=1.5\times10^{-5}$ 
(or $\Mp=5 \,\MEarth$), it is expected that the initial evolution 
of the planet will follow Type~I migration, since the usual
gap opening criteria are not satisfied.
In the linear theory of \citet{tanaka2002}, the rate of migration 
resulting from the action of both Lindblad and (unsaturated) 
coorbital corotation torques is given by
\begin{equation}
\frac{da_{\mathrm{I}}}{dt}=-\left(2.73+1.08\,s\right)%
     \left(\frac{\Mp}{\Ms} \frac{a}{H}\right)^2%
     \frac{\Sigma_{p}}{\Mp}\,a^3\,\Omega_{p},
\label{eq:tanakadotaI}
\end{equation}
where $s$ is the slope of the unperturbed surface density. 
For the case of saturated (zero) coorbital corotation
torques, the migration rate is given by
\begin{equation}
\frac{da_{\mathrm{I}}}{dt}=-\left(4.68- 0.20 \,s\right)%
     \left(\frac{\Mp}{\Ms} \frac{a}{H}\right)^2%
     \frac{\Sigma_{p}}{\Mp}\,a^3\,\Omega_{p}.
\label{eq:tanakadotaIsat}
\end{equation}
The conditions for saturation are discussed in section~\ref{sec:sat}.
For higher planet masses that arise in the later stages of the 
simulations, the torques are expected to be saturated.

In the presence of a sufficiently clean density gap and for a 
planet whose mass is less than the local disk mass, 
the rate of migration 
follows Type~II theory that is dictated by disk viscous inflow 
\begin{equation}
\frac{da_{\mathrm{II}}}{dt} = - \zeta \frac{\nu}{a}.
\label{eq:dotaII}
\end{equation}
Note that if there is residual material in the horseshoe orbit region,
the migration rate can differ from that in equation~(\ref{eq:dotaII}).
The coefficient $\zeta$ on the right-hand side of 
equation~(\ref{eq:dotaII}) is of order unity and also depends on the 
evolutionary state of the disk. For a steady-state disk, the coefficient 
is $3/2$. But for nonsteady disks where $\nu \Sigma$ varies in radius, 
as in our initial states, the coefficient may differ by order unity 
amounts.

In the unsaturated case, some nonlinear effects of the corotation 
resonance can cause migration rates to differ from those predicted 
by equation~(\ref{eq:tanakadotaI}) \citep{masset2006}. 
For $s=1/2$, $H/r=0.05$, these effects occur in the range 
of masses is between $\approx10\,\MEarth$ and $\approx20\,\MEarth$. 
However, in the models presented here, the planet grows too quickly
through this mass range (taking less than a few tens of orbits)
to significantly affect migration (see Fig.~\ref{fig:bid1} in
Appendix~\ref{sec:MVV}).

When the amount of material in the horseshoe orbit region is larger
than the planet's mass, a regime of fast migration known as Type~III 
may occur. The origins of such a regime are not yet entirely clear. 
The model of MP03 suggests that it is driven by strong 
corotation torques originating from material that streams past the
planet, while the planet is moving in the radial direction. However, 
an analytic model of OL06 suggests that such
torques could originate from trapped librating gas.
A somewhat similar model was developed by \citet{pawel2004}.

\subsubsection{Orbital Radius Evolution}
\label{sec:OrbitalRadiusEvolution}

\begin{figure*}
\centering%
\resizebox{\linewidth}{!}{%
\includegraphics{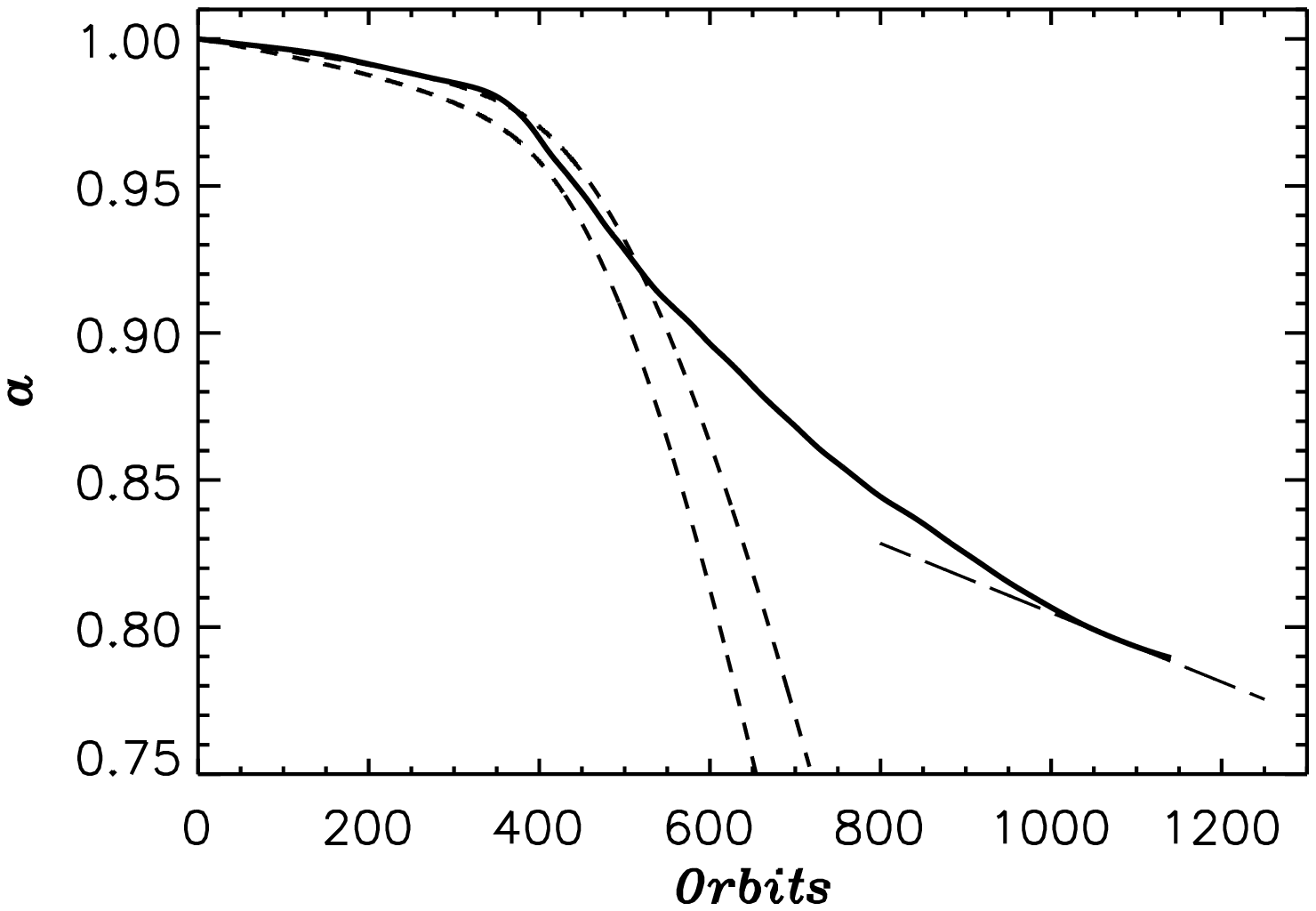}%
\includegraphics{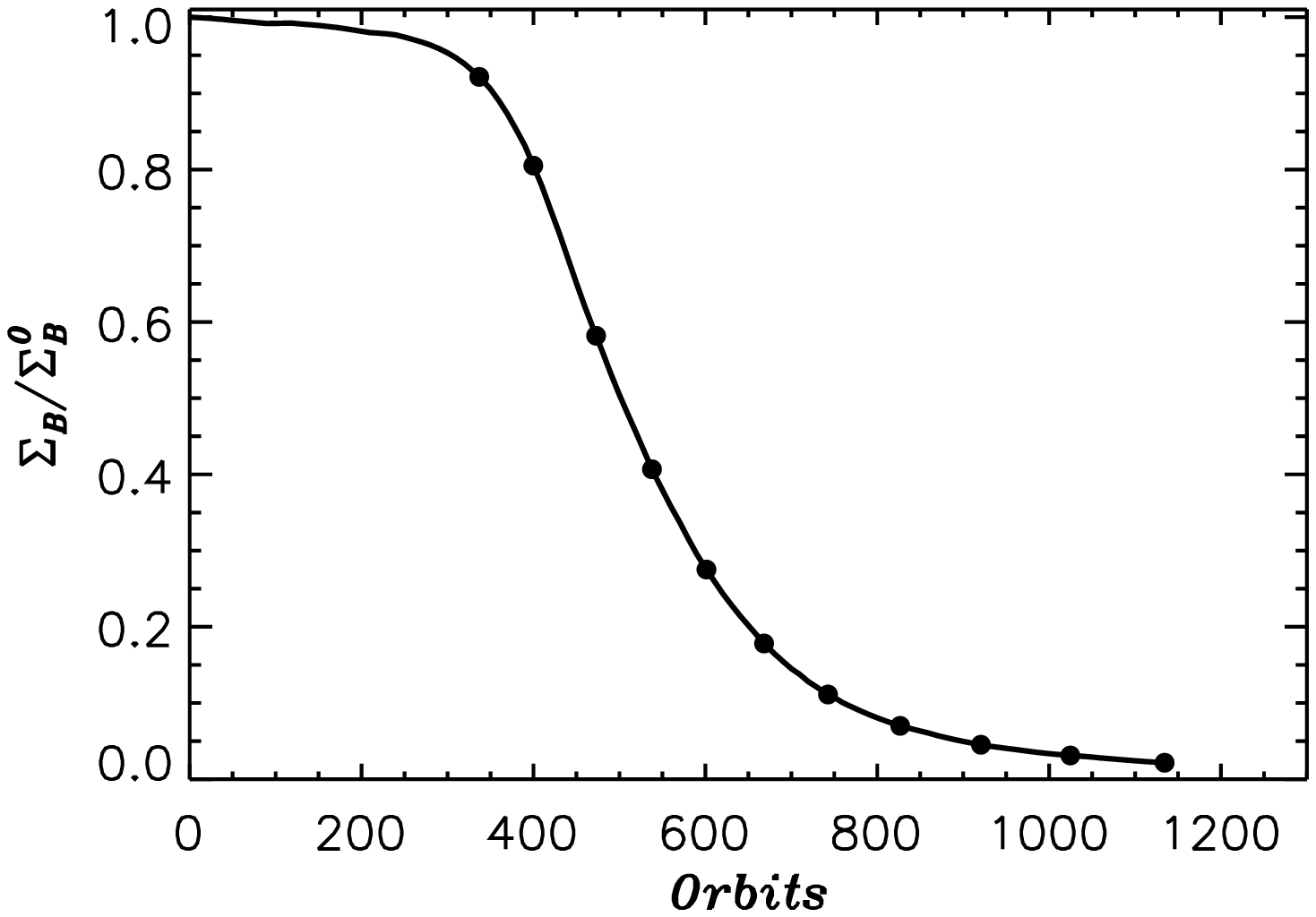}}
\caption{Orbital migration of a planet undergoing run-away gas 
         accretion.
 \textit{Left}: Orbital radius in units of $a_{0}$ 
         ($5.2\,\AU$), as a function of time in units of the initial
         orbital period ($\approx 12$ years).
         The initial planet mass is 
         $5\,\MEarth$. The initial surface density is 
         $\Sigma_{p}=3\times10^{-4}\,\densu%
         \approx 100\,\mathrm{g}\,\mathrm{cm}^{-2}$
         at the planet's initial orbital radius and $H/r=0.05$. 
         \textit{Solid curve}: 
         Results from the three-dimensional numerical simulation of a 
         migrating, gas-accreting planet.   
         \textit{Short-dashed curves}: 
         Predictions based on Type~I 
         migration theory, obtained by solving 
         equations~(\ref{eq:tanakadotaI}) and (\ref{eq:tanakadotaIsat}),
         for a planet that undergoes the mass growth given 
         by the solid line in Figure~\ref{fig:mp_sig} and is embedded
         in a disk with the initial unperturbed density distribution. 
         The upper (lower) curve is for migration with 
         unsaturated (saturated) coorbital torques. 
         \textit{Long-dashed line}: 
         Consistent with Type~II migration,
         the line has slope $-1.5\,\nu/a$ and passes through
         $a\approx 0.8\,a_{0}$ when $\Mp\approx0.9\,\MJup$.
 \textit{Right}: Average disk density near the planet relative 
         to the local initial value as a function of time.
         The density is averaged over a band of radial width 
         $2\,H$ centered on the orbit of the planet (see text 
         for details). 
         Solid circles mark times when the mass ratio $\Mp/\Ms$ 
         is equal to $5\times10^{-5}$ ($\Mp=16.7\,\MEarth$) and
         when it is an integer multiple of $1\times10^{-4}$
         ($\Mp=33.3\,\MEarth$).
         }
\label{fig:a_sn}
\end{figure*}
We evaluate quantities $\Sigma_{p}$, $H$, and $\Omega_{p}$  
at the planet's orbital radius, $a$. Surface density 
$\Sigma_{p}=\Sigma_{p}(a)$ is evaluated according to its initial value 
$\Sigma_{p}(a) \propto (a_{0}/a)^{s}$, and so ignores evolutionary 
effects and tidal gap formation. 
The planet mass $\Mp$ is regarded as a function of time that 
we obtain from our simulations, via piecewise polynomial fits.
For the numerical models we consider, $s=-d\ln{\Sigma_{p}}/d\ln{a}=1/2$.
Equations~(\ref{eq:tanakadotaI}) and (\ref{eq:tanakadotaIsat}) are then 
solved numerically, providing the migration tracks 
$a_{\mathrm{I}}=a_{\mathrm{I}}(t)$. 

In the left panel of Figure~\ref{fig:a_sn}, we compare such tracks 
with outcomes from our simulations.
For the first $400$ orbits, while $\Rhill\lesssim 0.9\,H$ and 
$\Mp\lesssim 0.27\,\MJup$, the orbital radius (i.e., semi-major axis) 
evolution is in good agreement with the results of Type~I migration. 
The unsaturated coorbital torques appear to give a better fit than 
the saturated ones. But this is not always the case, as we see later 
when different disk parameters are considered. 
The right panel of Figure~\ref{fig:a_sn} plots the density
evolution of the gas near the planet, $\Sigma_{B}$, computed 
as ratio of the disk mass in the radial band 
$|r-a|/a\le H/r$ to the area of the band
($\Sigma^{0}_{B}$ is the local initial value of $\Sigma_{B}$).
It shows that the migration rate follows the Type~I tracks on 
the left while the disk density near the planet remains close 
to the local initial disk value, assumed in 
equations~(\ref{eq:tanakadotaI}) and (\ref{eq:tanakadotaIsat}). 
Up to a time of about $400$ orbits, the density near the planet 
is reduced below its local initial value by less than $20$\%. 
At time of about $600$ orbits, the density near the planet's 
orbit is reduced by about a factor of $3$, and we should expect 
the migration rates deduced from the simulation to be 
substantially slowed below the rates based on Type~I theory, 
in accord with the results on the left panel.
After about $1000$ orbits, when $\Mp\gtrsim 0.9\,\MJup$,
the migration rate in the simulation becomes comparable to the 
(local) viscous inflow rate (\textit{long-dashed line}). 
At this point, the disk density near the planet is depleted
by a factor of about $30$.

\begin{figure}
\centering%
\resizebox{\linewidth}{!}{%
\includegraphics{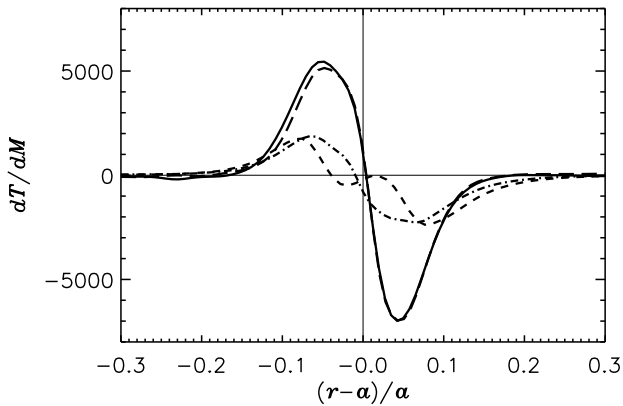}}
\caption{Torque per unit disk mass on the planet as a function of 
         normalized distance from the migrating and growing 
         planet plotted in Figure~\ref{fig:a_sn}
         ($\Sigma_{p}=3\times10^{-4}\,\densu%
         \approx 100\,\mathrm{g}\,\mathrm{cm}^{-2}$ at the planet's 
         initial orbital radius and $H/r=0.05$). 
         The vertical scale is in units of $G\Ms (\Mp/\Ms)^2/a$, 
         where $a=a(t)$.   
         The solid, long-dashed, dot-dashed, and short-dashed curves 
         refer to times when $\Mp=6.0\,\MEarth$, $9.3\,\MEarth$,
         $0.36\,\MJup$, and $1.0\,\MJup$, respectively.
         }
\label{fig:dTdM_sn_mig}
\end{figure}
The torque per unit disk mass as a function of distance from
the planet for the case in Figure~\ref{fig:a_sn} is plotted in 
Figure~\ref{fig:dTdM_sn_mig}. The plot shows very similar behavior
to the case of a stationary, nongrowing planet seen in
Figure~\ref{fig:dTdM}. Therefore, there is no evidence that planet 
migration or growth substantially affects the disk-planet torques
for these model parameters. In particular, there is no evidence
for strong coorbital torques.

\begin{figure*}
\centering%
\resizebox{\linewidth}{!}{%
\includegraphics{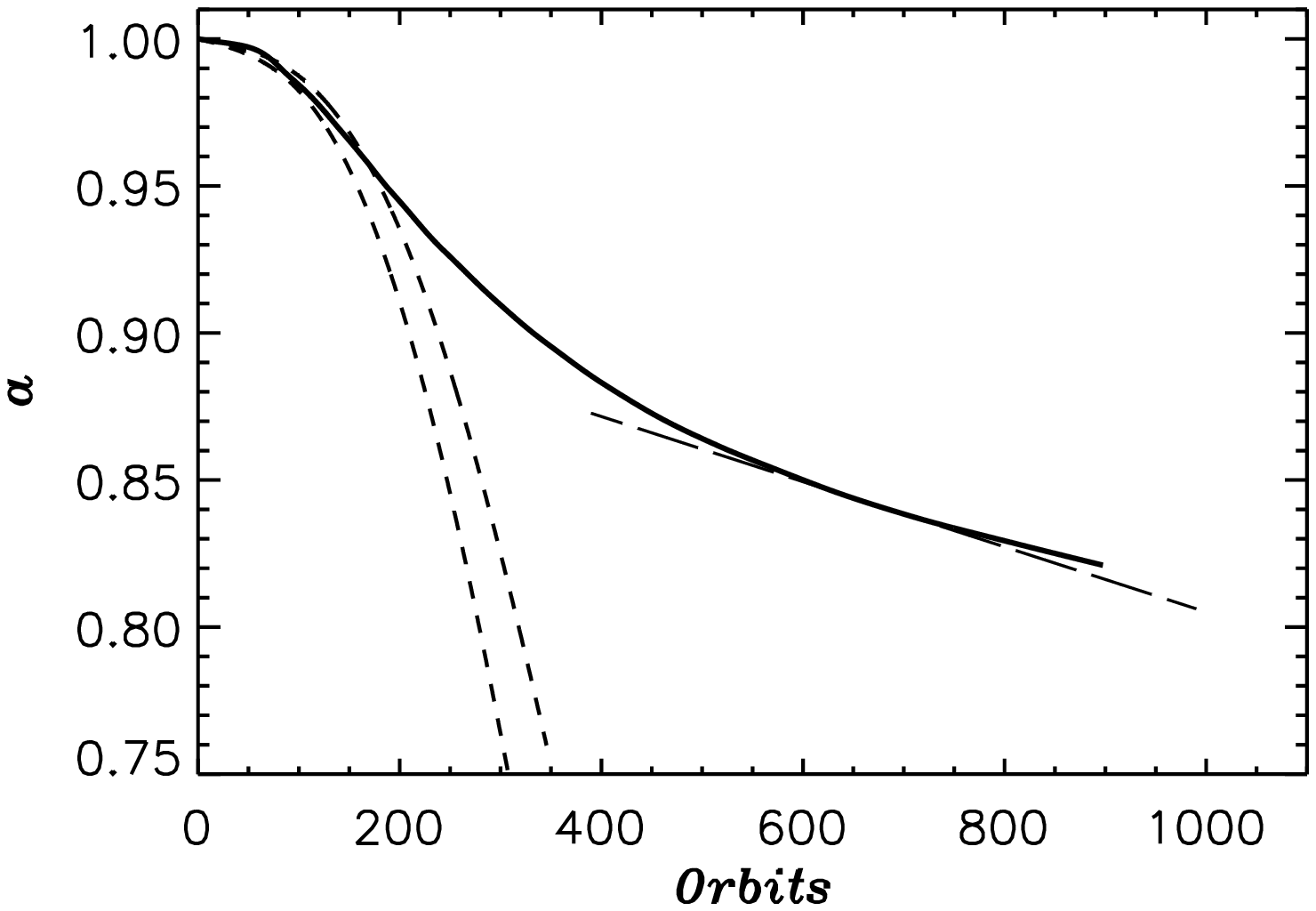}%
\includegraphics{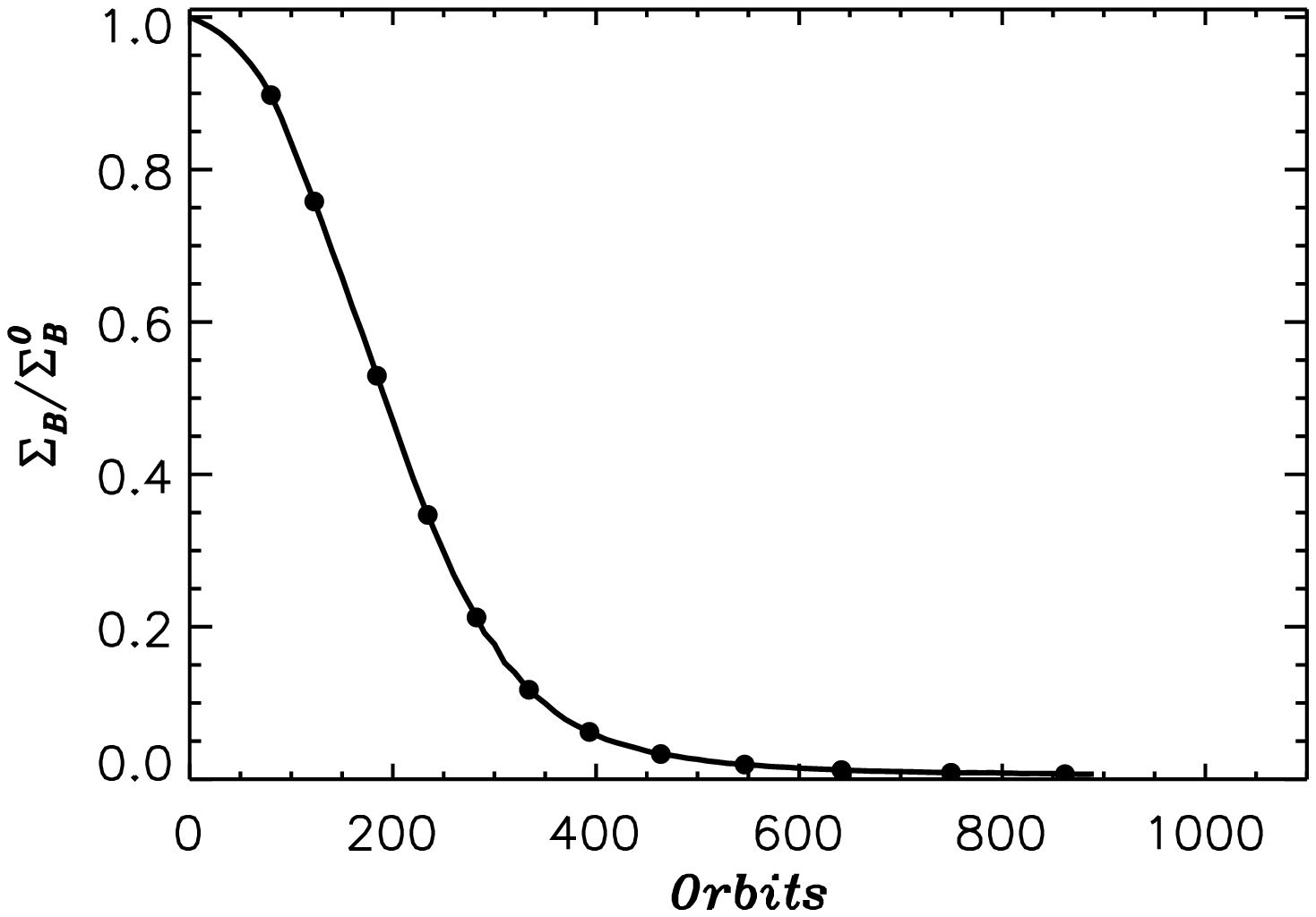}}
\caption{Orbital migration of a planet undergoing run-away gas
         accretion. Same as Figure~\ref{fig:a_sn}, but for a 
         cooler disk with aspect ratio  $H/r=0.04$ (the same  
         $\nu$ and initial $\Sigma_{p}$). 
 \textit{Left}: The theoretical Type~I migration tracks 
         (\textit{dashed curves}) use the mass evolution shown 
         as a long-dashed curve in Figure~\ref{fig:mp_hnu}. 
         As in Figure~\ref{fig:a_sn}, the upper (lower) 
         short-dashed curve is for unsaturated (saturated) 
         coorbital torques.
         The long-dashed line, representing Type~II migration, 
         has a slope equal to $-1.5\,\nu/a$ and passes through 
         $a\approx 0.85\,a_{0}$, when $\Mp\approx 0.9\,\MJup$.
 \textit{Right}: Normalized disk density near the planet as 
         a function of time, as described on right panel of 
         Figure~\ref{fig:a_sn} (see also text).
         Solid circles mark times when $\Mp/\Ms$ is
         $5\times10^{-5}$ ($\Mp=16.7\,\MEarth$) or an integer 
         multiple of $1\times10^{-4}$ ($\Mp=33.3\,\MEarth$).
         }
\label{fig:a_sn_h}
\end{figure*}
The results obtained from a model with $H/r=0.04$ (i.e., with a 
lower disk temperature compared to the model in 
Fig.~\ref{fig:a_sn}) are shown in the left panel of 
Figure~\ref{fig:a_sn_h}. 
As in the case of the warmer disk, the Type~I migration tracks 
(\textit{short-dashed curves}) reproduce reasonably well the 
radial migration from the simulation (\textit{solid line}) while 
$\Mp\lesssim 0.14\,\MJup$ (see \textit{long-dashed line} in 
Fig.~\ref{fig:mp_hnu}) or $\Rhill\lesssim 0.9\,H$.  
As before, the right panel of Figure~\ref{fig:a_sn_h} shows that
the migration rate follows the Type~I tracks while the disk density 
near the planet remains close to the local unperturbed value.
Again, when $\Mp\gtrsim 0.75\,\MJup$, 
$\Sigma_{B}/\Sigma^{0}_{B}\lesssim 0.03$ and $|da/dt|$ is on the 
order of the viscous inflow velocity (\textit{long-dashed line}).

\begin{figure*}
\centering%
\resizebox{\linewidth}{!}{%
\includegraphics{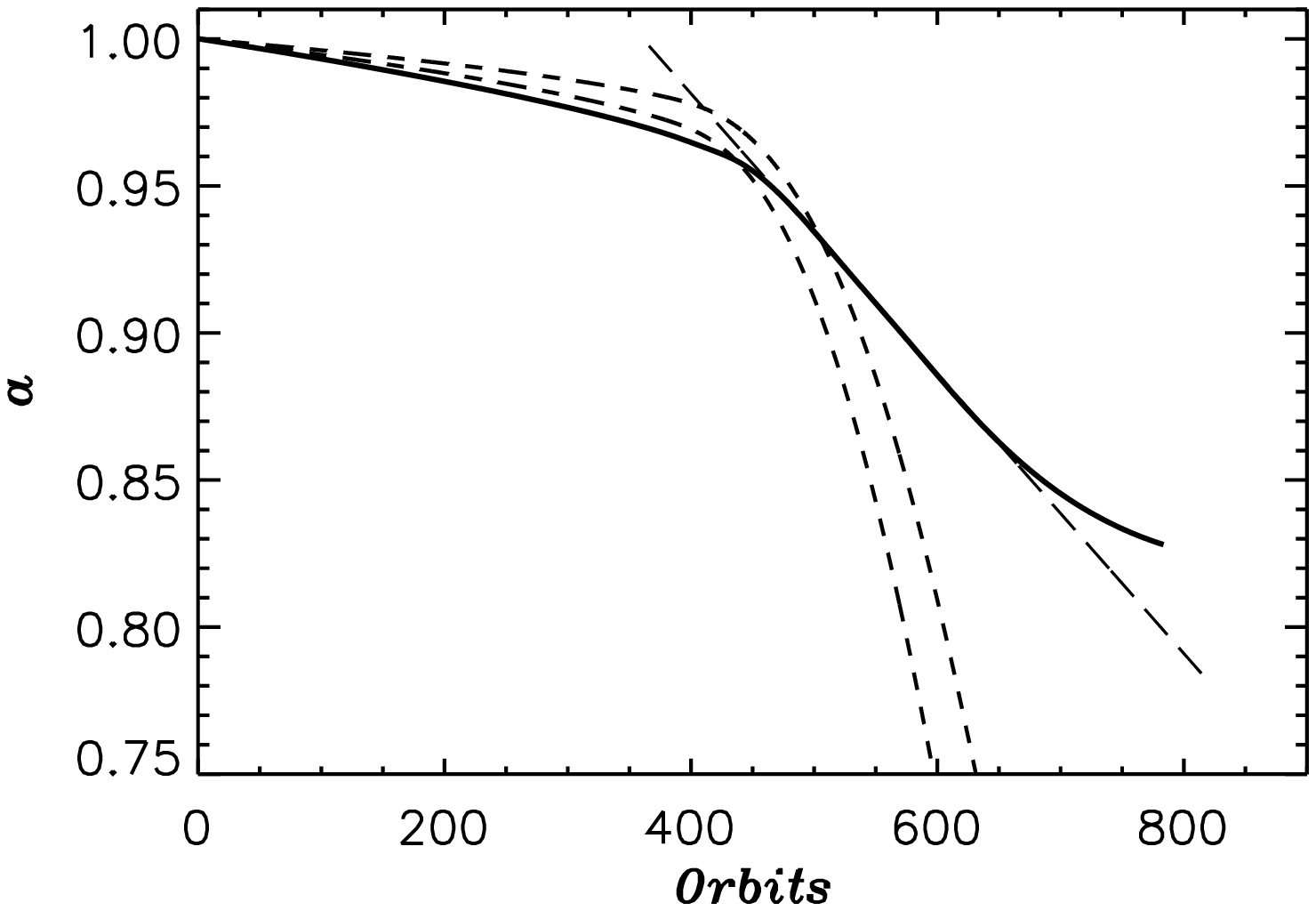}%
\includegraphics{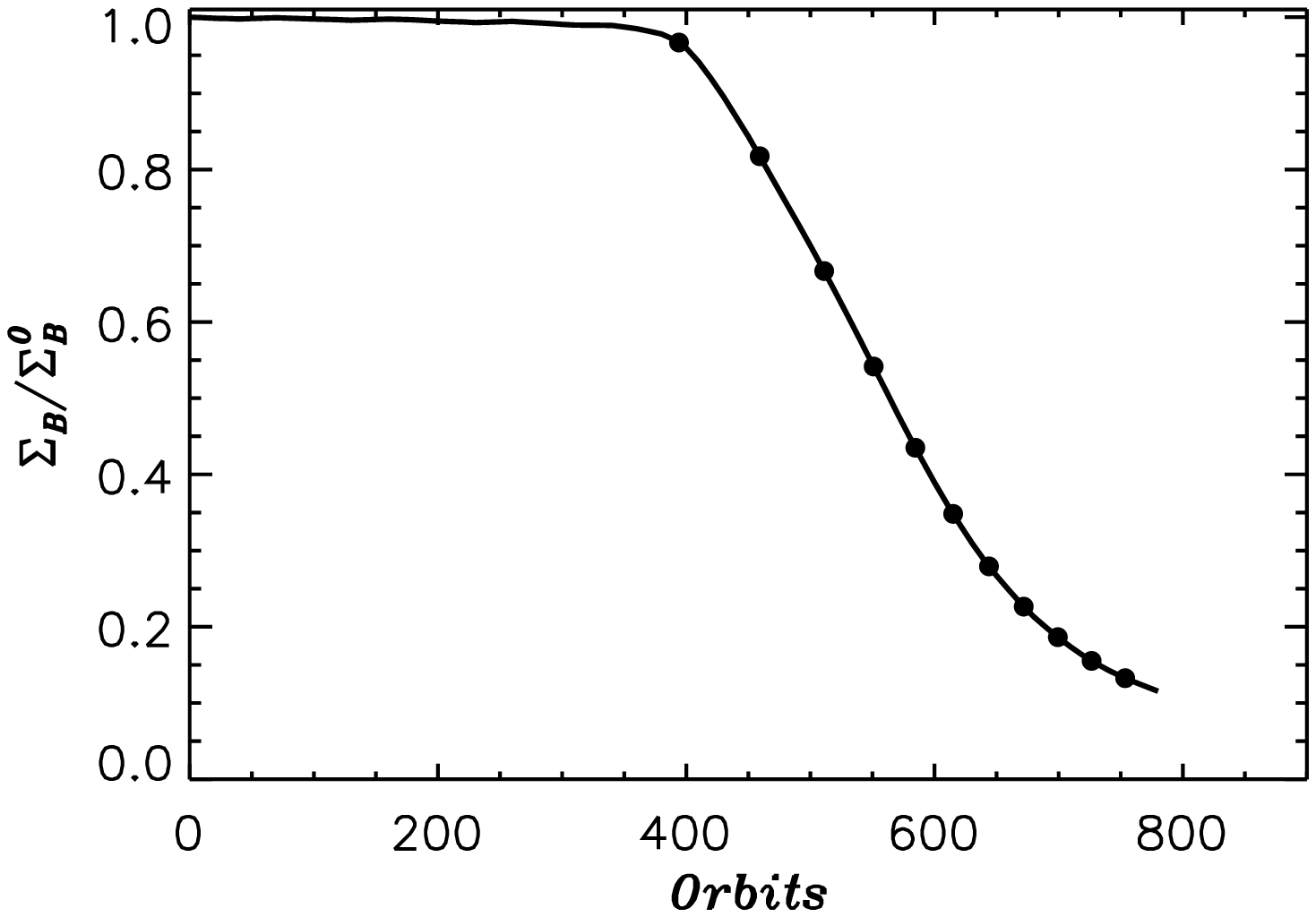}}
\caption{%
 \textit{Left}: 
         Same as left panel of Figure~\ref{fig:a_sn}, but for 
         a disk with ten times the turbulent kinematic viscosity
         ($\nu=1\times10^{-4}\,\viscu$ or $\alpha=0.04$), same 
         $H/r$, and initial $\Sigma_{p}$. 
         As in Figure~\ref{fig:a_sn}, the upper 
         (lower) short-dashed line is for unsaturated (saturated) 
         coorbital torques, using the mass evolution shown as a 
         short-dashed curve in Figure~\ref{fig:mp_hnu}.
         The long-dashed line representing Type~II migration has 
         a slope equal to $-0.7\,\nu/a$. 
 \textit{Right}: Average disk density near the planet relative 
         to the local initial (unperturbed) value as a function 
         of time, as in the right panel of Figure~\ref{fig:a_sn}.
         Solid circles mark times when $\Mp/\Ms$ is $5\times10^{-5}$
         ($\Mp=16.7\,\MEarth$) or an integer multiple of
         $2\times10^{-4}$ ($\Mp=66.6\,\MEarth$).
         }
\label{fig:a_sn_nu}
\end{figure*}
The dependence of migration on viscosity was investigated by 
running a simulation with kinematic viscosity 
$\nu=1\times10^{-4}\,\viscu$ ($\alpha=0.04$), ten times the value 
in Figure~\ref{fig:a_sn} with all other parameters being the same.
The results are shown in Figure~\ref{fig:a_sn_nu}.
The left panel shows the orbital migration from the simulation
as a solid curve and the Type~I migration based on 
equations~(\ref{eq:tanakadotaI}) and (\ref{eq:tanakadotaIsat}) 
as dashed curves. In this case, the relation $\Mp=\Mp(t)$ 
represented by a short-dashed line in Figure~\ref{fig:mp_hnu} 
is used in equations~(\ref{eq:tanakadotaI}) and 
(\ref{eq:tanakadotaIsat}).
The long-dashed line indicates a migration at a constant rate of 
$|\dot{a}|\approx 0.7\,\nu/a$, with $a\approx 0.9 \, a_0$. 
The long-dashed line passes through a range of masses that spans 
from $\approx 0.2\,\MJup$ to $\approx 1.2\,\MJup$.
However, at $\Mp\approx 1\,\MJup$ ($t\approx 600$ orbits), the 
density gap along the planet's orbit has not yet fully formed.
This can be observed on the right panel of 
Figure~\ref{fig:a_sn_nu}, which displays the averaged disk 
density near the planet normalized to the local unperturbed 
(initial) disk value. 
There is a drop of only a factor of $2.5$ in the disk 
density near the planet by the time $\Mp\approx 1\,\MJup$. 
The reason is that one of the conditions for steady-state 
gap formation,
$\Mp/\Ms>40\,\nu/(a^2 \Omega)\sim 4\times10^{-3}$ 
\citep{lin1993}, is not fulfilled in this higher viscosity 
case until $\Mp\gtrsim 4\,\MJup$.
At about $780$ orbits, $\Sigma_{B}/\Sigma^{0}_{B}\sim 0.1$
but the planet mass has reached beyond $2\,\MJup$ and is 
therefore more massive than the local disk mass. At those 
stages of the orbital evolution, inertia effects and further 
gap clearing are likely playing an important role in reducing 
the migration rate, as demonstrated in the next paragraph.

\begin{figure}
\centering%
\resizebox{\linewidth}{!}{%
\includegraphics{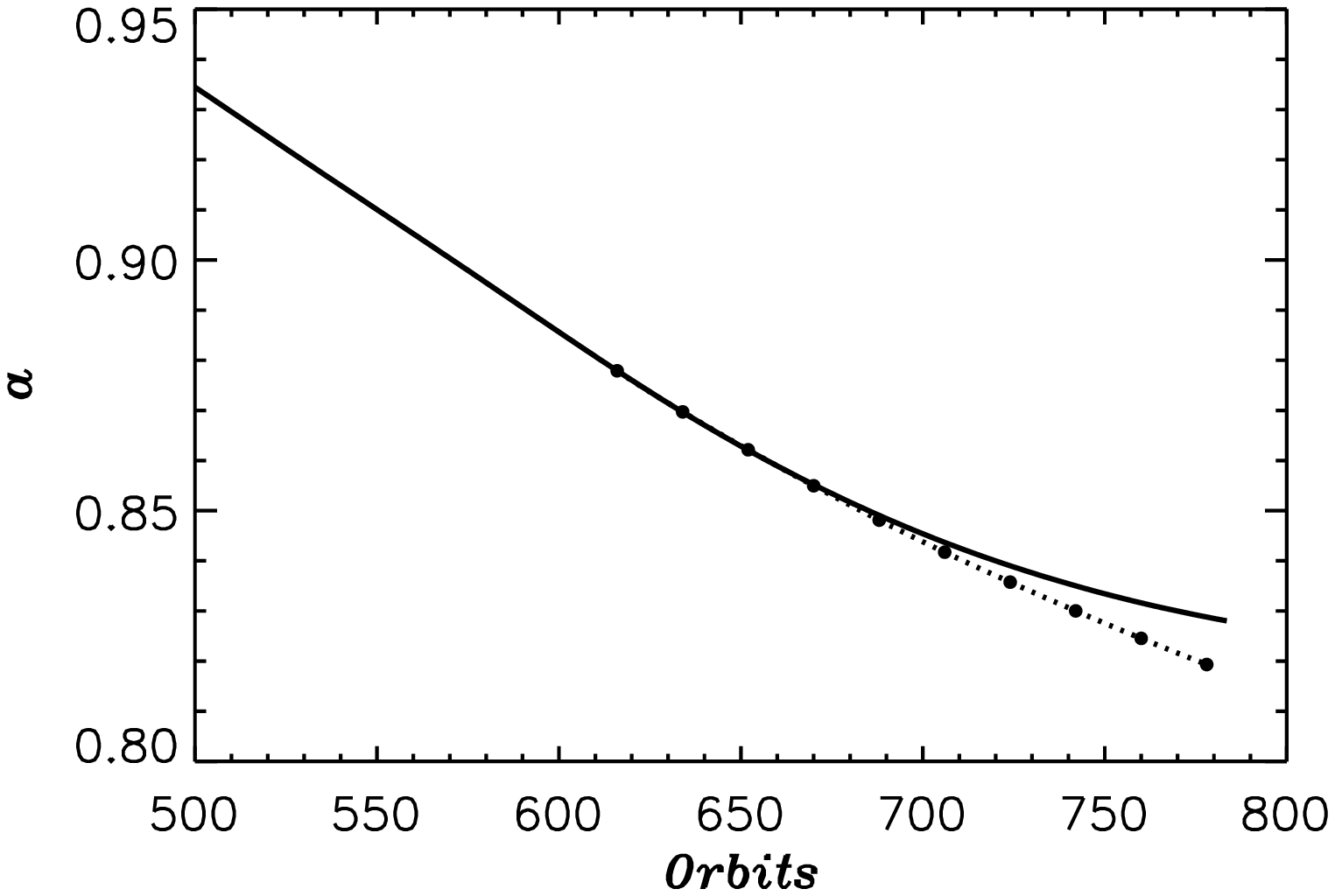}}
\caption{Comparison of radial migration obtained from the 
         simulation on the left panel of Figure~\ref{fig:a_sn_nu}
         (\textit{solid line}) with that obtained from a 
         similar three-dimensional simulation 
         (\textit{dotted line with solid circles}) 
         with a fixed mass planet $\Mp=1\,\MJup$ (see 
         text for further details).
         }
\label{fig:a_sn_nu_cmp}
\end{figure}
Figure~\ref{fig:a_sn_nu_cmp} displays a comparison of the orbital 
radius evolution from two calculations. 
The solid line is the same as that in the left panel of 
Figure~\ref{fig:a_sn_nu}. 
The dotted line with solid circles is the outcome of a 
three-dimensional simulation
in which the planet mass is fixed at $\Mp=1\,\MJup$.  
Material is removed from the vicinity of the planet according to 
usual the procedure we apply (see section~\ref{sec:GasAccretion}), 
but in this case it is not added to the mass of the planet.
The planet's orbit is held fixed for the first $100$ orbital 
periods, after which time it is allowed to evolve under the 
action of disk torques.  
The plot shows that there is general agreement, while 
$\Mp\sim 1\,\MJup$, with the variable mass model and that
the effect of adding mass to the planet in this regime is 
to slow its migration rate.

The local viscous timescale, $t_{\nu}=r^2/\nu$, in the models 
presented in Figures~\ref{fig:a_sn_nu} and \ref{fig:a_sn_nu_cmp} 
is about $1600$ orbital periods at $r=a_{0}$. 
Therefore, one might wonder whether the viscous evolution of 
the disk at radii larger than the outer grid boundary has 
any significant impact on the orbital evolution of the planet.
We address this issue in Appendix~\ref{sec:MVV} and show that
extending the disk further out at larger radii does not affect 
the migration tracks shown in Figures~\ref{fig:a_sn_nu} 
and \ref{fig:a_sn_nu_cmp}.
In Appendix~\ref{sec:MVV}, we also present results for cases
with viscosity parameter $\alpha=0.2$ (kinematic viscosity 
$\nu=5\times10^{-4}\,\viscu$) that have $t_{\nu} \simeq 320$ 
orbits at $r=a_{0}$. This case also leads to inward migration 
that can be interpreted as a Type~I regime, partially modified
by the perturbed surface density of the disk.

\section{Type~III Migration}
\label{sec:typeiii}

Figures~\ref{fig:a_sn} and \ref{fig:a_sn_h} indicate that
a growing planet undergoes Type~I migration, as long the 
disk density near the planet remains undepleted. 
At higher planet masses where the gap opening sets in, 
there is a smooth transition towards Type~II migration with 
migration speeds that are on the order of the viscous inflow 
velocity. There is no evidence for another form of migration,
since the torque distributions are essentially the same in the 
migrating and nonmigrating cases explored thus far (compare 
Figures~\ref{fig:dTdM} and \ref{fig:dTdM_sn_mig}). 
Type~III migration was suggested to involve coorbital material
that provides a fast form of migration (MP03).
In this section we discuss planet migration for several variants 
on the models of section~\ref{sec:GrowingMigratingPlanets} that 
should be favorable for a Type~III regime of migration. 
We describe a case that appears to exhibit Type~III migration.

\subsection{Higher Disk Mass}
\label{sec:highdiskmass}

\begin{figure*}
\centering%
\resizebox{\linewidth}{!}{%
\includegraphics{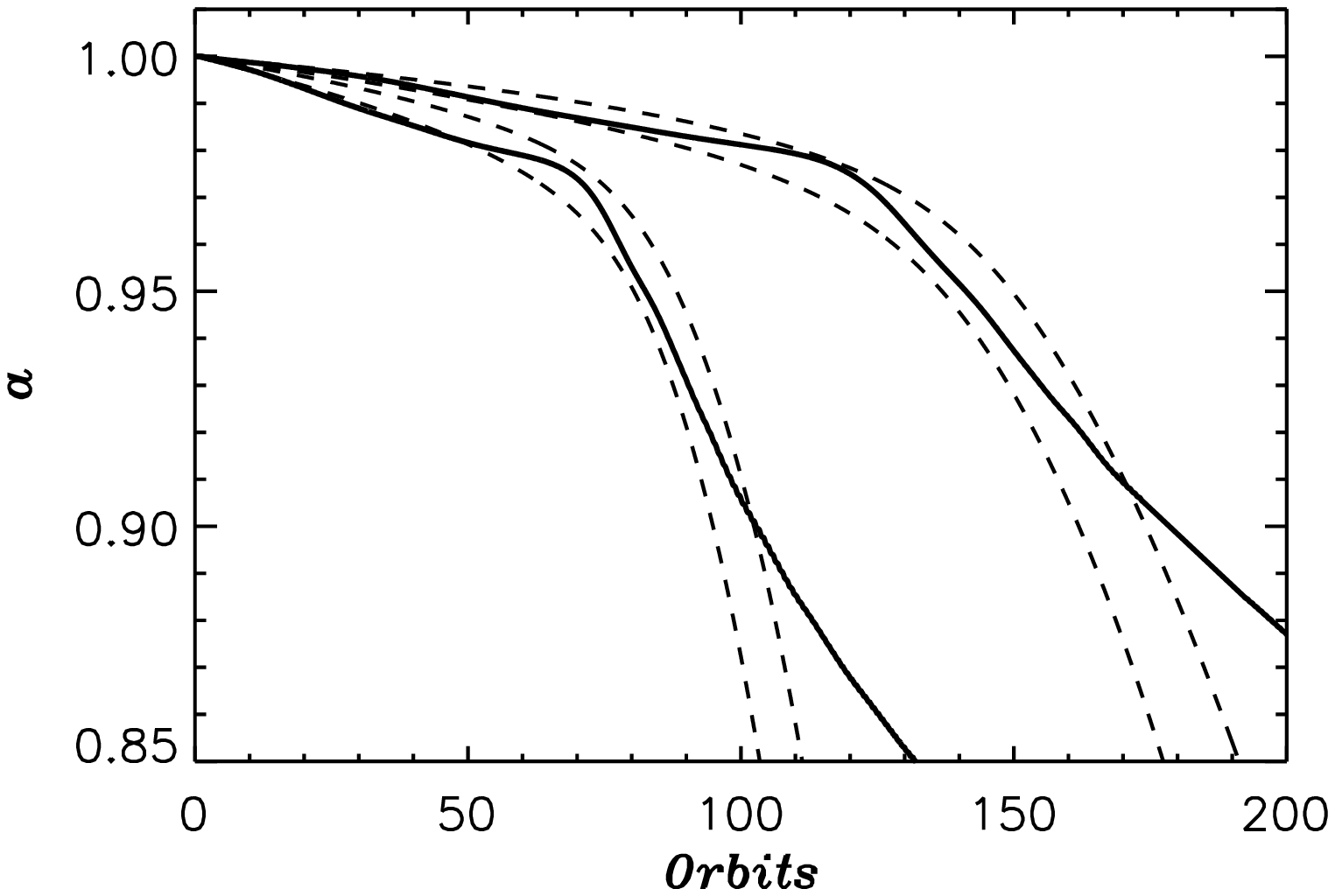}%
\includegraphics{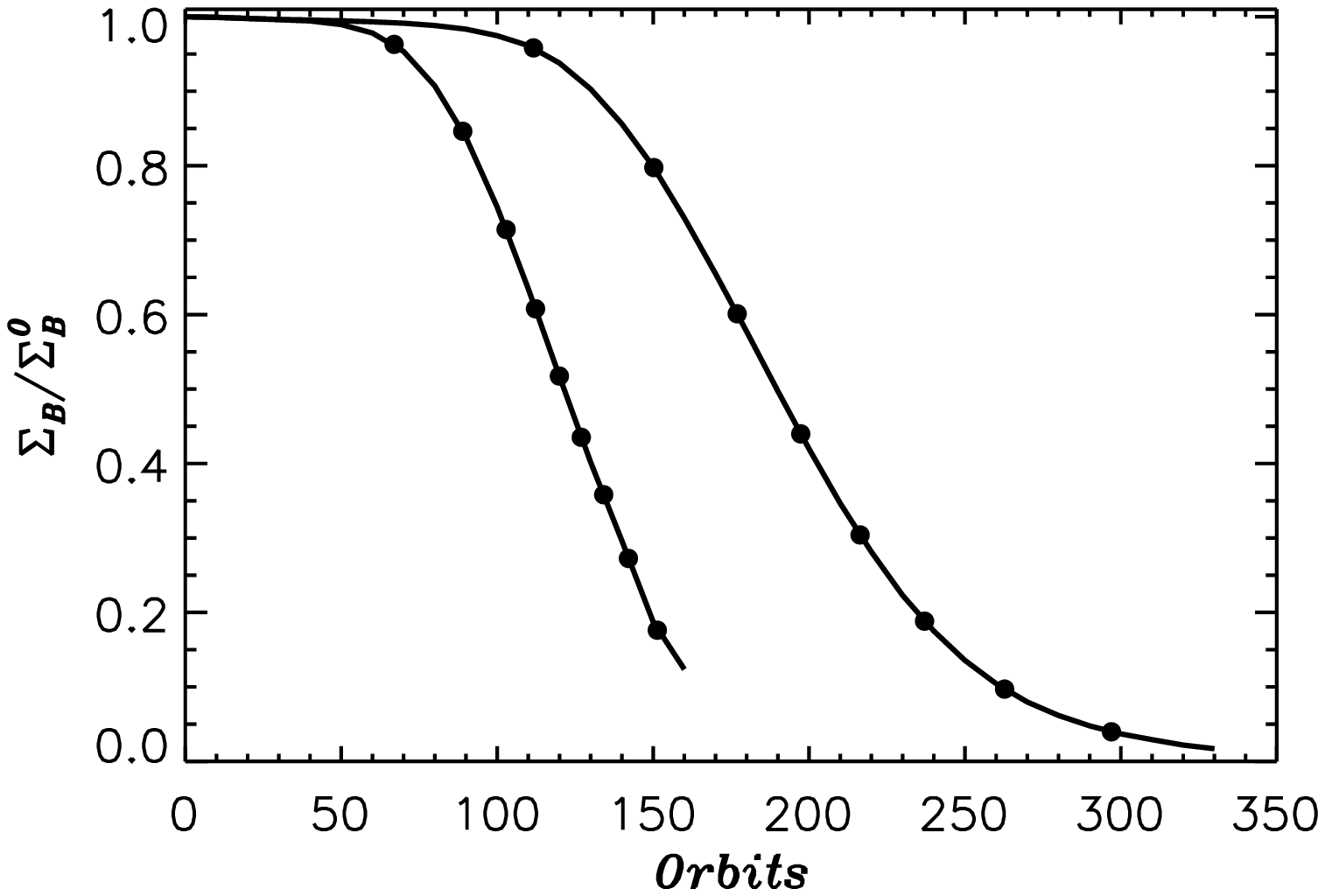}}
\caption{
 \textit{Right}:
         Orbital evolution under the same conditions as the 
         model in Figure~\ref{fig:a_sn}, but with higher disk 
         densities.
         \textit{Solid curves}: Simulation results for orbital 
         migration of a planet in a three-dimensional disk with
         initial surface density equal to 
         $\Sigma_{p}=9\times10^{-4}\,\densu$, or about 
         $300\,\mathrm{g}\,\mathrm{cm}^{-2}$ at $a_{0}=5.2\,\AU$
         (upper migration track), and
         $\Sigma_{p}=1.5\times10^{-3}\,\densu$, or about 
         $500\,\mathrm{g}\,\mathrm{cm}^{-2}$
         (lower migration track).
         \textit{Dashed curves}: Predicted orbital migration 
         according to 
         Type~I theory, equations~(\ref{eq:tanakadotaI}) 
         (\textit{upper curve} of pair for unsaturated coorbital 
         torques) and (\ref{eq:tanakadotaIsat})  
         (\textit{lower curve} of pair for saturated coorbital
         torques).
 \textit{Right}: Average disk density near the planet relative 
         to the local initial (unperturbed) value as a function 
         of time, as in the right panel of Figure~\ref{fig:a_sn}.
         Solid circles mark times when $\Mp/\Ms$ is $5\times10^{-5}$
         ($\Mp=16.7\,\MEarth$) or an integer multiple of
         $2\times10^{-4}$ ($\Mp=66.6\,\MEarth$).
         }
\label{fig:a_xsn}
\end{figure*}
Coorbital torques are stronger for higher mass disks. Masses in the 
coorbital region are on the order of $8\pi\,\Rhill\,a\,\Sigma(a)$. 
For the model presented in Figure~\ref{fig:a_sn}, involving disks of 
relatively low density, the coorbital disk mass is approximately equal 
to the planet mass when $\Mp\approx 0.2\,\MJup$.
We describe here results of three-dimensional calculations with
initial surface densities 
$\Sigma_{p}=9\times10^{-4}\,\densu\approx 300\,\mathrm{g}\,%
\mathrm{cm}^{-2}$ 
and
$\Sigma_{p}=1.5\times10^{-3}\,\densu\approx 500\,\mathrm{g}\,%
\mathrm{cm}^{-2}$
at the planet's initial orbital radius of $a_{0}=5.2\,\AU$.
The mass evolution in the former case is plotted as the dashed line in 
Figure~\ref{fig:mp_sig}. The mass evolution in latter case is similar, 
but the growth proceeds very rapidly reaching about $1\,\MJup$ within
$130$ orbital periods. 
The resulting orbital radius evolution for both simulations is plotted 
in Figure~\ref{fig:a_xsn} (\textit{left panel}) along with the average 
disk density near the planet normalized to the local unperturbed value
(\textit{right panel}).
For both cases presented in the figure, at earlier times 
($t\lesssim 170$ and $t\lesssim 100$ initial orbits, respectively),
the simulated migration rates are comparable to the Type~I rates. 
During that stage of the evolution, the coorbital region is 
more massive than the planet.
In the model with initial 
$\Sigma_{p}\approx 300\,\mathrm{g}\,\mathrm{cm}^{-2}$ at $5.2\,\AU$
(upper migration track in Fig.~\ref{fig:a_xsn}), for times 
$t\lesssim 170$ orbits ($\Mp\lesssim 0.3\,\MJup$)
the coorbital region mass to planet mass ratio is larger
than $2$. In the model with initial
$\Sigma_{p}\approx 500\,\mathrm{g}\,\mathrm{cm}^{-2}$  at $5.2\,\AU$
(lower migration track Fig.~\ref{fig:a_xsn}), for times
$t\lesssim 100$ orbits ($\Mp\lesssim 0.4\,\MJup$)
the ratio of coorbital region mass to planet mass is larger
than $3$.
However, during those stages, the results are generally consistent 
with the Type~I migration and some slowing at later times, with no 
indication of another form of migration.

\begin{figure*}
\centering%
\resizebox{\linewidth}{!}{%
\includegraphics{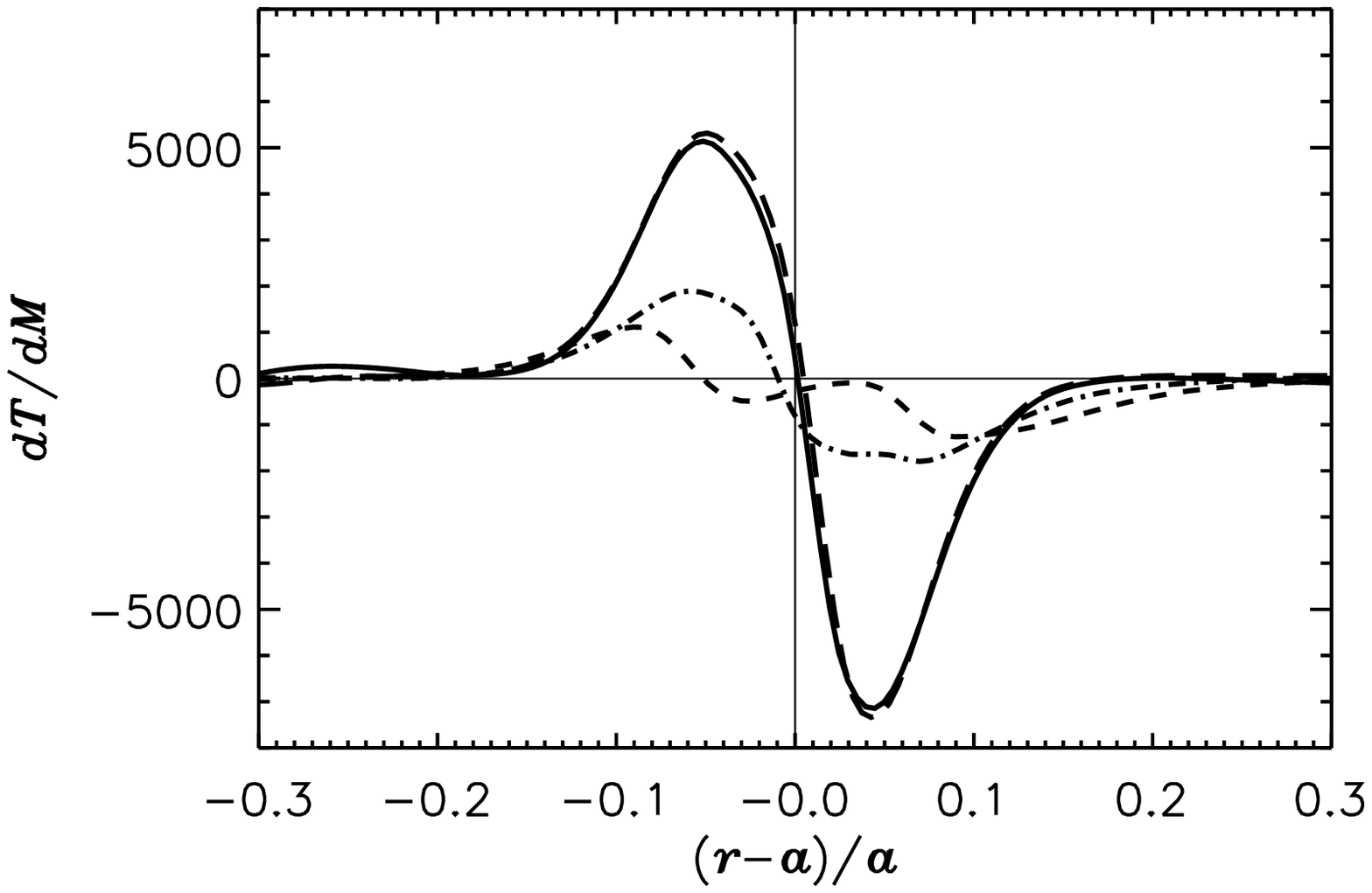}%
\includegraphics{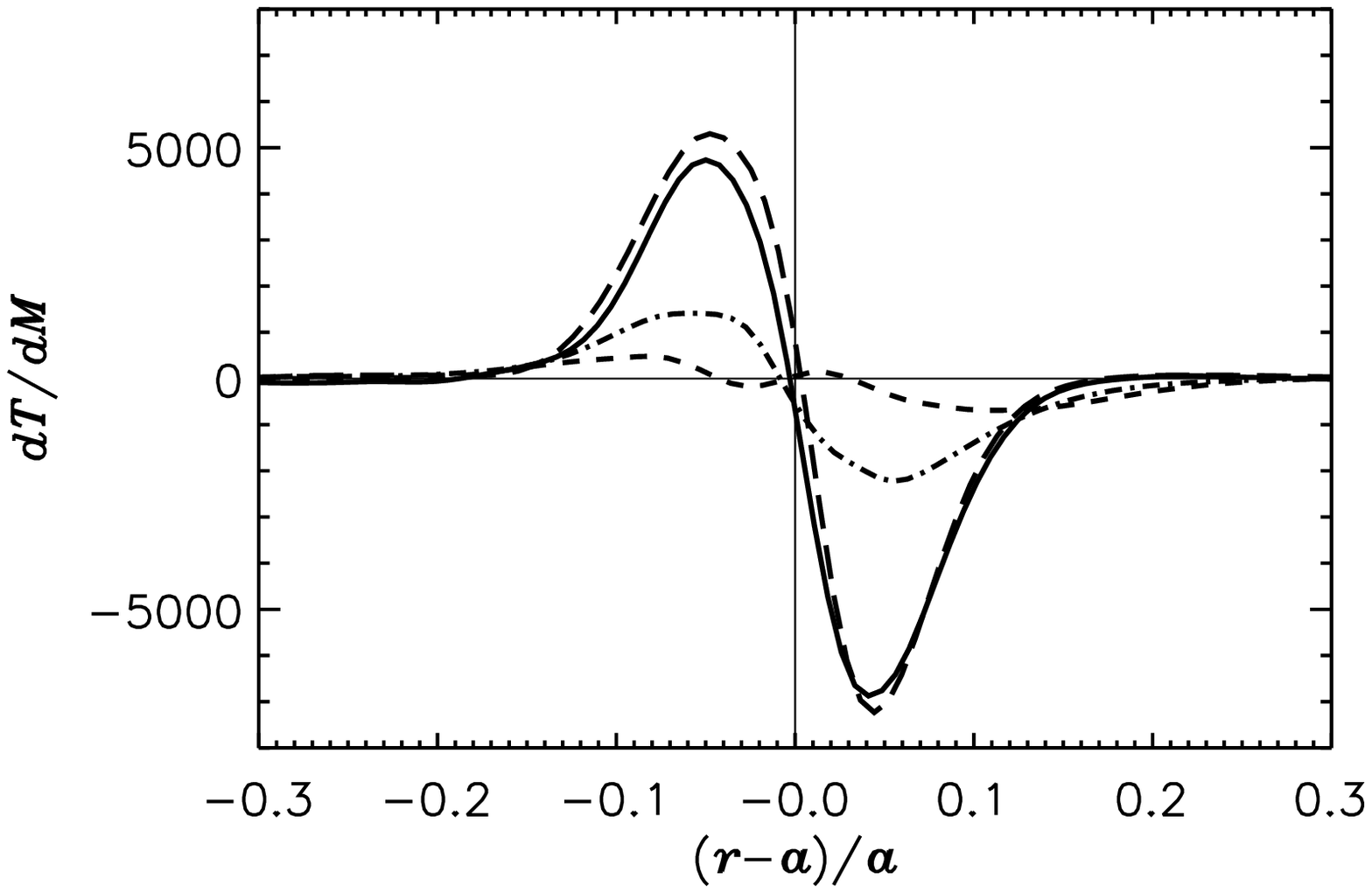}}
\caption{Torque per unit disk mass on the planet as a function of 
         normalized distance for the migrating and growing planets 
         plotted in Figure~\ref{fig:a_xsn}. 
 \textit{Left}: Case with initial surface density at the initial 
         orbit of the planet equal to 
         $\Sigma_{p} \approx 300\,\mathrm{g}\,\mathrm{cm}^{-2}$.
 \textit{Right}: Case with initial surface density at the initial 
         orbit of the planet equal to 
         $\Sigma_{p} \approx 500\,\mathrm{g}\,\mathrm{cm}^{-2}$.
         The vertical scale is in units of $G\Ms (\Mp/\Ms)^2/a$,
         where $a=a(t)$.
         The solid, long-dashed, dot-dashed, and short-dashed 
         curves refer to times when $\Mp=6.0\,\MEarth$, 
         $9.3\,\MEarth$, $0.36\,\MJup$, and $1.0\,\MJup$, 
         respectively.
         }
\label{fig:dTdM_xsn_mig}
\end{figure*}
To examine the situation in more detail, we plot the torque per unit 
disk mass as a function of distance from the planet in 
Figure~\ref{fig:dTdM_xsn_mig}. The plot shows very similar behavior
to the case of a nonmigrating, nongrowing planet seen in
Figure~\ref{fig:dTdM}, as well as to the case of a migrating, growing 
planet within a lower density disk presented in 
Figure~\ref{fig:dTdM_sn_mig}. Again, there is no evidence that planet 
migration or growth substantially affects the disk-planet torques for 
the parameters adopted in these models. Furthermore, there is no 
evidence for strong coorbital torques dominating planet's migration.

In carrying out calculations at higher disk masses, we have introduced
a possible inconsistency between the orbital motion of the disk and 
the planet. The orbital motion of the planet is affected by the 
axisymmetric gravitational force of the disk. On the other hand, the
motion of the disk near the planet is not affected by this force,
since disk self-gravity is ignored. This difference in rotation rates
can lead to an artificial increase in the planet migration rate 
\citep{pierens2005,baruteau2008}. This issue has some quantitative
effect on our results in this section. But, the qualitative results 
(approximately following the expectations of standard Type~I and II
theory) remain. We examine this issue further in 
Appendix~\ref{sec:diskgrav}.

\subsection{Higher Initial Planet Mass} 
\label{sec:highplanetmass}

We have shown that if a low mass protoplanet is allowed to rapidly 
grow in mass while it migrates, the orbital radius evolution begins
at the Type~I rate (eqs.~\ref{eq:tanakadotaI} and 
\ref{eq:tanakadotaIsat}) and approaches the Type~II migration rate 
as a clean gap develops. 
Since the evolving planet gains mass at the fastest possible rate, 
the run-away accretion rate, the time available for gap clearing
is relatively short. 
Such conditions should be favorable for migration dominated by 
coorbital torques. But as we saw in Figure~\ref{fig:a_xsn}, such 
situations only reveal Types~I  and II migration.
In this section, we explore a more extreme situation for providing 
coorbital material. We consider the case that a planet of higher 
initial mass (higher than the $5\,\MEarth$ considered thus far) 
is suddenly immersed in a smooth disk.  Gap clearing is then not 
initially present for the higher mass planets.
More coorbital gas is available for affecting migration.

\begin{figure}
\centering%
\resizebox{\linewidth}{!}{%
\includegraphics{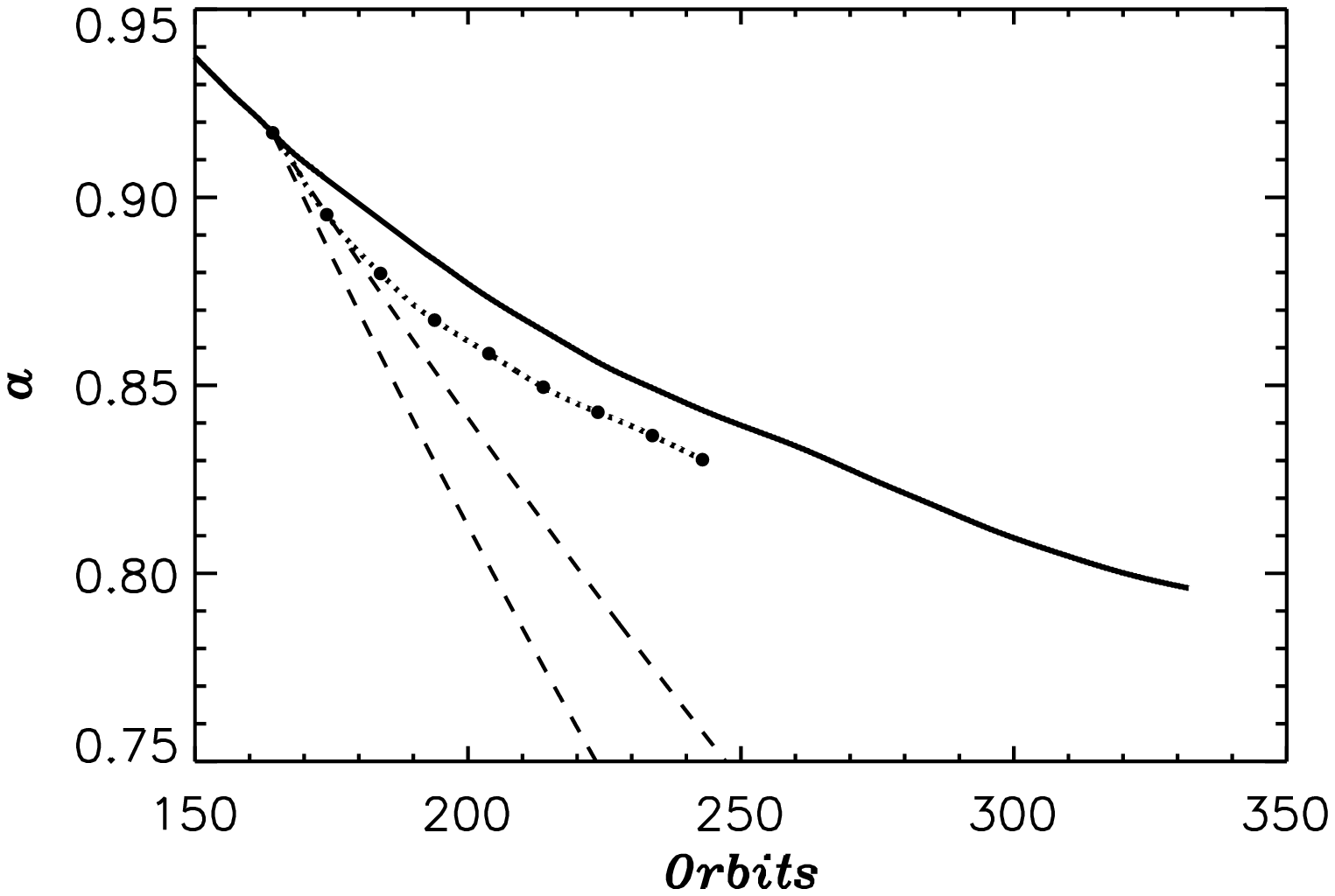}}
\caption{Migration with different initial conditions.
         \textit{Solid curve}:
         Orbital radius evolution of a planet with initial mass
         $\Mp=5\,\MEarth$ that interacts with a three-dimensional 
         disk having initial surface density at the planet's
         initial radial position
         $\Sigma_{p}\approx 300\,\mathrm{g}\,\mathrm{cm}^{-2}$ 
         at $a_{0}=5.2\,\AU$ (same as the upper migration 
         track plotted in Fig.~\ref{fig:a_xsn}).  
         It has mass $\Mp=0.3\,\MJup$ at a time of about $165$
         orbits (see Fig.~\ref{fig:mp_sig}, \textit{dashed line}),
         when $a\simeq 0.92\,a_{0}$.
         \textit{Dotted curve with solid circles}: 
         Orbital radius evolution of a planet with initial mass 
         $\Mp=0.3\,\MJup$ that interacts with the same initial 
         unperturbed disk density distribution as the solid 
         curve case has at time $t=0$. 
         The planet starts at the same radius ($a\simeq 0.92\,a_{0}$) 
         as the solid curve where that planet has acquired a 
         mass of $0.3\,\MJup$.
         The difference in the two cases is that the solid curve 
         case has a partially cleared gap when $\Mp=0.3\,\MJup$
         (see Fig.~\ref{fig:a_xsn}, \textit{right panel}), while
         the dotted curve case starts in a smooth unperturbed disk.
         \textit{Dashed curves}: Orbital radius evolution of a 
         planet according to Type~I theory (eq.~\ref{eq:tanakadotaI}
         and \ref{eq:tanakadotaIsat}) 
         for a planet of fixed mass $\Mp=0.3\,\MJup$ 
         (\textit{lower curve} of pair for saturated coorbital 
         torques) and disk density at $r=0.92\,a_{0}$ for the 
         unperturbed initial disk. 
         }
\label{fig:a_03sr}
\end{figure}
We consider a planet with initial mass $\Mp=0.3\,\MJup$ 
($\Mp/\Ms=3\times10^{-4}$) that is allowed to grow and migrate in 
a three-dimensional disk with initial density 
$\Sigma_{p}\approx 300\,\mathrm{g}\,\mathrm{cm}^{-2}$ at $a_{0}=5.2\,\AU$
(same as the lower initial density disk in Fig.~\ref{fig:a_xsn},
$H/r=0.05$, and $\nu=1\times10^{-5}\,\viscu$). 
Its orbital radius evolution is plotted as a dotted curve with
solid circles in Figure~\ref{fig:a_03sr}, together with the migration 
track of the model that starts with $\Mp=5\,\MEarth$ at $t=0$ 
(plotted as a \textit{solid curve}). 
For purposes of comparison, the initial orbital radius $a_0$ 
for the dotted curve case is chosen to be the $a$ value of 
the solid curve case when its planet mass is also $0.3\,\MJup$.
Given the large initial mass of the planet for the dotted curve 
case, the mass growth is very rapid: the planet gains about 
$0.7\,\MJup$ over the first $\sim 50$ orbits of evolution.
The figure shows that, with these disk conditions, migration
rates differ only for a brief period of time, but they soon 
converge to values compatible with orbital migration in the
more relaxed disk (compare slopes of \textit{solid} and 
\textit{dotted lines}).  
The dashed lines in the figure show that the planet initially 
migrates at the Type~I rate\footnote{Note that, for a constant
mass planet and $\Sigma_{p}\propto a^{-1/2}$, it follows that
$|\dot{a}_{\mathrm{I}}|\propto a$ (see eq.~\ref{eq:tanakadotaI}).}. 
But it later slows to nearly the same rate as the solid curve 
case. Again, there is no indication of Type~III migration.

\subsection{Nongrowing Planets in a Colder Disk}
\label{sec:NongrowingPlanets}

We consider the case of a nongrowing $0.3\,\MJup$ planet by 
removing gas mass near the planet without adding the mass of 
this material to the planet's mass.
This situation may mimic the effects of an efficient disk wind.
These models differ from those in MP03 and DBL05,
who considered nonaccreting planets, only with respect to the 
accretion boundary conditions near the planet and the time of 
planet release.

Unlike the mass removal case, the nonaccreting case may introduce 
a complication because of the buildup of gas within the planet's 
Hill sphere, which can become more massive than the planet. 
It has been argued that inertia effects from material close
to the planet could introduce complications in self-consistently
analyzing the dynamics of the system \citep{papa2007}. 
Appendix~\ref{sec:typeiii_app} describes some effects of 
the nonaccreting boundary condition.
To avoid this potential problem, we remove gas near the planet 
and ignore torques exerted by the gas on the planet within the 
inner half of the Hill sphere (by radius), where most of the 
bound gas resides (see Fig.~\ref{fig:boundtracers}).

We are interested in seeing whether these situations could give
rise to strong torques outside the Hill sphere that cannot be 
accounted for by Type~I or Type~II theory in the case of a planet
of fixed mass. As we show below, there are conditions under which
such strong torques occur in the coorbital region.

\subsubsection{Simulations Setup}
\label{sec:SimulationsSetup}

We consider a Saturn mass ($0.3\,\MJup$) planet in a disk having
$H/r=0.03$. The initial (unperturbed) disk surface density varies
as $\Sigma=\Sigma_{p}(a_0)\,(a_0/r)^{3/2}$, with 
$\Sigma_{p}(a_0)=2\times 10^{-3}\,\densu%
\approx 670\,\mathrm{g}\,\mathrm{cm}^{-2}$, 
and kinematic viscosity $\nu=1\times 10^{-5}\,\viscu$. 
Most of these calculations are carried out in two dimensions since
$\Rhill\simeq 1.5\,H$. However, we checked that results from 
three-dimensional models are in general agreement with those from
two-dimensional models 
(see \textit{dotted curves} in Fig.~\ref{fig:a_fast}).
Simulations in three dimensions use the grid system outlined in
section~\ref{sec:GrowingMigratingPlanets}. As in the 
three-dimensional case, the two-dimensional grid has a linear base 
resolution of $0.014\,a_0$. In the coorbital region around the 
planet, the linear resolution is $0.02\,\Rhill$. Since we intend to 
study some global properties of flow dynamics in the coorbital
region, three grid levels extend $2\pi$ in azimuth around the star. 
Convergence tests at these grid resolutions are presented in 
Appendix~\ref{sec:typeiii_app}.

As anticipated above, here we assume that some process removes
gas from the disk, according to the procedure detailed in 
section~\ref{sec:GasAccretion}, but that the planet mass remains 
constant. 
The migration rate of the fixed mass planet is
not substantially affected by the assumption that the gas is removed. 
In Appendix~\ref{sec:typeiii_app} we show that configurations
with a nonaccreting planet result in similar migration tracks. 
Hence, our conclusions would apply to nonaccreting planets as well.

In  MP03 and DBL05, the planet's orbital radius was 
initially fixed for over $470$ orbits, so that a time-steady disk gap
would form before it underwent migration. Here we reconsider that
configuration, but examine cases where the planet's initial orbital
radius is fixed for a shorter time, only $100$ orbits. 
This case is somewhat like that of section~\ref{sec:highplanetmass}
which has more gas in the coorbital region (\textit{lower curve} case
in Fig.~\ref{fig:a_xsn}), but instead has a fixed mass planet in a 
cooler disk. The following factors applied here should help increase 
the torques from the coorbital region: cooler disk, fixed planet mass,
higher disk density, and reduced time on initially fixed orbit. 

\subsubsection{Results}
\label{sec:typeiii_results}

\begin{figure}
\centering%
\resizebox{\linewidth}{!}{%
\includegraphics{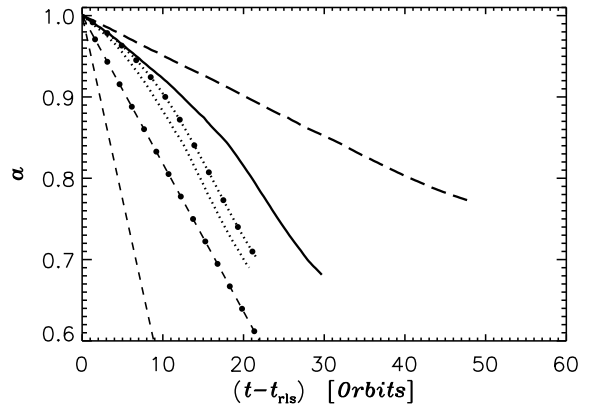}}
\caption{Orbital migration of a Saturn-mass planet of fixed mass
         ($\Mp=0.3\,\MJup$) in a cold ($H/r=0.03$) and high mass
         disk 
         ($\Sigma_{p}\approx 670\,\mathrm{g}\,\mathrm{cm}^{-2}$
         at the planet's initial position). Mass is removed from the
         disk near the planet to prevent a mass buildup there. 
         The planet is embedded in a two-dimensional disk and held 
         on a fixed orbit for $t_{\mathrm{rls}}=50$ 
         (\textit{dotted lines}), $100$ (\textit{solid line}), and
         $200$ (\textit{long-dashed line}) initial orbital periods.
         The dotted line with solid circles plots the migration track
         from a three-dimensional disk model with 
         $t_{\mathrm{rls}}=50$ orbits.
         The orbital radius is in units of $a_{0}$. 
         For $a_{0}=5.2\,\AU$, the unit of time is $\approx 12$ years.
         The predicted Type~I migration tracks, assuming the planet 
         does not open a gap, are plotted for a two-dimensional
         (\textit{short-dashed line}) and a three-dimensional
         (\textit{short-dashed line with solid circles}) disk.
         }
\label{fig:a_fast}
\end{figure}
\begin{figure}
\centering%
\resizebox{\linewidth}{!}{%
\includegraphics{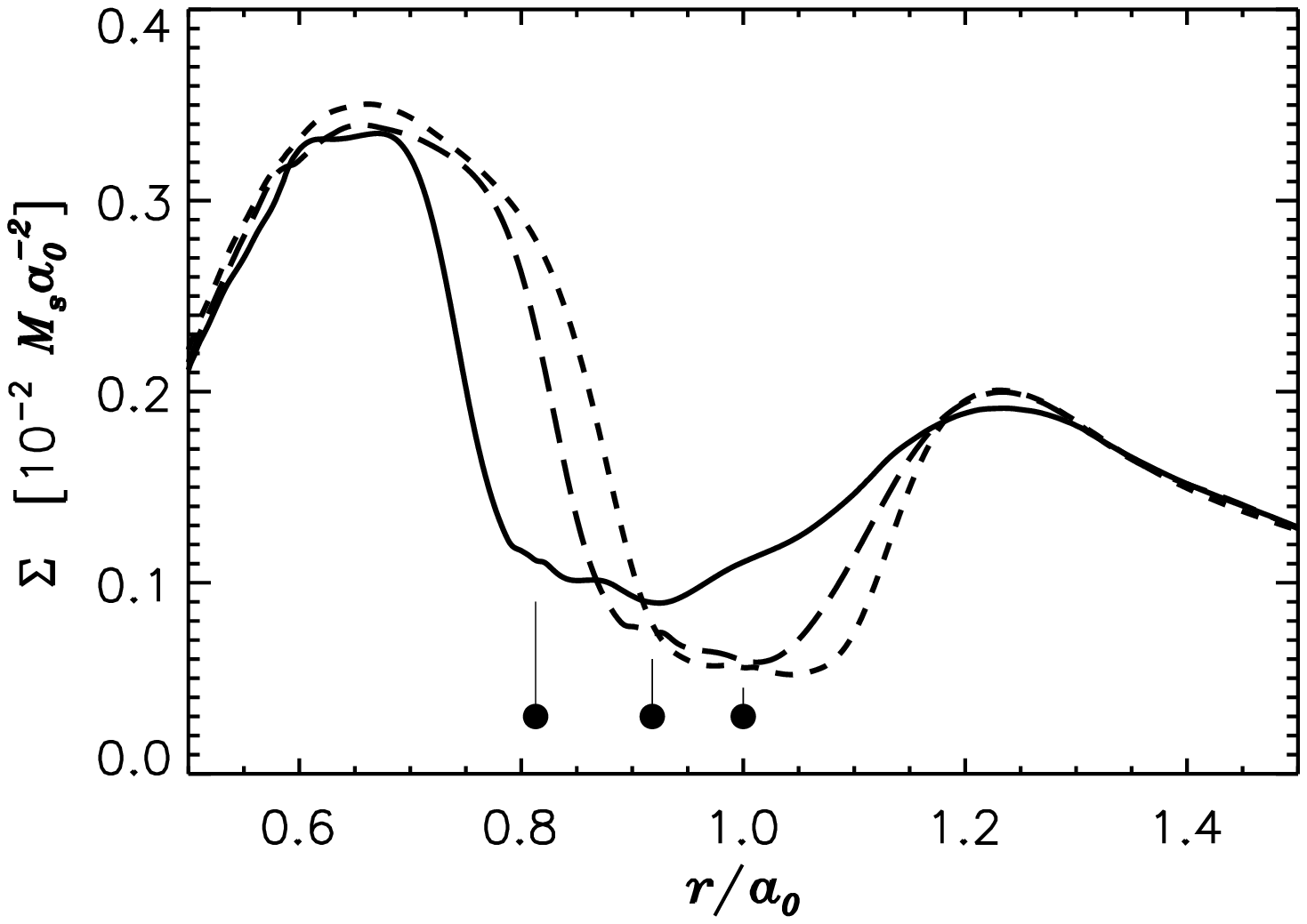}}
\caption{Axisymmetric radial density distribution $\Sigma(r,t)$,
         of a disk containing a Saturn-mass planet ($\Mp=0.3\,\MJup$),
         plotted as a function of radius at $3$ times: the time
         of planet release $t_{\mathrm{rls}}=100$ initial orbital
         periods (\textit{short-dashed line}), $t_{\mathrm{rls}}+10$ 
         initial orbital periods (\textit{long-dashed line}), and 
         $t_{\mathrm{rls}}+20$  initial orbital periods 
         (\textit{solid line}).  
         The solid circles mark the planet orbital radii at these
         times, as the planet migrates inward.
         }
\label{fig:sig_fast}
\end{figure} 
Figure~\ref{fig:a_fast} shows that the migration timescale of the 
planet is quite short and that it lengthens as the release time
increases (and the gap deepens). We focus on the case with
$t_{\mathrm{rls}}=100$ (\textit{solid line} case in the figure),
which has  a migration timescale of order $100$ initial orbital 
periods. Although short, this migration timescale is longer 
than the Type~I migration timescale that would be predicted if the 
planet did not open a gap (\textit{lower short-dashed curve} of 
pair\footnote{Note that, for a constant mass planet and 
$\Sigma_{p}\propto a^{-3/2}$, it follows that 
$\dot{a}_{\mathrm{I}}$ is a constant 
(see eq.~\ref{eq:tanakadotaI}).}). 
The planet does open a partial gap, as seen in 
Figure~\ref{fig:sig_fast}. So it might appear that a weakened form
of Type~I migration, due to partial gap opening, could explain the 
simulated migration rate. However, we demonstrate below that the 
migration cannot be explained by the usual Type~I theory.

\begin{figure}
\centering%
\resizebox{\linewidth}{!}{%
\includegraphics{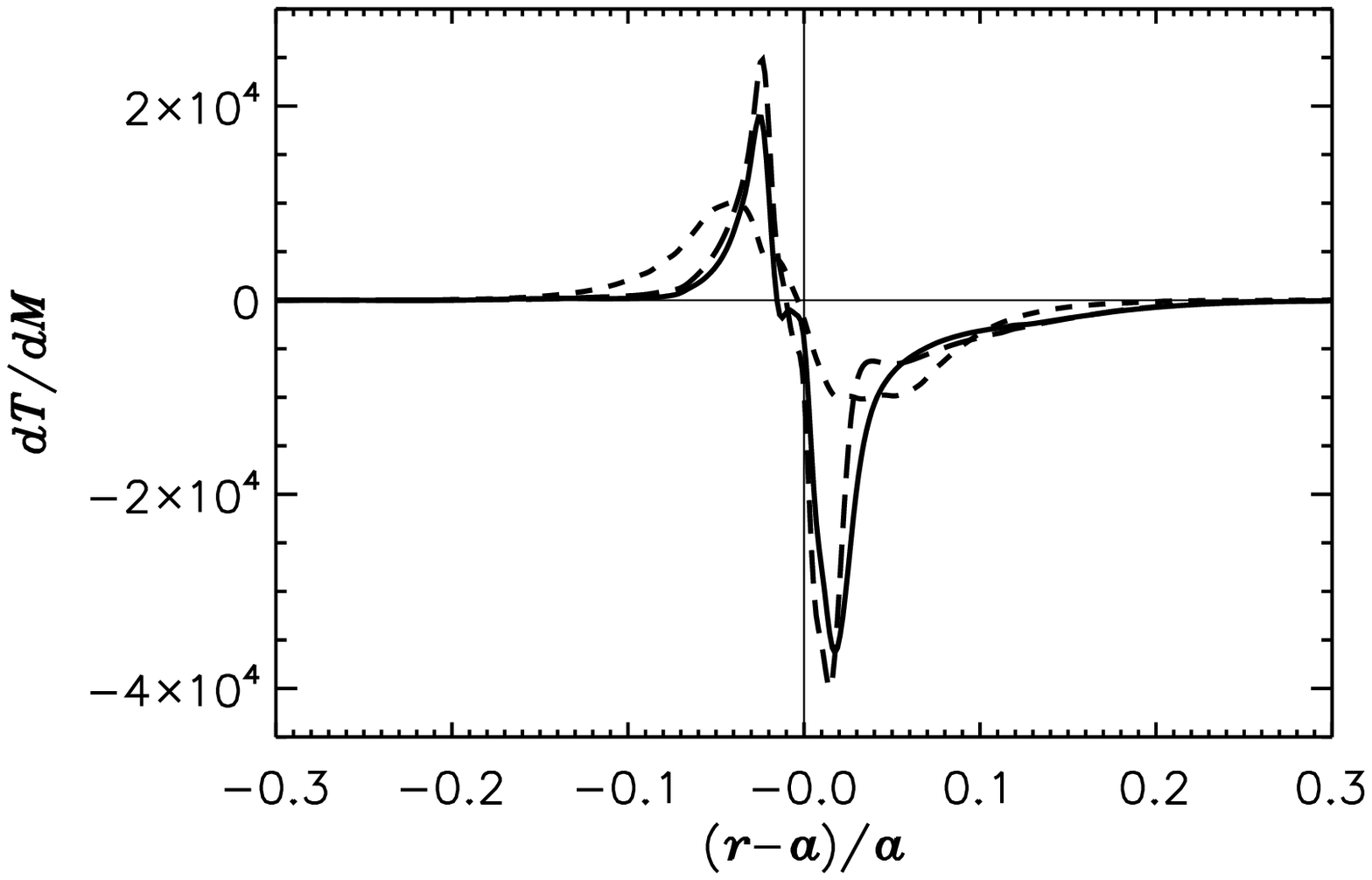}}
\caption{Torque per unit disk mass on a Saturn-mass planet 
         ($\Mp=0.3\,\MJup$) in units of $G\,\Ms\,(\Mp/\Ms)^2/a(t)$ 
         as a function of the normalized distance from the planet, 
         undergoing fast migration, at $3$ different times: 
         the time of planet release $t_{\mathrm{rls}}=100$ initial 
         orbital periods (\textit{short-dashed curve}), 
         $t_{\mathrm{rls}}+10$ initial orbital periods 
         (\textit{long-dashed curve}), and $t_{\mathrm{rls}}+20$
         initial orbital periods (\textit{solid curve}).
         } 
\label{fig:dTdM_fast}
\end{figure}
Figure~\ref{fig:dTdM_fast} shows the torque distribution per unit
disk mass exerted on the planet, as defined by equation~(\ref{eq:dTdM}).
The torque density distribution at the time of release of the planet,
$100$ orbits after the start of the simulation, reveals a curve 
characteristic of Type~I torques. The distribution is similar to 
the cases plotted in Figure~\ref{fig:dTdM}, although it is somewhat
larger in magnitude, as expected by the lower sound speed of the 
gas and the two (rather than three) dimensions of the simulation.
However, at later times the torque distribution changes character,
with much larger values in the coorbital zone, within radial 
distances of about $2\,\Rhill\simeq  0.1\,a(t)$ from the orbit of 
the planet. In particular, there is substantial torque occurring in 
the radial band $|r-a|<\Rhill$, where $\Rhill=0.046\,a$. 
We argued in section~\ref{sec:radoverlap} that this region involves 
only coorbital torques (not Lindblad torques).
We have verified that this torque is not originating from within the 
planet's Hill sphere (see Fig.~\ref{fig:a_fast_chk}). 
The contribution from within the Hill sphere is about $20$\% of the
net torque at release time and generally less than about $10$\% at 
later times.

Figure~\ref{fig:sig_fast} shows that the planet is migrating on a
shorter timescale than that of gap opening. At release, the planet 
is fairly symmetrically positioned in the gap. Later, the planet 
lies much closer to the inner edge of the gap than to the outer 
edge, and the gap is less deep. Such a situation would be 
expected to lead to slower or even outward migration according 
to Type~I theory of Lindblad resonances, since the inner 
resonances (which provide outward migration) are more strongly 
activated than the outer resonances (which provide inward 
migration) due to the asymmetric density distribution near 
the planet. If fact, the slowing/stalling of inward migration 
due to the feedback from the inward disk density of a migrating 
planet was envisioned by \citet{hourigan1984} and 
\citet{ward1989} in their consideration of the inertial limit 
to planet migration. 
To quantify the effects of standard Type~I torques, we apply 
the Type~I torque distribution taken at the time of planet 
release in Figure~\ref{fig:dTdM_fast}. In doing so, we are
ignoring pressure effects on $dT/dM$ due to the changing
gap shape.
We determine the Type~I torques at the later times by 
integrating this torque distribution (appropriately shifted 
to the instantaneous position of the planet) over the disk 
mass distributions in Figure~\ref{fig:sig_fast}. 
The torque is then given by 
\begin{equation}
T_{\mathrm{I}}(t) = 2 \pi\!\!%
                    \int\!\! \frac{dT}{dM}(x, t_{\mathrm{rls}})\,%
                             \Sigma(r,t)  \, r \, dr
\label{eq:dTdr_I}
\end{equation}
where $x=(r-a)/a$ and $a=a(t)$.
We find that the resulting Type~I migration rates at times 
of $10$ and $20$ orbits after release are \textit{outward} 
and equal to $\dot{a} = 2 \times 10^{-4}\, a_0\,\Omega_0$ 
and  $4 \times 10^{-5} a_0\,\Omega_0$, respectively.
Clearly, results from the simulation are not consistent 
with the expectations of the usual Type~I migration theory. 
Instead, we claim the effects of the corotation resonances 
are critical for migration here.
  
\begin{figure*}
\centering%
\resizebox{\linewidth}{!}{%
\includegraphics{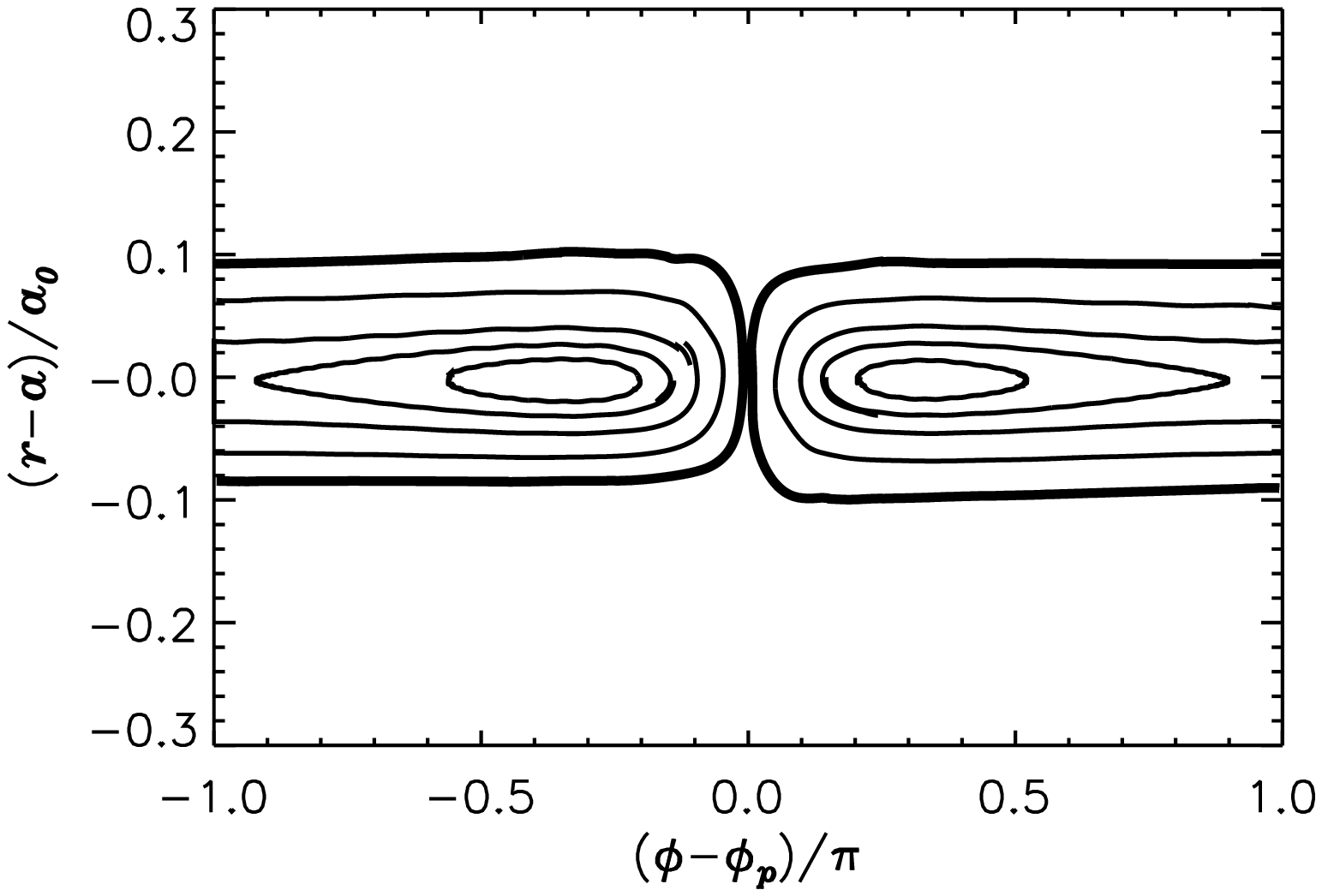}%
\includegraphics{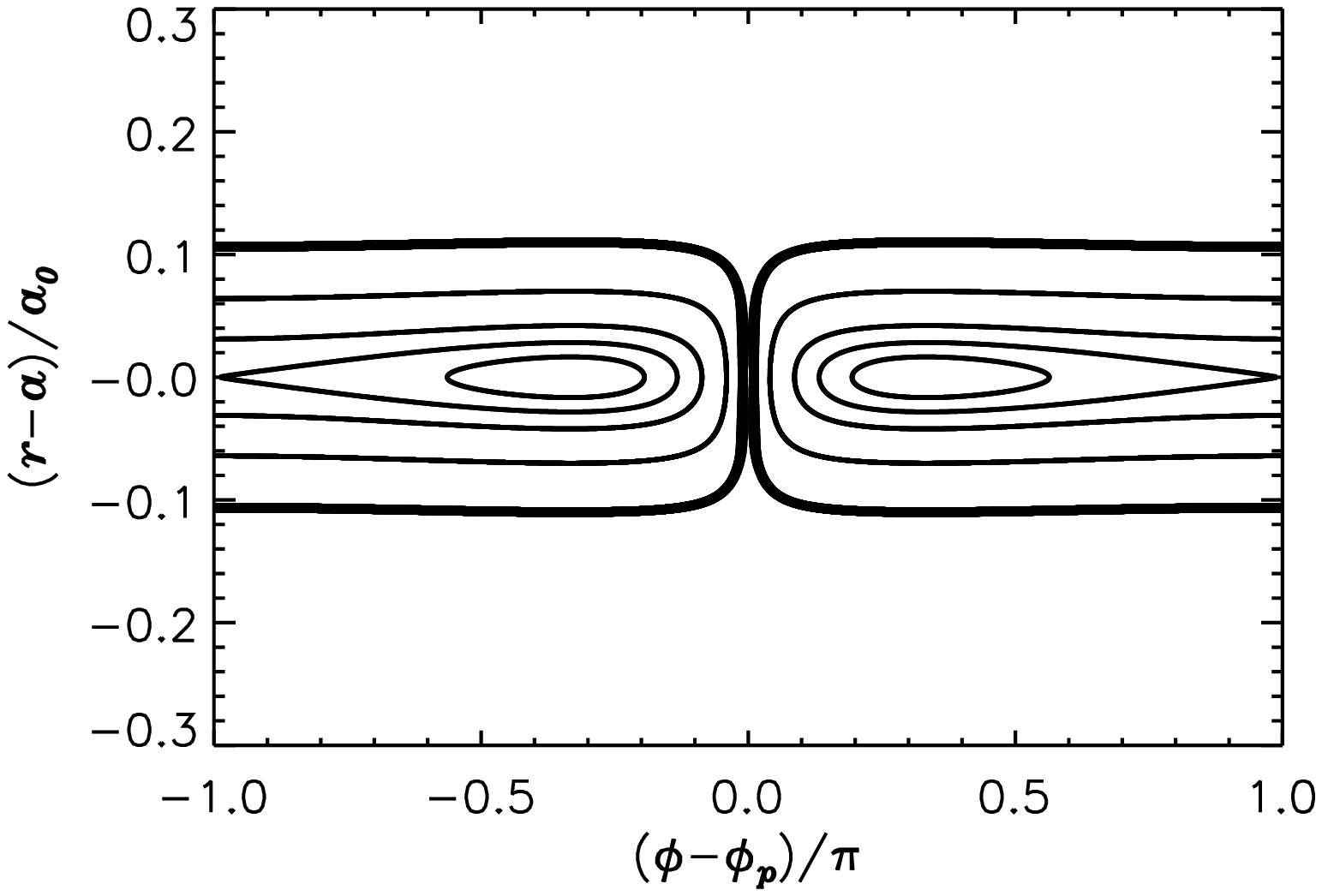}}
\caption{Trajectories of gas in the coorbital region near a 
         Saturn-mass planet ($\Mp=0.3\,\MJup$) on a fixed
         circular orbit with orbital radius $a=a_0$. 
         The trajectories are determined in the comoving frame
         of the planet, which is located at the origin. Disks 
         properties are the same as in Figure~\ref{fig:a_fast}.
         The left panel shows the results of our simulation 
         and the right panel shows results given by a 
         theoretical model (OL06). 
         }
\label{fig:str-nonmig}
\end{figure*}

\begin{figure*}
\centering%
\resizebox{\linewidth}{!}{%
\includegraphics{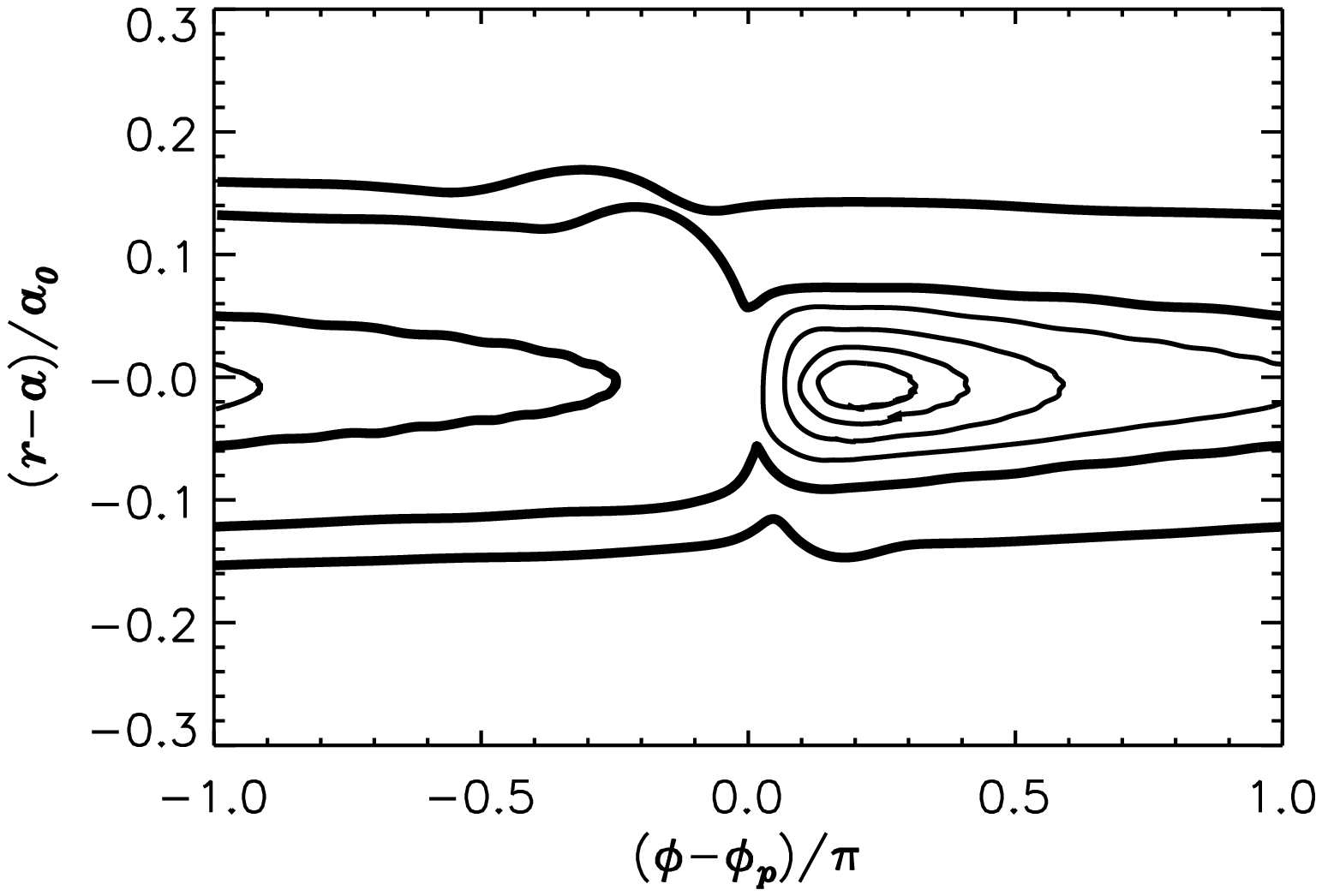}%
\includegraphics{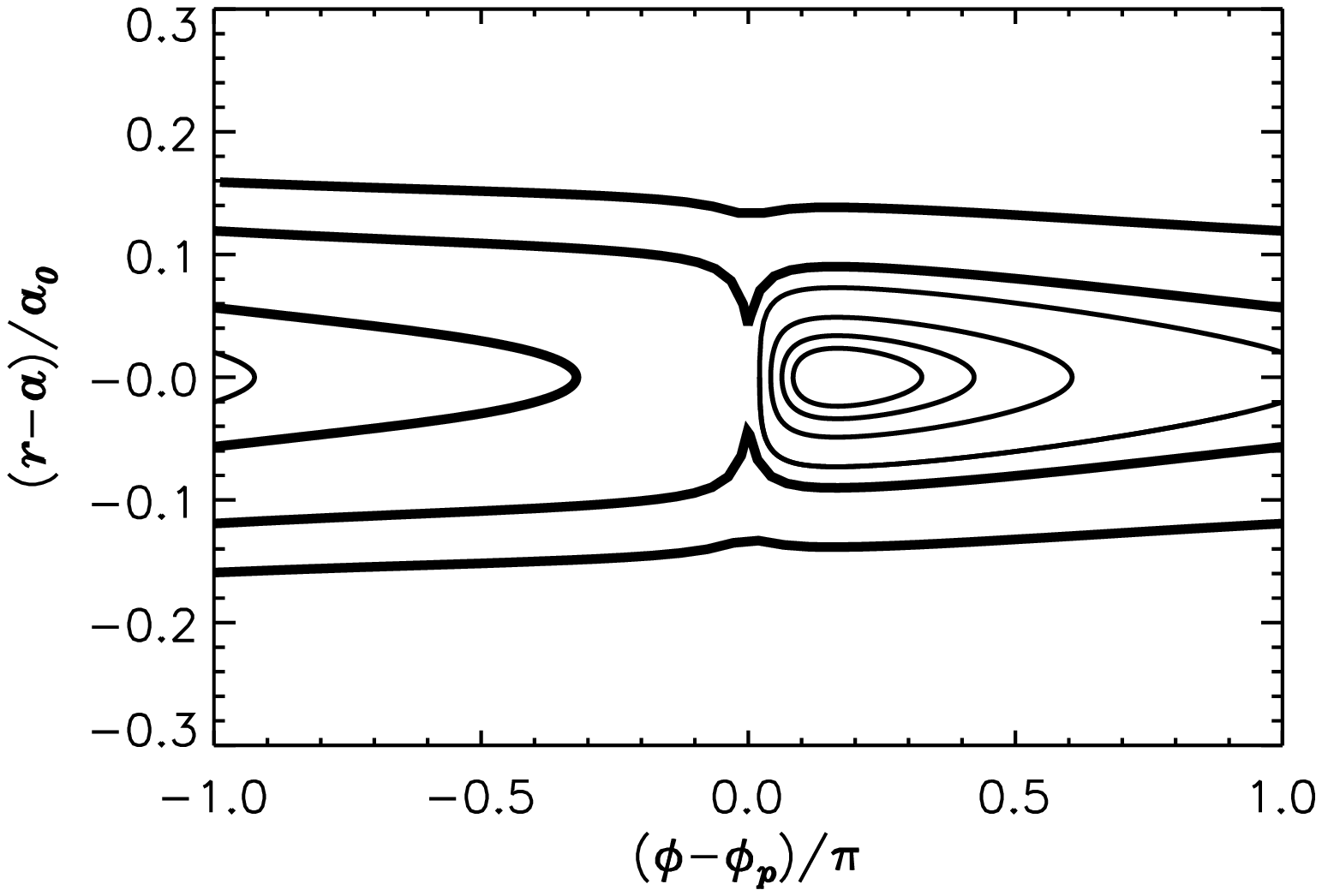}}
\caption{Trajectories of gas in the coorbital region near a
         Saturn-mass planet undergoing fast migration  
         (solid curve in Fig.~\ref{fig:a_fast}).
         The trajectories are determined in the comoving frame
         of the planet, located at the origin. 
         The thicker lines denote open trajectories that pass
         by the planet. The thinner lines denote closed 
         trajectories containing trapped gas.
         Results from the simulation are presented in the left 
         panel and results from a theoretical model (OL06) are
         displayed in the right panel.
         }
\label{fig:str-mig}
\end{figure*}

\begin{figure}
\centering%
\resizebox{\linewidth}{!}{%
\includegraphics{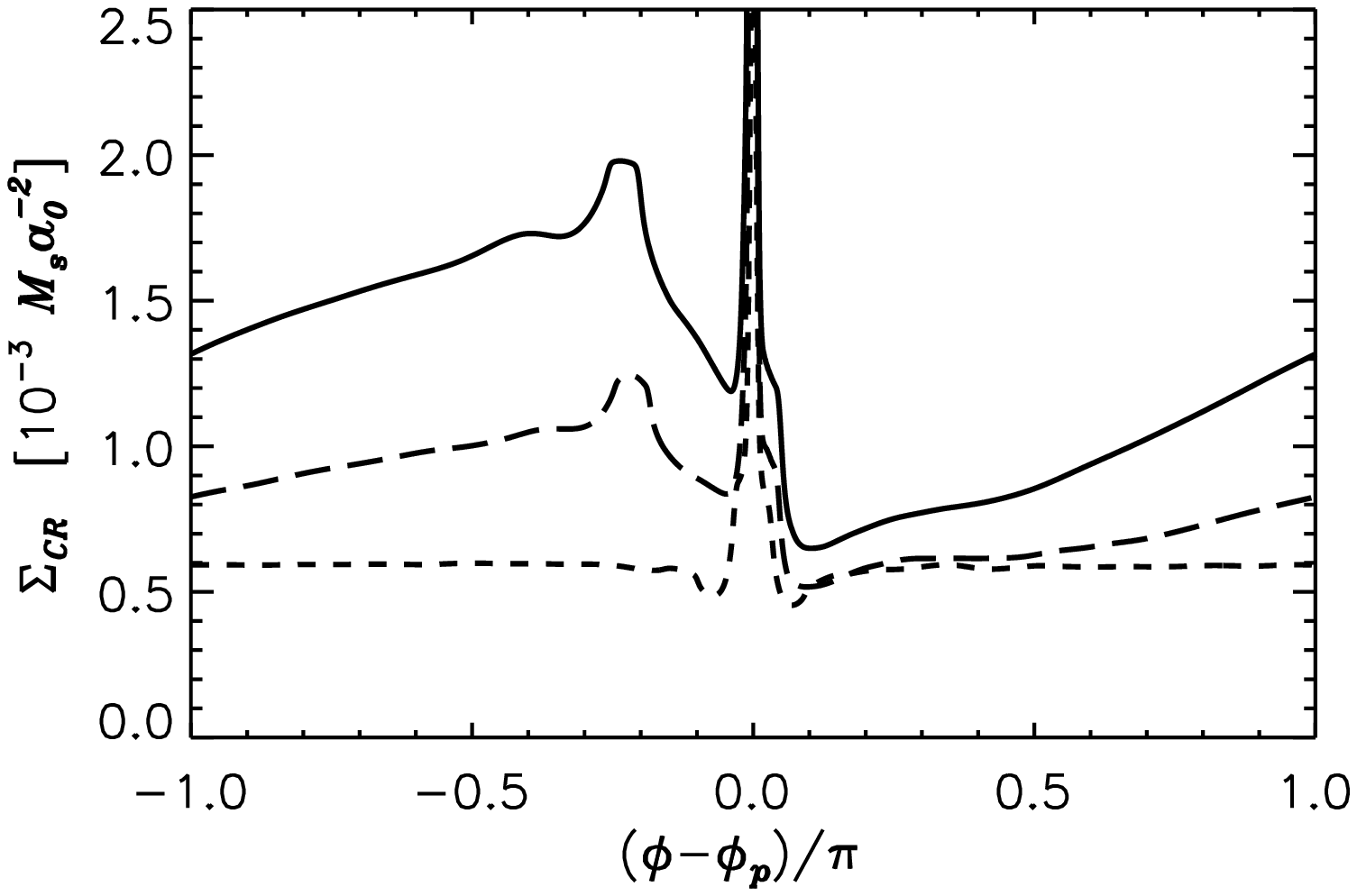}}
\caption{Radial average of the surface density within the coorbital
         region (defined as having radial extent 
         $2\,\Rhill\simeq 0.1\,a(t)$ of the planet's orbit)  as a 
         function of azimuth at $3$ different times: the time of 
         planet release $t_{\mathrm{rls}}=100$ initial orbital periods
         (\textit{short-dashed line}), $t_{\mathrm{rls}}+10$ initial 
         orbital periods (\textit{long-dashed line}), and 
         $t_{\mathrm{rls}}+20$ initial orbital periods 
         (\textit{solid line}). 
         }
\label{fig:sig_B_fast}
\end{figure}

In the model by OL06, fast migration is due to torques caused
by a density asymmetry in the coorbital region between gas on
the leading and trailing sides of the planet. The gas on the
leading side of the planet is trapped and contains gas acquired
at other radii, while the trailing side contains ambient gas near
the planet. The contrast between the trapped and ambient gas is
limited by viscous diffusion. The trapped gas is in a quasi-steady
advective-diffusive equilibrium. The density asymmetry and thus
the torque is caused by the motion of the planet. 
 
To test this model, we analyzed streamlines in the coorbital region
in the frame comoving with the planet. We determine the streamlines
in the simulations by following the motion of tracer particles that
move with the velocity of the gas (see Appendix~\ref{sec:BoundGas}).
In Figure~\ref{fig:str-nonmig} we plot coorbital streamlines near
before the planet is released, i.e., while the planet was on a fixed
stationary orbit. The figure shows good agreement between the 
simulation and theory. The streamlines are symmetric between the
leading ($\phi > \phi_p$) and trailing ($\phi < \phi_p$) sides of
the planet. 
Figure~\ref{fig:str-mig} shows the streamlines after the planet is
released, while the planet is migrating. Strictly speaking, these 
are not streamlines in the simulation case but trajectories, since 
the flow is not in a strict steady state in the comoving frame of 
the planet because the planet is migrating at a variable rate.
The theoretical streamlines depend on the planet-to-star mass ratio
and the migration rate of the planet. They are calculated assuming
a steady state and constant migration rate by means of the linear
perturbation model of OL06. The theoretical streamlines were 
calculated by using intermediate parameter values from the simulation
during the interval of planet migration: 
$\dot{a} =-0.002\,a_{0}\,\Omega_{0}$ and $r=0.85\,a_0$.
The simulated and theoretical streamlines in Figure~\ref{fig:str-mig} 
are in approximate agreement.
They show closed streamlines on the leading side of the planet's 
azimuthal motion. They contain the trapped gas described above. The 
open streamlines on the trailing side of the planet involve ambient
gas that streams outward past the planet.
The smaller closed streamlines are centered at about the same azimuth 
in the two plots, about $0.2 \, \pi$ ahead of the planet.
Figure~\ref{fig:sig_B_fast} shows that the gas density asymmetry in 
the coorbital region between the leading and trailing sides of the 
planet increases with time. The unperturbed background density
increases with time as the planet encounters higher density gas in its
inward migration. Notice that the density increase is higher on the
trailing side of the planet than on the leading side. This result
suggests that the trapped gas approximately retains its initial density
as the planet migrates. 
The gas on the trailing side more fully reflects the local density. 
The density asymmetry then gives rise to the dominant torque on the
planet. 

The OL06 model does not determine the value of coorbital corotational
torque for a migrating planet. It does provide a detailed analysis for
the noncoorbital corotational torque. In that case, the effect of 
migration is to amplify the standard coorbital torque for a 
nonmigrating planet \citep{gt1979}. 
By analogy, one might expect similar behavior in the coorbital case.
The standard torque is proportional to the radial derivative of
$\Sigma(r)/B(r)$, and hence of the disk vortensity $-2 B(r)/\Sigma(r)$,
where $\Sigma(r)$ is the disk axisymmetric surface density
(azimuthally averaged surface density) and $B(r)$ is the Oort 
constant. In the unperturbed disk model 
considered here, the vortensity is constant, and so the coorbital
torque is predicted to be zero. However, the gap structure in the
disk modifies the surface density $\Sigma(r)$ near the planet (see 
Fig.~\ref{fig:sig_fast}), thereby providing a vortensity gradient.

\subsection{Conditions for Type~III Migration}
\label{sec:typeiii_conditions}
 
The results in section~\ref{sec:typeiii_results} provide evidence 
for migration dominated by coorbital torques, or Type~III migration. 
Within the framework of the OL06 model we generalize the results
of the simulations and describe some conditions that are favorable
for Type~III migration. 

As in section~\ref{sec:typeiii_results}, we consider a planet of 
fixed mass. For a planet whose mass is large enough to open a gap,
we apply the initial condition that the planet undergoes migration
before steady gap formation completes, as was the case in 
section~\ref{sec:typeiii_results}.

We require that the planet does not strongly deplete the gas in 
the coorbital region as it migrates. This requirement implies that 
the migration timescale across the coorbital region be shorter 
than the timescale to clear a gap over that region.
Figure~\ref{fig:sig_fast} demonstrates that this condition holds for
the model in section~\ref{sec:typeiii_results}.  

To derive a crude estimate for this condition, we assume that the 
torques exerted by the planet lead to a local change in disk angular  
momentum over a region whose size is comparable to the coorbital 
region. Each one-sided torque (interior and exterior 
to the orbit of the planet) on the gas is capable of clearing a gap,
while the net effect of both interior and exterior torques results 
in migration. The condition that the gap clearing timescale is 
longer than the migration timescale becomes
\begin{equation}
M_c \ga \frac{\Mp}{A},
\label{eq:cond-gap}
\end{equation}
where $M_c$ is the mass of the coorbital region and $A$ is the 
dimensionless torque asymmetry
\begin{equation}
A = \frac{|T_e| - |T_i|}{|T_e|},
\end{equation}
where $T_i$ and $T_e$ denote the torques interior and exterior to 
the planet's orbit, respectively.

Another condition is that the migration rate be large enough that
there a strong asymmetry in streamlines between the leading and
trailing sides of the planet, as seen in Figure~\ref{fig:str-mig}.
According to OL06, the asymmetry is strong for migration rates 
greater than $|\dot{a}_{A}| = 1.45\,\Omega\,a\,\Mp/\Ms$.
Since the migration begins as Type~I migration (see 
\textit{short-dashed line} in Fig.~\ref{fig:dTdM_fast}), the condition
is that $|\dot{a}_{\mathrm{I}}| \ga |\dot{a}_{A}|$. 
This condition is approximately
\begin{equation}
\Sigma_{p}\ga \frac{\Ms}{a^2} \left( \frac{H}{a} \right)^2.
\label{eq:cond-mig}
\end{equation}
Notice that the condition is independent of planet mass, since
both $\dot{a}_{\mathrm{I}}$ and $\dot{a}_{A}$ are linear in the
planet mass.

We now apply these conditions to the model simulated in 
section~\ref{sec:typeiii_results}.
The asymmetry parameter is estimated as $A \approx 0.3$,
from the initial torque distribution in Figure~\ref{fig:dTdM_fast}.
Condition~(\ref{eq:cond-gap}) is then satisfied for this model,
since $M_c \approx 7\,\Mp$.
Condition~(\ref{eq:cond-mig}) is also (marginally) satisfied, since
the initial density is $\Sigma_{p} a^2/\Ms = 2\times 10^{-3}$ 
and $(H/a)^2 = 9 \times 10^{-4}$. 
None of the other models discussed in this paper satisfy both 
conditions. 

It also appears that the condition that the planet mass is fixed
(or slowly increasing) is important. The dashed curve on the right 
panel of Figure~\ref{fig:a_fast_chk} in 
Appendix~\ref{sec:typeiii_app} suggests that a planet that grows 
in  mass at the run-away rate would not undergo rapid migration 
long enough to move very far. The slow-down is partly due to gap
opening that reduces the torques.

The picture is then that a fixed mass planet, initially undergoing 
sufficiently fast Type~I migration, develops strong coorbital torques 
due to asymmetric trapped gas. The situation is not simple, however.
We saw in section~\ref{sec:typeiii_results}, that the disk's 
feedback to the planet's motion might slow or halt Type~I migration,
but the coorbital torques allow the inward migration to continue.
We found that the rate of the resulting migration is actually slower
than Type~I migration for a smooth disk.  
So the conditions~(\ref{eq:cond-gap}) and (\ref{eq:cond-mig}) 
are suggestive only at this point and require further testing.

\section{Summary and Discussion}
\label{sec:summary}
 
We have analyzed the evolution of migrating planets that undergo
run-away gas accretion by means of multi-dimensional numerical
simulations. The results agree with the predictions of Type~I
and Type~II migration (see Figures~\ref{fig:a_sn} through 
\ref{fig:a_sn_nu}) for a planet of time-varying mass that we 
obtain from simulations. The set of simulations include cases 
with disk densities as low as the minimum mass solar nebula value
and as high as $5$ times that value, viscosities 
$\alpha\approx 0.004$ and $10$ times that value (also $50$ times 
that value, $\alpha\approx 0.2$, in a test reported in 
Appendix~\ref{sec:MVV}), disk temperatures corresponding to 
$H/r = 0.05$ and a colder case of $H/r=0.04$.
The mass accretion rates onto the planet 
(see Figures~\ref{fig:mp_sig} and \ref{fig:mp_hnu}) are in 
general agreement with previous determinations based on fixed 
mass planets on fixed circular orbits, when comparing cases with
the same planet mass and disk properties for which the growth 
timescale is longer than the gap opening timescale. 
Planet mass growth rates can be understood in terms of accretion
within the Bondi radius at lower planet masses, accretion within
the Hill radius at intermediate masses, and accretion through the
gap at higher planet mass (see Fig.~\ref{fig:tauG_sn_fr}).
Mass growth rates typically peak at the intermediate planet masses, 
a few tenths of a Jupiter mass.

An important diagnostic for the nature of the disk-planet torques
is the scaled torque density distribution per unit disk mass as a
function of the scaled radial distance from the planet.
In the linear regime of the standard theory of disk-planet 
resonances, for a fixed form of the disk density 
distribution $d \ln{\Sigma}/d\ln{r}$ and gas properties (sound 
speed and viscosity), this scaled torque distribution should be 
universal, independent of planet mass and disk density value.
We verified this universality for low mass planets, although the 
distribution varies somewhat with planet mass for larger mass planets
that open gaps (compare low mass cases in Figures~\ref{fig:dTdM}, 
\ref{fig:dTdM_sn_mig}, and \ref{fig:dTdM_xsn_mig}).

\begin{figure}
\centering%
\resizebox{\linewidth}{!}{%
\includegraphics{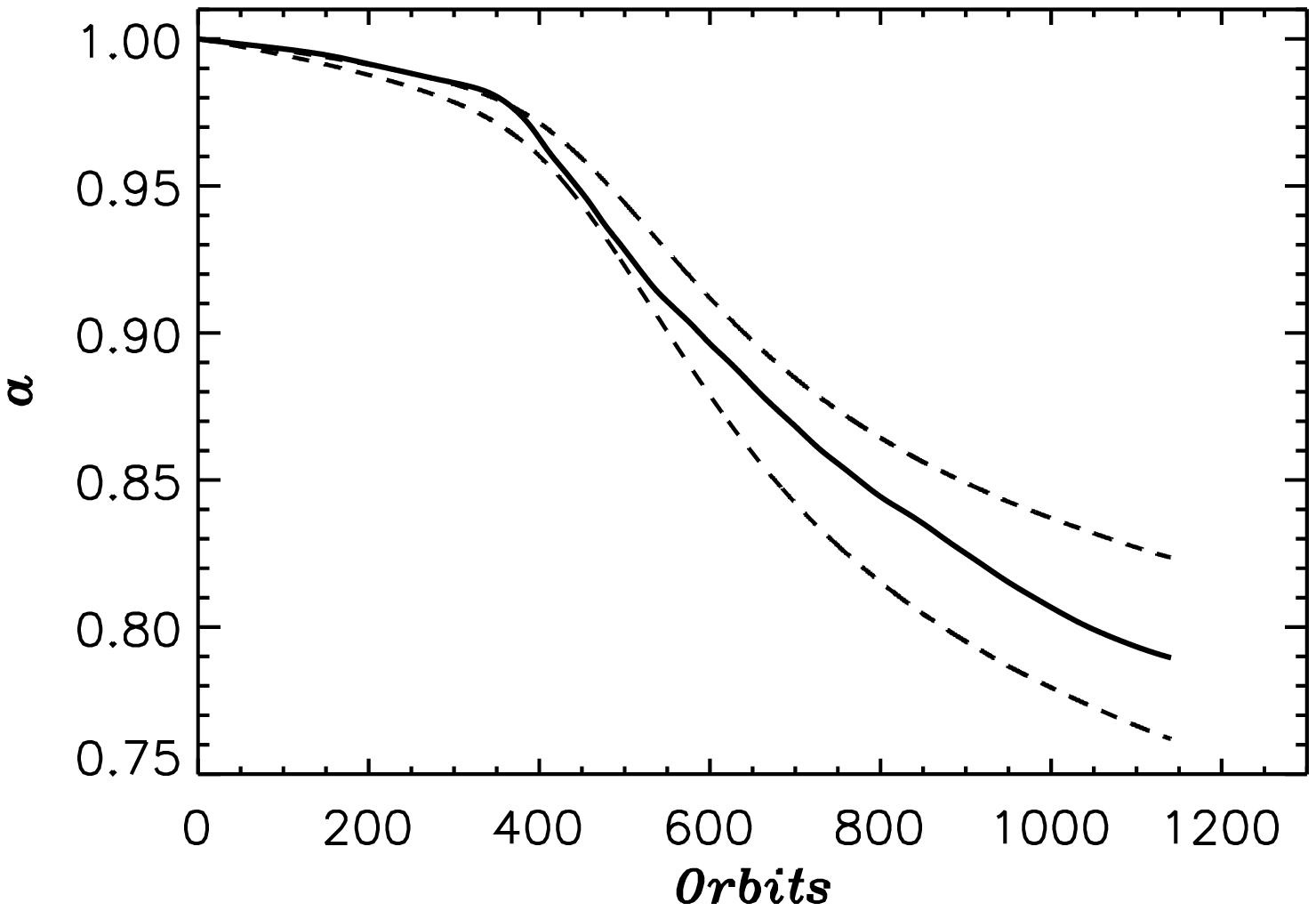}}
\caption{Orbital migration of a planet undergoing run-away gas 
         accretion. Orbital radius in units of $a_{0}$ 
         ($5.2\,\AU$), as a function of time in units of the
         initial orbital period ($\approx 12$ years).
         The initial planet mass is $5\,\MEarth$. The initial
         surface density is $\Sigma_{p}=3\times10^{-4}\,\densu%
         \approx 100\,\mathrm{g}\,\mathrm{cm}^{-2}$ at the 
         planet's initial orbital radius and $H/r=0.05$. 
         \textit{Solid curve}: 
         Results from the same simulation as in 
         Figure~\ref{fig:a_sn} for a migrating, mass-gaining
         planet. 
         \textit{Short-dashed curves}: 
         Predictions based on Type~I migration theory, obtained
         by solving equations~(\ref{eq:tanakadotaI}) and 
         (\ref{eq:tanakadotaIsat}), taking into account the time
         variable mass of the planet (Fig.~\ref{fig:mp_sig}, 
         \textit{solid line}) and roughly accounting for the gas
         depletion near the planet (see text). The upper (lower)
         dashed curve is for migration with unsaturated (saturated)
         coorbital torques.
         }
\label{fig:a_snc}
\end{figure}
There is no fundamental distinction between the torques involved
in Type~I and Type~II migration. This follows from the near
independence of the scaled torque density distribution with planet
mass. Previous concepts of Type~II suggested that a planet in a
clean gap would migrate inward like a test particle that follows
the disk accretion. Here we describe a view for cases where the gap
is not completely clear of material. The difference in Type~I and 
Type~II rates is due to the mass density distribution of the gas
that multiplies torque density in determining the torque on the
planet. In Type~II migration, the density distribution adjusts so
that the net torque on the planet causes it to migrate at
approximately the viscous evolution rate of the disk.
The transition between the two forms of migration is quite smooth.
To illustrate this point, we plot in Figure~\ref{fig:a_snc}
the orbital evolution of the planet through the phase where gap
formation sets in. The figure shows that the migration rate can 
be accounted for by standard Type~I theory, corrected for the gas
depletion in the gap region, although there is no unique 
prescription to do this. In the figure, the theoretical curves 
(\textit{dashed lines}) are determined by Type~I migration theory,
equations~(\ref{eq:tanakadotaI}) and (\ref{eq:tanakadotaIsat}),
with the density $\Sigma$ taken to be the average value in a 
radial band of half-width $0.15\,a$ (a typical gap width) centered
on the planet.

For a given planet mass, the torque density diagnostic reveals
that the distribution is not strongly affected by migration or 
accretion (Figure~\ref{fig:dTdM} is quite similar to 
Figures~\ref{fig:dTdM_sn_mig} and \ref{fig:dTdM_xsn_mig} for 
similar planet masses). 
In particular, there is no evidence for strong coorbital torques.
However, in a certain case, we do find evidence for strong coorbital
torques, or Type~III migration. This case has a planet of fixed mass, 
$0.3 \MJup$, that is immersed into a cold, smooth disk. 
The planet is held at a fixed orbit for relatively short time 
($\sim 100$ orbits) before being released, so that gap clearing is 
incomplete. 
The torque distribution at the time the planet is released follows
the expectations of the standard theory for nonmigrating planets. 
This can be seen by comparing the curve for the $0.3\,\MJup$ case in 
Figure~\ref{fig:dTdM} (\textit{dot-dashed curve}) with the 
short-dashed curve (time $t=t_{\mathrm{rls}}$) in 
Figure~\ref{fig:dTdM_fast}.
(There are differences in the magnitude of the distributions 
because of differences in gas sound speeds and dimensionalities 
of the calculations, but the forms of the distributions are similar).
At later times, Figure~\ref{fig:dTdM_fast} reveals a transition to 
a completely different torque distribution, where coorbital torques 
play a critical role. 

The strong coorbital torque can be understood in terms of an 
asymmetry in the streamlines between the leading and trailing 
sides of the planet, in accord with the analytic model of OL06 
(see Figures~\ref{fig:str-nonmig} and \ref{fig:str-mig}) and also
along the lines of \citet{pawel2004}. 
This asymmetry causes trapped material to persist on the leading 
side of the planet which has a different density from the ambient 
gas that flows on the trailing side (see Fig.~\ref{fig:sig_B_fast}).
The asymmetry gives rise to the coorbital torque. We suggest some 
criteria for this form of migration 
(see section~\ref{sec:typeiii_conditions}). More exploration is 
needed to test them.

Although we find evidence for Type~III migration, the conditions 
required appear somewhat artificial, i.e., incomplete gap clearing
(nonequilibrium gap) of a cool disk with a planet of fixed mass.
It is not clear 
whether and/or how conditions for Type~III could arise in a more 
plausible evolution scenario. 

We have generally assumed that the planet is able to accrete almost
all gas the disk is able to provide (so-called run-away gas accretion).
For a disk viscosity $\nu\gtrsim 1\times 10^{-5}\,\viscu$, the mass
accretion rates are large.
The time to build a $1\,\MJup$ planet starting with a $5\,\MEarth$ 
planet in a minimum mass solar nebula is shorter than $\sim 10^5$
years, substantially less than the observationally determined disk
lifetimes of $\sim 10^6$ years \citep{haisch2001,flaherty2008}.
Over this $10^5$ year time interval, the planet has radially migrated
inward by only $\sim 20$\% of its initial radius. In other words, for
these models, the mass doubling timescale for a $\Mp \la 1\,\MJup$
planet is short compared to the migration timescale and the disk
lifetime.
This situation stands in strong contrast to the earlier phases of
planet formation where the migration timescales are shorter than 
the planet mass doubling timescales and disk lifetimes  
\citep[e.g.,][]{ward1997,hubickyj2005}.

The run-away accretion rates pose some challenges for explaining the
mass distribution of planets \citep{butler2006}.  
Typical accretion rates in T Tauri stars are 
$\sim 1\times 10^{-8}\,M_{\odot}$ per year \citep{hartmann1998}. For
a steady-state unperturbed disk (without a planet), the accretion rate
is given by $3 \pi \nu \Sigma$. The initial disks considered in this paper
are not in a steady state, but this accretion rate provides a reasonable
estimate. For the minimum mass nebula model (Fig.~\ref{fig:mp_sig}), the
kinematic  turbulent viscosity we typically adopt, 
$\nu\sim 10^{-5}\,\viscu$, was chosen so that the accretion rate evaluates
to about this same value, $\sim 1\times 10^{-8}\,M_{\odot}$ per year.
In the case of a planet embedded in a disk, if there is a comparable 
accretion rate onto a planet of mass $\Mp \la 1 \MJup$ 
\citep[as found by][]{lubow2006}, then the mass doubling timescale for 
a Jupiter-mass planet is about $10^5$ years, consistent with what we 
found in the simulations in this paper.
But then it is not at all clear why planets would not almost always
achieve masses higher than $\sim 1\,\MJup$, in contradiction with the
observed mass distribution of extra-solar planets and the case of 
Saturn. Special timing for disk dispersal could be invoked, but may
be artificial.

There are a few possible explanations. A colder disk ($H/r < 0.05$)
would experience stronger tidal truncation effects from a 
Jupiter-mass planet, as $H$ becomes significantly smaller than 
$\Rhill\simeq 0.07\,a$. This effect could certainly reduce the accretion
rate by a large enough factor (say $10$), so that the mass doubling
timescale for a $\sim 1 \MJup$ planet would become of order $10^6$ years, 
consistent with the suggestion by \citet{dobbs-dixon2007}. 
The issue then is how cold the disk would need to be. The model in this
paper with $H/r =0.04$ (Fig.~\ref{fig:mp_hnu}, \textit{long-dashed line})
has the same unperturbed overall disk accretion rate as the $H/r =0.05$ 
case (Figures~\ref{fig:mp_sig} and \ref{fig:mp_hnu}, \textit{solid line}), 
since $\nu$ and the initial surface density $\Sigma(r)$ are the same. 
The accretion rate onto the planet from the colder disk ($H/r=0.04$) 
differs from the case with $H/r=0.05$ by only $10$\%. 
Model $h$ in \citet{lubow2006} for a nonmigrating planet of fixed mass, 
having $H/r=0.03$, has an accretion rate that is $\sim 25$\% less than 
model $b$, which has $H/r=0.05$ and the same unperturbed overall disk 
accretion rate. 
Consequently, in this disk thickness range ($0.03\le H/r \le 0.05$), 
we find that the reduction in the accretion rate onto the planet due 
to cooler disks is not significant.
For higher mass planets ($5$--$10\,\MJup$), we expect that the accretion
rate will be reduced by tidal truncation effects to a level where the
planet mass doubling timescale is comparable to the disk lifetime, as
found in previous studies of planets on fixed orbits
\citep{lubow1999,bate2003,gennaro2003b}.
This effect  may set the upper limit to planet masses.

Another possibility is that there is a feedback effect that limits the
gas accretion  rate. Perhaps the heating of the protoplanet envelope by
impacting solids continues to later times than is assumed in the 
standard core accretion model. Depletion of disk solids near the planet
occurs in the standard core accretion model, when planet migration is 
not included.
With migration, it is possible that continued accretion of disk solids
would occur \citep[e.g.,][]{alibert2005a}, resulting in continued heating
that could limit the gas accretion rate further. It is not clear
how well this possibility works, since planetesimals will get trapped
into resonances as the planet migrates \citep{zhou2007}.
Having a higher mass solid core is problematic in the case of Jupiter,
whose solid core mass is thought to be a small fraction of the total
mass 
(see \citealp{guillot2005} and discussion in \citealp{lissauer2007}).
It is also possible that winds emanating from the circumplanetary disk
within the planet's Hill sphere could reduce the accretion rate onto 
the planet. Magnetically driven winds are believed to play an important
role in the case of young stars \citep{blandford1982,pudritz1986,shu1994}.
There are likely differences in the flow properties from the stellar
outflow case \citep{fendt2003}.
However, it is not clear that the winds would be able to expel a large
enough fraction (say $90$\%) of the accreting gas to sufficiently reduce
the accretion rate onto the planet.

\acknowledgments

We thank Hui Li and Doug Lin for discussions about Type~III migration.
We also thank the referee for providing prompt and useful comments.
GD is supported through the NASA Postdoctoral Program.
SL acknowledges support from NASA
Origins of Solar Systems grants NNG04GG50G and NNX07AI72G.
We acknowledge computational facilities supported by the NASA 
High-End Computing Program systems under grants SMD-07-0372
and SMD-08-0582.


%% in parentheses, after the keyword, but they will not be verified.

\textit{Facilities:} \facility{%
NAS: NASA Advanced Supercomputing Division (Columbia/Schirra); 
NCCS: NASA Center for Computational Sciences (Palm/Exp\-lore)}.



\appendix

\section{Numerical Sensitivity Study}
\label{sec:NumericalSensitivityStudy}

We conducted several tests to assess the sensitivity of the results 
presented in section~\ref{sec:GrowingMigratingPlanets} to the 
choice of various numerical parameters.
Since the main objective of that section is the mass and
orbital evolution of an embedded planet, we present here
quantitative comparisons of planetary masses and orbital radii 
as function of time.

\subsection{A Resolution Test}
\label{sec:ResolutionTest}

For purposes of a resolution study, we performed a three-dimensional
calculation in which the grid resolution is raised by a factor of
$3/2$ over the standard resolution 
(see section~\ref{sec:GrowingMigratingPlanets})
in each coordinate direction, throughout the entire disk domain, 
and on all grid levels. 
Note that such an increase implies an overall refinement gain of 
a factor $(3/2)^3\simeq 3.4$, in terms of \textit{volume} 
resolution of the system or number of grid elements. 
Nested grids cover extended disk regions, so that the planet always 
remains in the domain described by the most refined grid during the
calculation.
We focus on the disk model with initial surface density at the 
planet's initial position $\Sigma_{p}=9\times10^{-4}\,\densu$
($300\,\mathrm{g}\,\mathrm{cm}^{-2}$), 
$H/r=0.05$, and $\nu=1\times10^{-5}\,\viscu$. 
The planet's mass and orbital 
radius evolution, obtained at standard grid resolution, are 
shown in Figure~\ref{fig:mp_sig} (\textit{dashed curve}) 
and Figure~\ref{fig:a_xsn} (\textit{top-most solid curve}), 
respectively. 
In order to carry out a quantitative comparison, results 
(for both $\Mp$ and $a$) from the two calculations, which will 
be labelled as $1$ and $2$, are averaged over time intervals of 
half of the (initial) orbital period and then relative differences 
are computed as
\begin{equation}
\frac{\Delta X}{\bar{X}} =2\left(\frac{X_{1}-X_{2}}{X_{1}+X_{2}}\right),
\label{eq:Xdiff}
\end{equation}
where $X$ is either $\Mp$ or $a$. In order to give time-averaged
estimates of the relative differences, over the course the 
calculations, we perform a running-time average 
of quantity $\Delta X/\bar{X}$, which is defined by
\begin{equation}
\left\langle\frac{\Delta X}{\bar{X}}\right\rangle_{t}=%
     \frac{1}{t}\!\int_0^{t}\!\frac{\Delta X}{\bar{X}}\,dt'.
\label{eq:Xrunav}
\end{equation}
 
\begin{figure*}
\centering%
\resizebox{\linewidth}{!}{%
\includegraphics{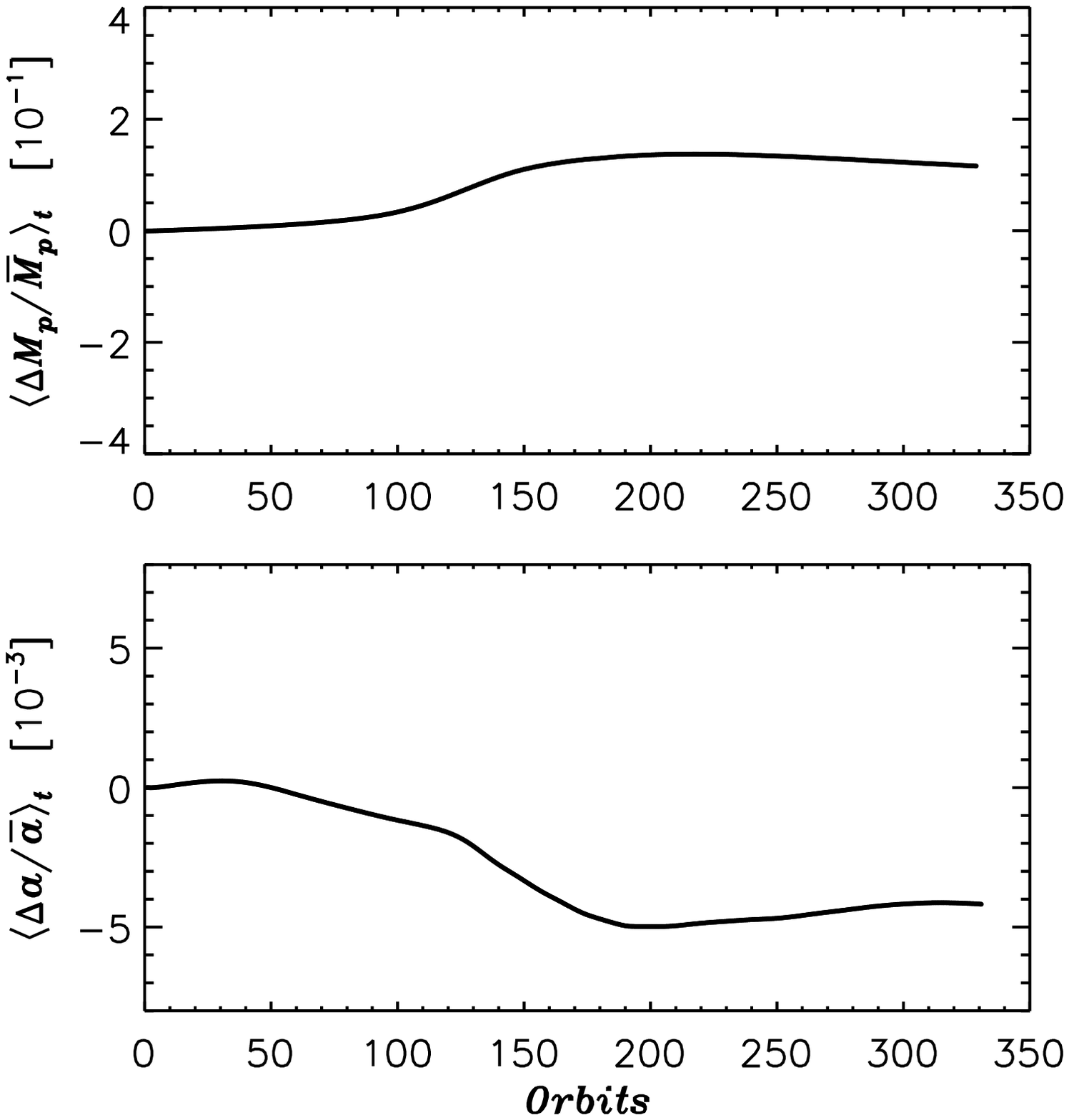}%
\includegraphics{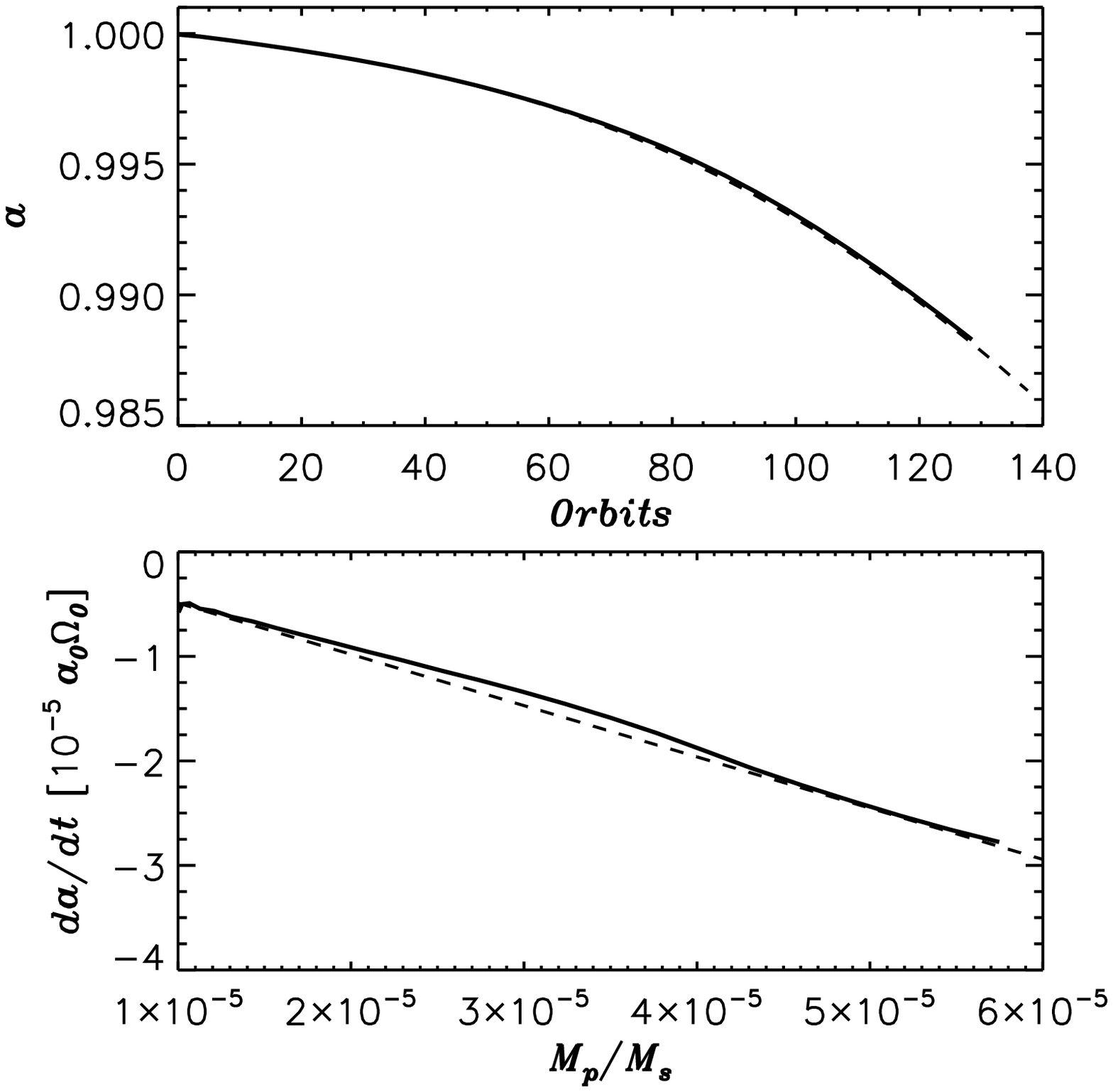}}
\caption{
 \textit{Left}:
         Running-time average, defined by equation~(\ref{eq:Xrunav}), 
         of the relative differences between two three-dimensional
         calculations,
         whose numerical resolutions differ by a factor of $3.4$
         in terms of grid elements (see text for details). Most of
         the difference in the mass growth of the planet 
         (\textit{top})
         is accumulated between about $100$ and $150$ orbits,
         during the early phases of the rapid mass growth, when
         $\Mp\sim 1\times 10^{-4}$ (see \textit{dashed curve} in
         Fig.~\ref{fig:mp_sig}). Overall, the running-time 
         average of $\Delta \Mp/\bar{M}_{p}$ is in the range
         $10$--$15$\%.
         Relative differences between the evolution of the orbital 
         radii (\textit{bottom}) remain negligible over the course
         of the calculations and the running-time average is 
         contained within $1$\%.
 \textit{Right}:
         Orbital migration of a planet that grows at a prescribed rate
         in a three-dimensional disk whose initial (unperturbed)
         surface density has slope $s=-d\ln{\Sigma_{p}}/d\ln{a}=3/2$.
         Results from a calculation (\textit{solid curve}) are compared
         to predictions of Type~I theory (\textit{dashed curve}, see
         section~\ref{sec:RegimesofMigration}). The top panel shows
         the orbital radius evolution, whereas the bottom panel shows 
         the migration rate as a function of the planet mass.
         }
\label{fig:res_test}
\end{figure*}
Results are shown in the left panels of Figure~\ref{fig:res_test}. 
The largest differences are observed in the results for the mass 
evolution (\textit{top-left}), after the onset of the rapid accretion 
phase (see \textit{dashed line} in Fig.~\ref{fig:mp_sig}). 
The average difference, over the 
entire evolution, stays within $10$--$15$\%. The average relative 
difference between the evolution results of the orbital radii 
(\textit{bottom-left}) is much smaller and remains well within $1$\%.
 
In order to test whether the orbital evolution is consistent
with Type~I migration theory of \citet{tanaka2002} at our 
standard resolution, we set up a three-dimensional disk model with
a planet that grows at a \textit{prescribed} rate and whose initial
mass is $\Mp=1\times 10^{-5}\,\Ms$ (about $3\,\MEarth$).
The initial (unperturbed) surface density has slope $s=3/2$, so that
$\dot{a_{\mathrm{I}}}(t)\propto \Mp(t)$.
In the right panels of Figure~\ref{fig:res_test},
results from the simulation (\textit{solid line}) are compared
against predictions of Type~I theory (\textit{dashed line}, see
section~\ref{sec:RegimesofMigration})
for both the orbital radius evolution (\textit{top panel}) and 
the migration rate as a function of $\Mp$ (\textit{bottom panel}).

\subsection{Boundary Condition Effects}
\label{sec:BoundaryConditionEffects}

\begin{figure*}
\centering%
\resizebox{\linewidth}{!}{%
\includegraphics{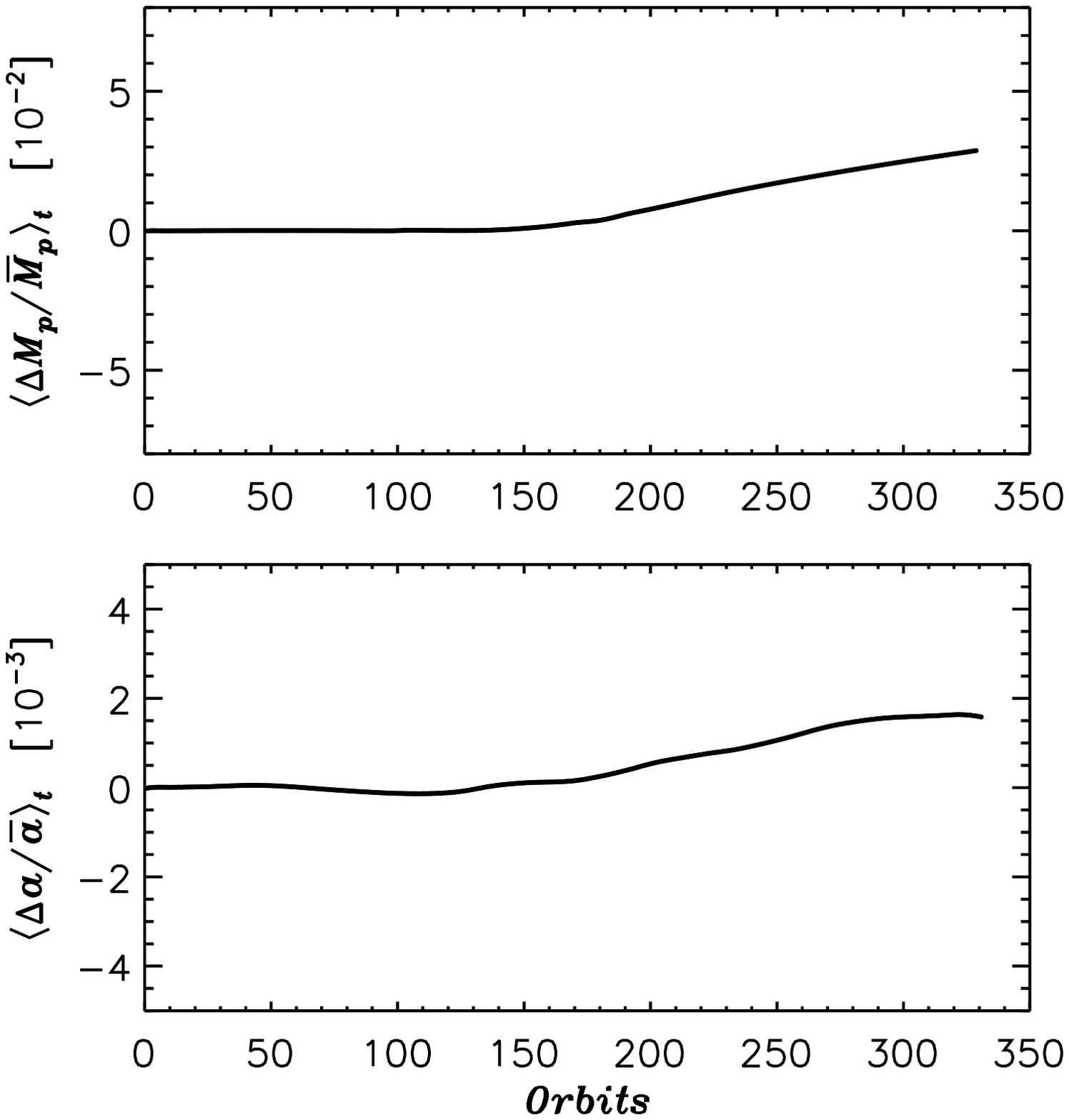}%
\includegraphics{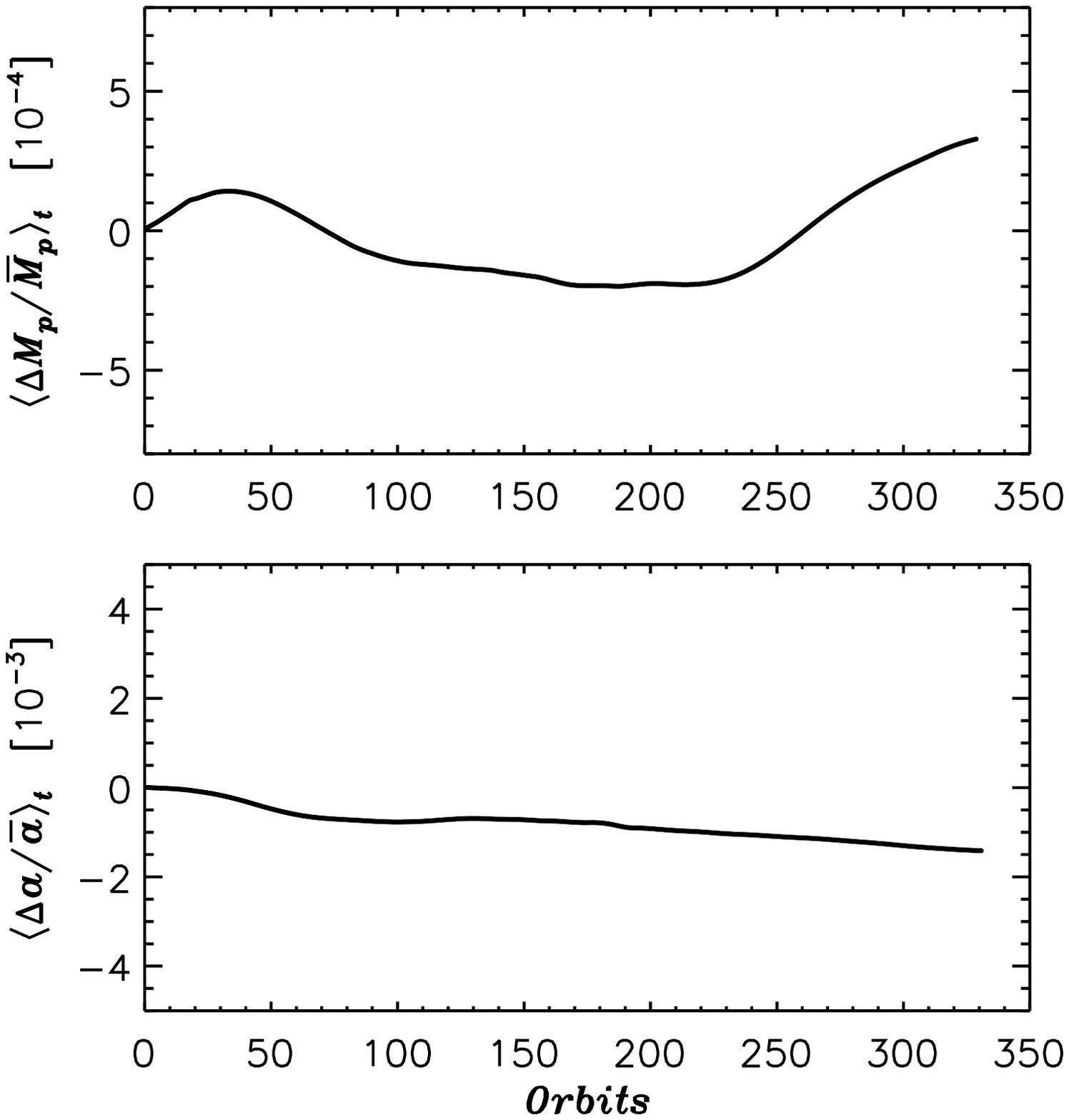}}
\caption{%
 \textit{Left}:
         Running-time average of relative differences between 
         results obtained from two three-dimensional models:
         one with inner grid boundary at radius
         $R_{\mathrm{min}}=0.4\,a_{0}$ and nonreflecting
         boundary conditions \citep{godon1996,godon1997} and
         the other 
         with $R_{\mathrm{min}}=0.19\,a_{0}$ and outflow
         boundary conditions (see section~\ref{sec:NumericalMethod}). 
         Average relative differences in
         planet's mass (\textit{top}) and orbital radius evolution 
         (\textit{bottom}) are around $3$\% and $\ll\sim 1$\%,
         respectively.
 \textit{Right}:
         Same comparison between two three-dimensional calculations
         with outer grid boundary at radii $R_{\mathrm{max}}=2.5\,a_{0}$
         and $4.9\,a_{0}$, respectively.
         Relative differences much smaller than
         $1$\% are observed in the evolution of both the planet's
         mass and orbital radius.
         }
\label{fig:OB_test}
\end{figure*}
Boundary conditions may play some role and affect the late stages of
the system's evolution, especially when $\Mp$ becomes on the order 
of a Jupiter mass.
In our situation, the major concerns that may arise are related to
the positions of the grid radial boundaries. The finite extent of
the inner radius of the grid ($R_{\mathrm{min}}$) may lead to an 
augmented depletion of the disk within the orbit of the planet. 
At the outer radial boundary ($R_{\mathrm{max}}$), reflection of 
waves may affect torques exerted on the planet. The first effect 
can be mitigated by reducing the inner radius of the grid or 
adopting nonreflective boundary conditions 
\cite[e.g.,][]{godon1996,godon1997}, whereas the second can be 
largely suppressed by choosing $R_{\mathrm{max}}\gg a$. 
However, while increasing the outer grid radius possibly 
lengthens the computing time only because a larger number of grid 
elements in the radial direction is required (for a given value of 
$\Delta R$), decreasing the inner grid radius directly affects
the time-step of the calculation, which is proportional
to $R^{3/2}_{\mathrm{min}}$ because of the stability criterion 
imposed by the Courant-Friedrichs-Lewy condition.

The left panels of Figure~\ref{fig:OB_test} shows a comparison of 
results obtained from a three-dimensional calculation 
($\Sigma_{p}=9\times10^{-4}\,\densu$ at $t=0$, $H/r=0.05$, and 
$\nu=1\times10^{-5}\,\viscu$) with nonreflective boundary 
conditions at $R_{\mathrm{min}}=0.4\,a_{0}$ 
and one with $R_{\mathrm{min}}=0.19\,a_{0}$ but outflow 
boundary conditions, as outlined in 
section~\ref{sec:NumericalMethod}. The position of the inner grid
boundary affects the density distribution interior to the
planet's orbit. The effect on the planet's mass is at the $3$\%
level, on average
over the entire course of the simulation (\textit{top-left panel}). 
The orbital radius evolution (\textit{bottom-left panel}) displays 
average relative differences much smaller than $1$\%.
In the right panels of Figure~\ref{fig:OB_test}, results from
a three-dimensional calculation with outer grid radius at 
$R_{\mathrm{max}}=2.5\,a_{0}$ are compared to those from a calculation
in which $R_{\mathrm{max}}=4.9\,a_{0}$.  
Both the evolution of planet mass (\textit{top-right panel}) and orbital 
radius (\textit{bottom-right panel}) are hardly affected by the position 
of the outer grid boundary.

\subsection{Effects of Excluded Torques}
\label{sec:ExcludedTorquesEffects}

The region of space in which material is gravitationally bound to 
the planet depends on several disk parameters (including $H/r$ and 
$\nu$) and on the planet's mass.
Calculations with fixed mass and fixed orbit planets indicate that,
for $H/r\approx0.05$ and $\nu\approx 1\times10^{-5}\,\viscu$, 
only material within about $0.3\,\Rhill$ is gravitationally bound 
to the planet when $5\,\MEarth\lesssim\Mp\lesssim 40\,\MEarth$
\citep{hubickyj2007}. 
Analytical and numerical models of disk formation around a 
Jupiter-mass planet suggest that such disks may extend over a 
distance of about $\Rhill/4$ (or less) around the planet.
In the calculations presented in
section~\ref{sec:GrowingMigratingPlanets} and \ref{sec:typeiii},
torques originating
within $\Rhill/2$ of the planet are not taken into account, which
may include nonzero net torques exerted by unbound material. 
In order to estimate how this choice affects the evolution of the 
orbital radius, we also considered cases in which only torques 
from within $0.3\,\Rhill$ are neglected. 
The models with initial surface density 
$\Sigma_{p}=3\times10^{-4}\,\densu$ 
and $9\times10^{-4}\,\densu$ at $r=a_{0}$ 
($H/r=0.05$, $\nu=1\times10^{-5}\,\viscu$), discussed in 
sections~\ref{sec:MassEvolution} and \ref{sec:OrbitalRadiusEvolution},
are restarted at regular time intervals, covering entirely their
respective mass range. 
The evolution is then integrated for time periods of $\gtrsim 100$ 
orbits at each restart.
We compare the evolution of orbital radii by measuring the relative 
differences $\Delta a/\bar{a}$ and find that for none of the cases 
considered $|\Delta a/\bar{a}|$ exceeds $1$\%. 
We therefore conclude that excluded torques from unbound
material have only marginal effects on the results presented in
section~\ref{sec:PlanetMigration} and \ref{sec:highdiskmass}.

\section{Migration in High Viscosity Disks}
\label{sec:MVV}

Results presented in section~\ref{sec:PlanetMigration} 
indicate that the orbital radius evolution of a growing planet
can be described reasonably well in terms of standard Type~I
and Type~II regimes of migration (as long as the local disk 
mass is comparable or larger than the planet's mass). 
This conclusion holds when the disk's kinematic viscosity is 
$\nu\sim 10^{-5}\,\viscu$ (see Fig.~\ref{fig:a_sn}) as well
as when $\nu\sim 10^{-4}\,\viscu$ (see Fig.~\ref{fig:a_sn_nu}),
which brackets a range of $\alpha$-parameters between
$4\times10^{-3}$ and $4\times10^{-2}$ at the location of the
planet.
Here we present further analysis of a case with 
$\nu=1\times 10^{-4}\,\viscu$ and examine cases with
$\nu=5\times 10^{-4}\,\viscu$ ($\alpha\sim 0.2$), which still
result in a form of Type~I migration modified by the perturbed
surface density of the disk.

\subsection{An Additional Model With $\nu=1\times 10^{-4}\,\viscu$}
\label{sec:nu1e-4}

As anticipated in section~\ref{sec:OrbitalRadiusEvolution},
some concern may arise when the viscous timescale, 
$t_{\nu}=r^2/\nu$, at 
$R_{\mathrm{max}}$ becomes comparable to the length of time
over which the orbital evolution of the planet is calculated.
However, it is unlikely that the disk's viscous evolution
at $R>R_{\mathrm{max}}$ has a large impact on the results 
displayed in Figure~\ref{fig:a_sn_nu} since $t_{\nu}$ at
$R=R_{\mathrm{max}}$ is more than $6$ times as long as the
viscous timescale at $a$, about $10^{4}$ (initial) orbital 
periods of the planet. Therefore, at this viscosity level, 
we may experience some effects only over simulations covering 
timescales longer than $10^4$ orbits (note that inward 
migration will increase even further this timescale).

\begin{figure*}
\centering%
\resizebox{\linewidth}{!}{%
\includegraphics{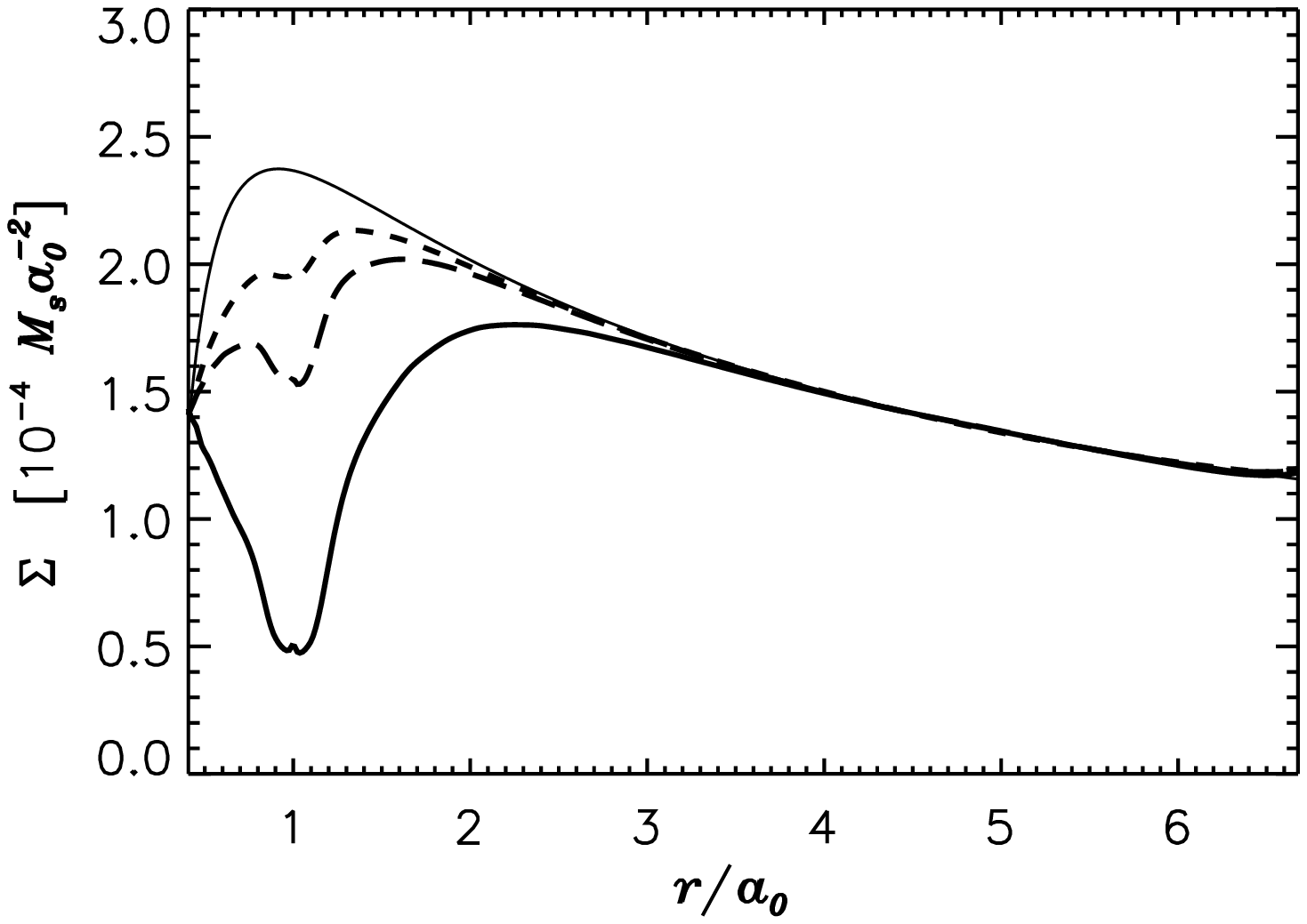}%
\includegraphics{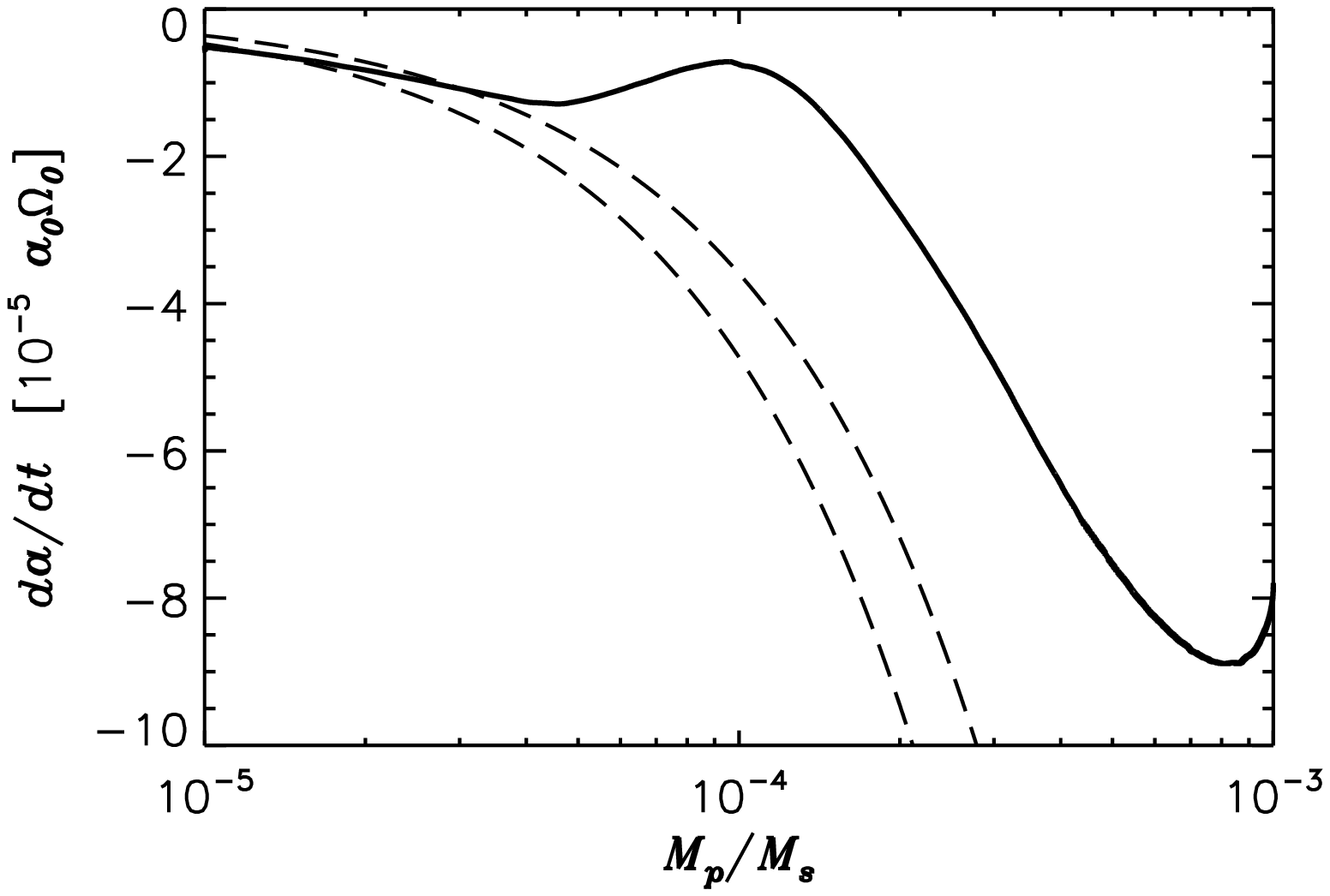}}
\caption{%
 \textit{Left}:
         Azimuthally averaged surface density of a 
         three-dimensional disk with 
         $\nu=1\times10^{-4}\,\viscu$ and $H/r=0.05$ at time
         $t=0$ (\textit{thinner solid line}) and at times when 
         $\Mp=\Mp(t)$ is equal to $0.1\,\MJup$ 
         (\textit{short-dashed line}), $0.3\,\MJup$ 
         (\textit{long-dashed line}), and $1\,\MJup$ 
         (\textit{thicker solid line}).
 \textit{Right}:
         Migration rate ($da/dt$) as a function of the planet mass
         ($\Mp$) over the first $330$ orbital periods of the 
         simulation, during which time the planet's orbit is held 
         fixed.
         The planet mass growth is prescribed. Migration rates are
         evaluated by means of Gauss perturbation equations.
         The upper and lower dashed curves indicate Type~I migration
         rates predicted by equation~(\ref{eq:tanakadotaI}) and 
         (\ref{eq:tanakadotaIsat}), respectively.
         }
\label{fig:bid1}
\end{figure*}

In order to address this issue more in detail, we set up a 
three-dimensional model with $\nu=1\times10^{-4}\,\viscu$
and $H/r=0.05$. We adopt the same parameters, numerical 
resolution ($0.014\, a_0$ of linear base resolution and 
$9\times10^{-4}\, a_0$ of resolution in the coorbital region 
around the planet), and boundary conditions, as those of the
simulations discussed in 
section~\ref{sec:GrowingMigratingPlanets}. In this model, 
however, the grid radial boundary extends out to 
$R_{\mathrm{max}}=6.7\,a_{0}$.
Therefore, the viscous timescale at $R_{\mathrm{max}}$ is a 
factor of at least $6.7^2\simeq 45$ as long as $t_{\nu}$ at 
the orbital radius of the planet. Thus, we could in principle
follow the planet's orbital evolution for tens of local 
viscous timescales.
Additionally, to monitor the sensitivity of our results to 
boundary and initial conditions at $R_{\mathrm{min}}$,
we use the initial surface density
represented as a thin solid line in the left panel of 
Figure~\ref{fig:bid1}. The initial mass density, $\rho$, is 
related to $\Sigma$ at $t=0$ as described in 
section~\ref{sec:DiskPlanetParameters}.
Material near the planet is removed from the disk, but its 
mass is not added to the planet's mass. 
Instead, for the purposes 
of this test, the planet mass is increased, at a prescribed 
rate, from $\Mp=1\times10^{-5}\,\Ms$ ($3\,\MEarth$) to 
$1\times10^{-3}\,\Ms$ ($1\,\MJup$) over about $330$ orbital 
periods. The orbit of the planet is held fixed ($a=a_{0}$) 
during this period of time. 

The left panel of Figure~\ref{fig:bid1} shows the azimuthally 
averaged surface density at times when $\Mp=0.1\,\MJup$ 
(\textit{short-dashed line}), 
$0.3\,\MJup$ (\textit{long-dashed line}), and $1\,\MJup$ 
(\textit{thick solid line}). 
By means of Gauss perturbation equations \citep[e.g.,][]{beutler2005}, 
we measure migration rates $da/dt$, which result from disk's 
gravitational forces (while $a=a_{0}$), as a function of time 
and hence of $\Mp=\Mp(t)$.
The right panel of Figure~\ref{fig:bid1} displays these \textit{static} 
migration rates (\textit{solid line}) compared to Type~I rates 
(\textit{dashed curves}) yielded by equation~(\ref{eq:tanakadotaI}) 
and (\ref{eq:tanakadotaIsat}).
The upper (lower) dashed curve refers to the unsaturated 
(saturated) coorbital corotation torques in the linear theory
(see section~\ref{sec:RegimesofMigration}). 
As observed in the figure (\textit{right panel}), initial
migration rates agree with those predicted by linear theory.
The reduction of $|\dot{a}|$, which peaks at $\Mp/\Ms\sim 10^{-4}$,
is likely related to the onset of nonlinear effects \citep{masset2006}. 

\begin{figure*}
\centering%
\resizebox{\linewidth}{!}{%
\includegraphics{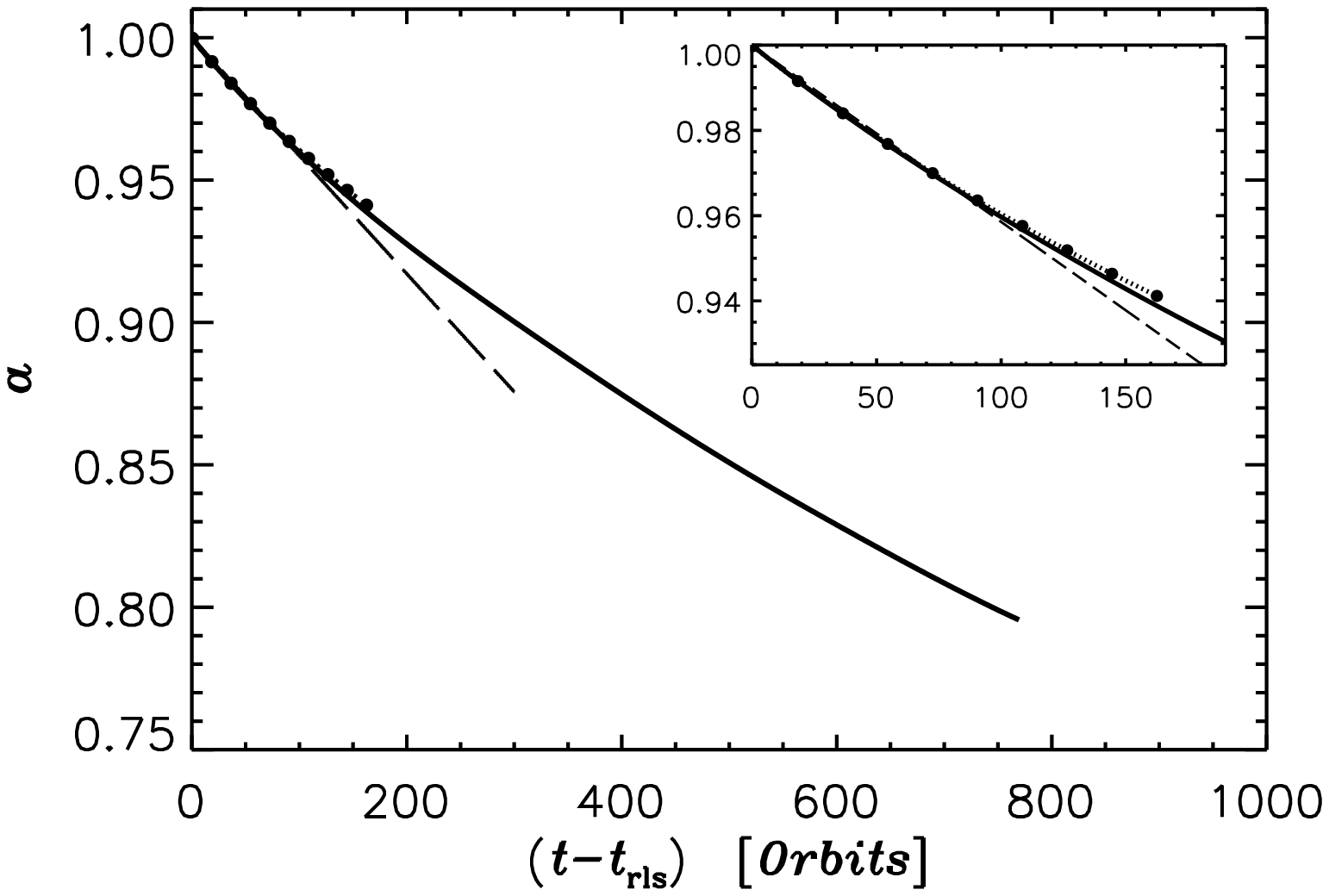}%
\includegraphics{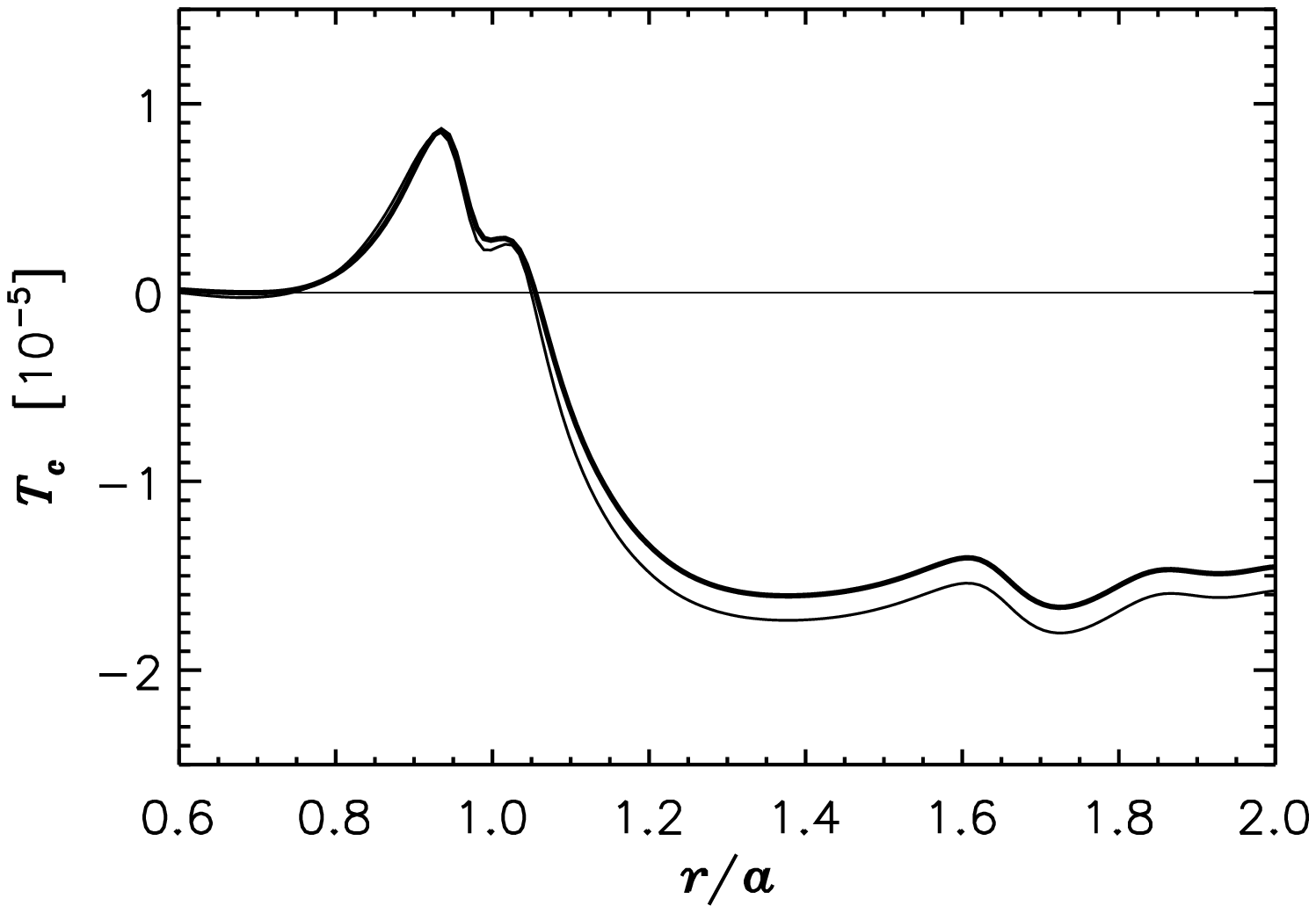}}
\caption{%
 \textit{Left}:
         Orbital radius evolution of a $1\,\MJup$ planet in a disk 
         whose kinematic viscosity is $\nu=1\times10^{-4}\,\viscu$
         ($H/r=0.05$).
         The solid line represents the model discussed in this 
         Appendix while the dotted line with solid circles is the
         same as Figure~\ref{fig:a_sn_nu_cmp}.
         Note that the two models produce very similar migration 
         tracks (see inset), although they use grids with different 
         outer radial boundaries, respectively at $6.7\,a_{0}$ and
         $2.5\,a_{0}$, and although the surface density profiles 
         at $r\lesssim a$ are different.
         In both calculations, the planet's orbit is held fixed 
         over the first $t_{\mathrm{rls}}$ (initial) orbital 
         periods (see text for details). The dashed (straight) 
         line has a slope about equal to
         $-7 \times 10^{-5}\,a_{0}\,\Omega_{0}$.
 \textit{Right}:
         Cumulative torque at time $t=t_{\mathrm{rls}}+600$ initial
         orbital periods (\textit{thicker curve}), in units of 
         $G\Ms\Mp/a$, where $a=a(t)$.
         See text for an explanation of the thinner curve.
         Nearly all torque is exerted by material within a radial
         distance of $0.25\,a$ from the planet's orbit. 
         }
\label{fig:bid2}
\end{figure*}
At time $t_{\mathrm{rls}}=340$ orbital periods, the planet is released 
and allowed to change its orbit in response to disk's torques. 
We recall that, at this stage, the planet's mass is constant and equal 
to $1\,\MJup$. The migration track is displayed as a solid line in 
Figure~\ref{fig:bid2} (\textit{left panel}). The planet's orbit is 
integrated for about $0.5\,t_{\nu}$ at $a_{0}$.
The dashed (straight) line has a slope about equal to 
$-7\times 10^{-5}\,a_{0}\,\Omega_{0}$ or $-0.7\,\nu/a$.
For comparison purposes, also plotted in
the left panel of Figure~\ref{fig:bid2}, as a dotted line with solid
circles, are results shown in the right panel of Figure~\ref{fig:a_sn_nu} 
(\textit{dotted line with solid circles}) and obtained with the model
discussed towards the end of section~\ref{sec:OrbitalRadiusEvolution},
which has a different disk density profile inside the planet's orbit
and a different radial coverage of the disk. Despite these differences,
the two models produce consistent and very similar migration tracks, 
indicating that disk regions at radii much smaller and much larger 
than the planet's orbital radius are not playing a determinant role.

The thick curve in the right panel of Figure~\ref{fig:bid2}
plots the cumulative torque (in units of $G\Ms\Mp/a$), i.e., 
the torque per unit radius $dT/dr$ integrated outward over radius,
at $t=t_{\mathrm{rls}}+600$ orbits. As also noted for the $1\,\MJup$
case in Figure~\ref{fig:dTdr}, almost all the torque is due to material
within a radial distance of $0.25\,a$ from the orbit of the planet.
Figure~\ref{fig:bid2} (\textit{left panel}) indicates that the migration
rate decreases over the course of the simulation. We find that this is 
not caused by a changing character of the torque per unit disk mass, 
$dT/dM$, but rather by a changing (azimuthally averaged) surface 
density profile around $r\sim a$.
We calculate $dT/dM$ at release time and the averaged surface density 
$600$ orbits after release. 
We then use equation~(\ref{eq:dTdr_I}) to obtain the 
\textit{expected} cumulative torque, at time $t=t_{\mathrm{rls}}+600$ 
orbits, under the assumption that the intrinsic character of $dT/dM$ 
remains unchanged over time. 
This is plotted as a thin curve in the right panel of 
Figure~\ref{fig:bid2}, along with the \textit{actual} cumulative torque
(\textit{thick curve}).
The difference between the two curves is less than $10$\%.

\subsection{A Model With $\nu=5\times 10^{-4}\,\viscu$}
\label{sec:nu5e10e-4}

\begin{figure*}
\centering%
\resizebox{\linewidth}{!}{%
\includegraphics{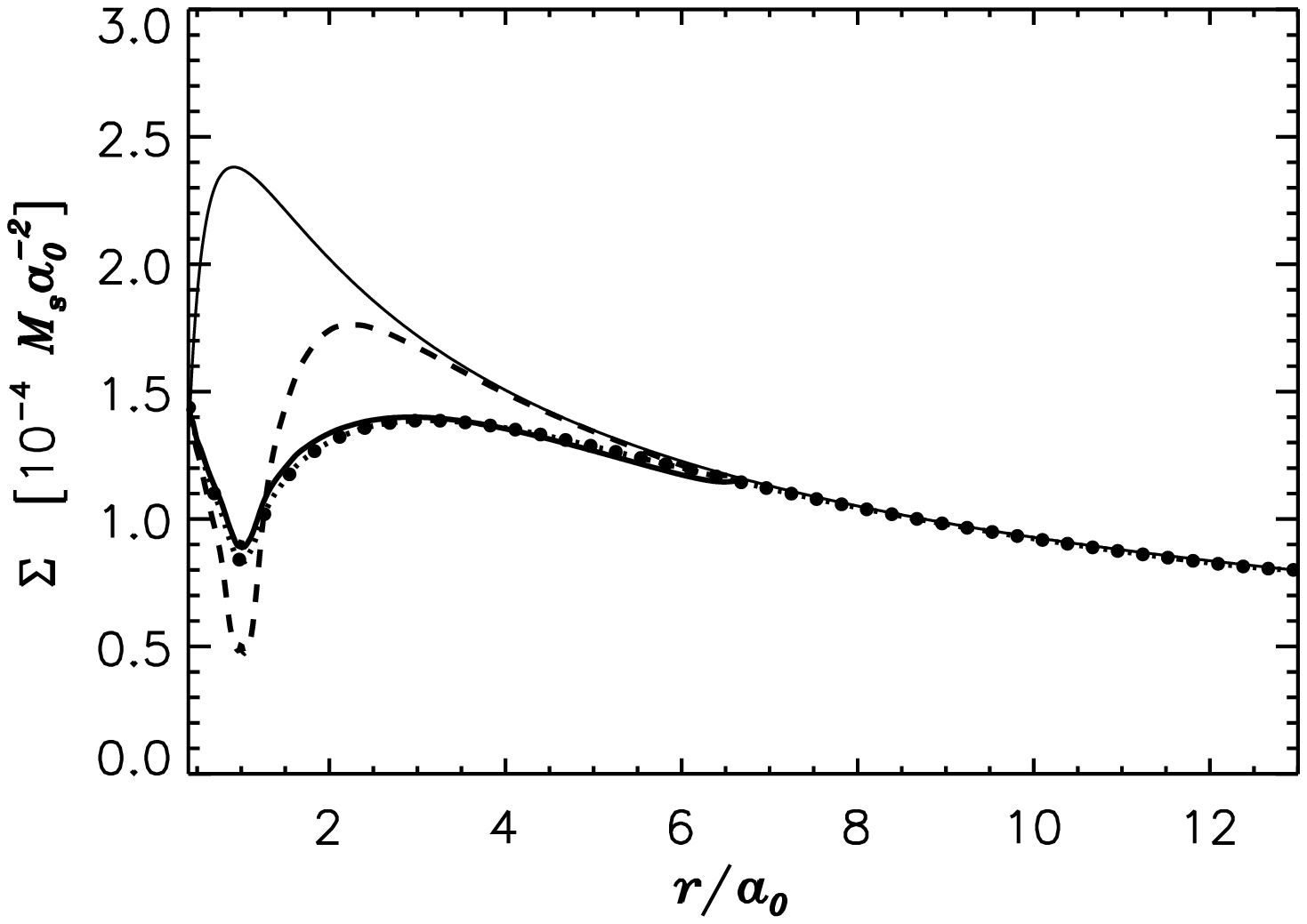}%
\includegraphics{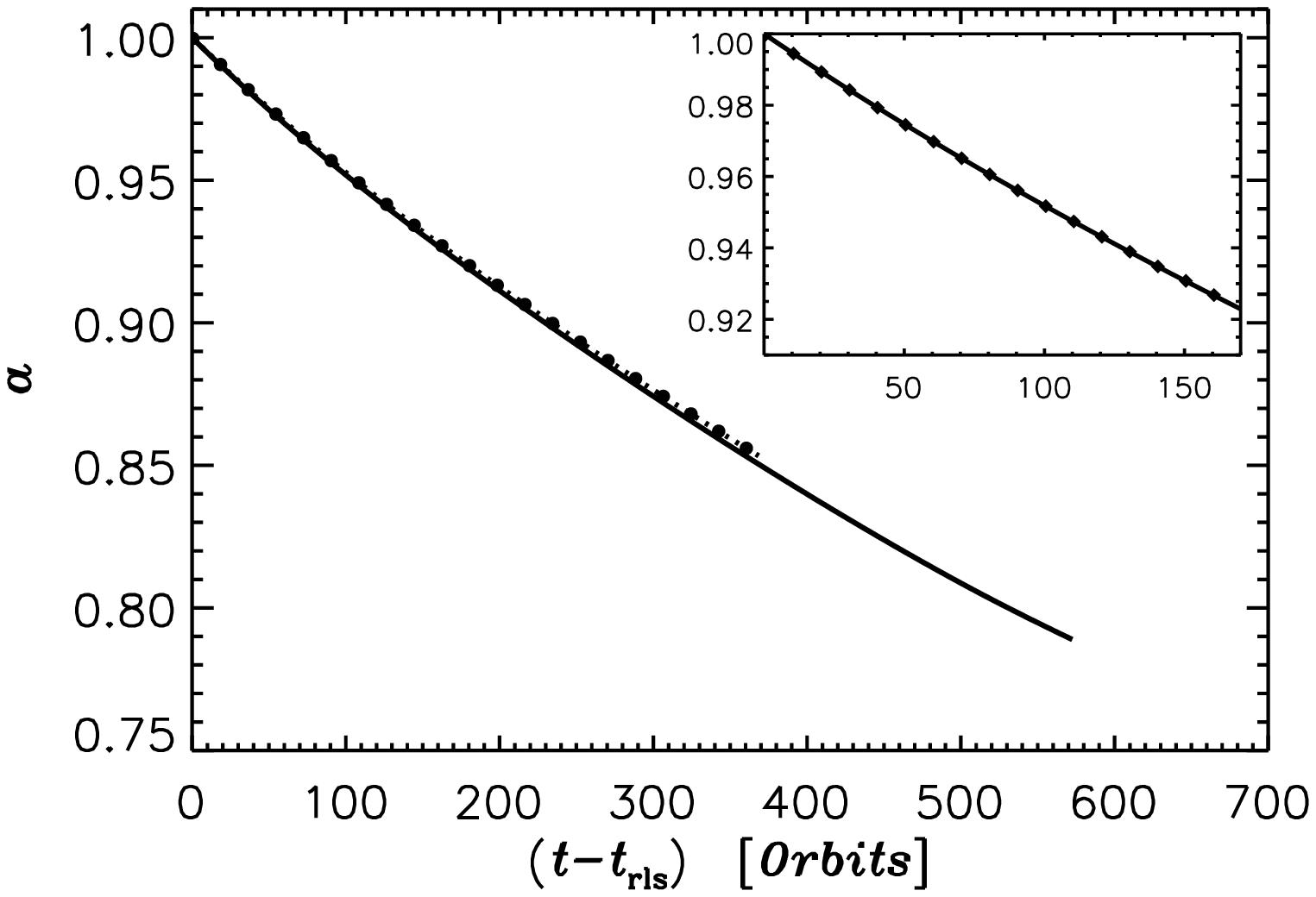}}
\caption{%
 \textit{Left}:
         Azimuthally averaged surface density of simulations with
         disk kinematic viscosity $\nu=5\times10^{-4}\,\viscu$ and 
         $H/r=0.05$. 
         The planet mass grows from $\Mp\simeq 3\,\MEarth$ to 
         $1\,\MJup$, at a prescribed rate, over roughly $330$ initial 
         orbital periods. The planet's orbit is fixed during this 
         time interval.
         \textit{Thin solid line}: Initial surface density
         distribution (same as the \textit{thin solid line} in the 
         \textit{left panel} of Fig.~\ref{fig:bid1}).
         \textit{Thick solid line}: Averaged surface density at time
         when $\Mp=1\,\MJup$ from a three-dimensional model whose
         outer radial boundary is $R_{\mathrm{max}}=6.7\,a_{0}$.
         \textit{Dotted line with solid circles}: Averaged surface 
         density when $\Mp=1\,\MJup$ from a two-dimensional model 
         with outer radial boundary at $13\,a_{0}$.
         \textit{Dashed line}: Case of a disk with $\Mp=1\,\MJup$ 
         and $\nu=1\times 10^{-4}\,\viscu$ (same as the 
         \textit{thick solid curve} in the \textit{left panel} of 
         Fig.~\ref{fig:bid1}).
 \textit{Right}:
         Orbital radius evolution after release, 
         $t_{\mathrm{rls}}=340$ initial orbits, obtained from the 
         three-dimensional model with $R_{\mathrm{max}}=6.7\,a_{0}$
         (\textit{solid line}) and the two-dimensional model 
         (\textit{dotted line with solid circles}).
         Note that the orbital evolution covers more than $1.8$
         local viscous timescales.
         The inset shows migration tracks obtained from the
         three-dimensional calculations over $t_{\nu}/2$ viscous
         timescales at the initial orbital radius of the planet.
         The solid diamonds represent data from the case with 
         $R_{\mathrm{max}}=13\,a_{0}$. The solid line represents
         the same model as in the main panel.
         }
\label{fig:bid3}
\end{figure*}
We wish to examine here whether the migration trend observed in raising 
the viscosity from $\nu\sim 10^{-5}\,\viscu$ to $\nu\sim 10^{-4}\,\viscu$ 
persists at larger viscosity.
As shown in the left panel of Figure~\ref{fig:bid1} 
(\textit{solid line}), when $\nu=1\times 10^{-4}\,\viscu$ a Jupiter-mass
planet is able to open only a shallow gap along its orbit (there is a
drop in density of about a factor $3$ relative to the value
just outside the gap). This is because gap opening conditions are 
not satisfied (see section~\ref{sec:OrbitalRadiusEvolution}).
At larger disk viscosity we therefore expect an even shallower gap.

We perform two three-dimensional simulations with the same setup as 
that outlined above but with $\nu=5\times 10^{-4}\,\viscu$ 
($\alpha=0.2$ at $r=a_{0}$) and outer radial boundaries at 
$R_{\mathrm{max}}=6.7\,a_{0}$ and $13\,a_{0}$, respectively.
The linear base resolution is 
$\Delta R=a_{0}\,\Delta\theta=a_{0}\,\Delta\phi=0.014\,a_0$
while the resolution in the coorbital region around the planet 
is $9\times10^{-4}\,a_0$.
The viscous diffusion timescale, $t_{\nu}$, at $r=a_{0}$ is 
approximately $320$ (initial) orbital periods whereas 
$t_{\nu}$ at $R_{\mathrm{max}}=6.7\,a_{0}$ is over $1.4\times 10^{4}$ 
orbits (and $5.4\times 10^{4}$ orbits at $R_{\mathrm{max}}=13\,a_{0}$).
We also consider a two-dimensional version of such models, 
having the same grid structure and resolution in the $r$-$\phi$ plane, 
and outer grid boundary located at $r_{\mathrm{max}}=13\,a_{0}$
(nearly $68\,\AU$ from the central star).
The planet mass grows, at a prescribed rate, from 
$\Mp\simeq 3\,\MEarth$ to $1\,\MJup$ over about $330$ periods
(which is similar to $t_{\nu}$ at the planet position),
while the planet's orbit is held fixed.
The left panel of Figure~\ref{fig:bid3} shows the initial surface
density (\textit{thin solid line}) and the azimuthally averaged 
density profile when $\Mp=1\,\MJup$ for the three-dimensional 
model with $R_{\mathrm{max}}=6.7\,a_{0}$ (\textit{thick solid line})
and the two-dimensional model 
(\textit{dotted line with solid circles}).
As reference, the azimuthally averaged density for 
the case with viscosity $\nu=1\times 10^{-4}\,\viscu$ and 
$\Mp=1\,\MJup$ is also plotted as a dashed line (same as the 
\textit{thick solid curve} in the \textit{left panel} of 
Fig.~\ref{fig:bid2}). 
In the overlapping disk region, two- and three-dimensional 
calculations give consistent results. No significant deviations 
from the initial density distribution are observed at $r\gg a$.

The viscosity condition for gap opening requires that 
$\Mp/\Ms\gtrsim 0.02$ (see section~\ref{sec:OrbitalRadiusEvolution}),
or $\Mp\gtrsim 20\,\MJup$, in order for gravitational torques
to overcome viscous torques. In fact, the density distribution in 
the left panel of Figure~\ref{fig:bid3} (\textit{thick solid line} and 
\textit{dotted line with solid circles}) shows a form of rather shallow
gap. Hence, Type~II migration should not be expected.

The planet is released from its fixed orbit at time 
$t_{\mathrm{rls}}=340$ ($\Mp=1\,\MJup$ for $t\ge t_{\mathrm{rls}}$) 
and the orbit is integrated for $1.8\,t_{\nu}$ viscous timescales 
at $r=a_{0}$. Figure~\ref{fig:bid3}, (\textit{right panel}) displays 
the orbital radius evolution for the two-dimensional 
(\textit{dotted line with solid circles}) 
and three-dimensional (\textit{thick solid line}) simulations. The two
migration tracks in the main panel closely follow one other. 
A comparison between the results obtained from the three-dimensional
models, over $t_{\nu}/2$ at the initial orbital radius of the planet,
is shown in the inset. 
Again, there is no indication that disk's evolution at $r\gg a$ has
a significant influence on planet's migration.
As for the case discussed above, nearly all the torque is accumulated by 
material within a radial band $|r-a|\lesssim 0.25\,a$ centered on the 
planet's orbit. 

The rate of migration after release time is approximately 
$-8\times 10^{-5}\,a_{0}\,\Omega_{0}$,
which is similar to that of the solid curve in the left panel of 
Figure~\ref{fig:bid2}. This near equality is expected for Type~I
migration, since it is independent of the level of disk viscosity.
We use the torque per unit disk mass, $dT/dM$, at $t=t_{\mathrm{rls}}$ 
from the model with $\nu=1\times 10^{-4}\,\viscu$ discussed above
in Appendix~\ref{sec:nu1e-4}
and the averaged surface density profile in the left panel of 
Figure~\ref{fig:bid3} (\textit{thick solid line}). By applying 
equation~(\ref{eq:dTdr_I}), we estimate the total torque expected under 
the assumption that $dT/dM$ has similar shapes in the two models.
There will be some dependence of $dT(r)/dM$ on viscosity, since 
viscosity affects the resonance widths.
This estimate yields a migration rate that agrees within a factor 
of $1.7$ with the value stated above, indicating that the intrinsic 
character of the torque per unit disk mass is roughly similar in 
these two cases. 

\section{Corrections for disk gravity}
\label{sec:diskgrav}

\begin{figure*}
\centering%
\resizebox{0.6\linewidth}{!}{%
\includegraphics{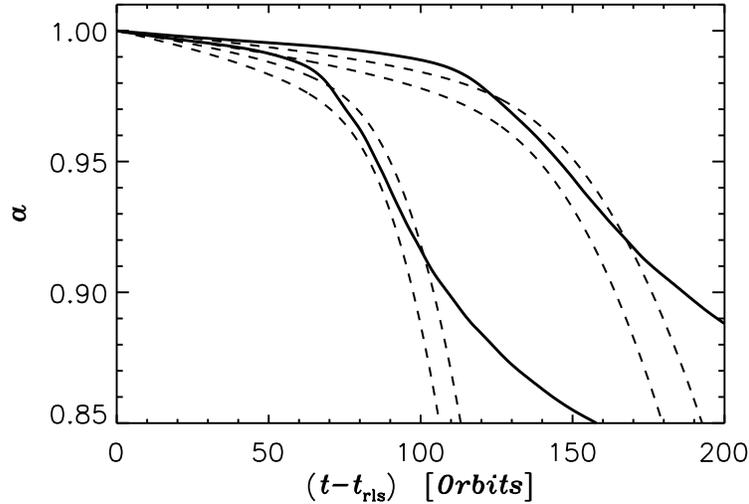}}
\caption{
         Orbital evolution under the same disk conditions as for 
         models in Figure~\ref{fig:a_xsn}. But, the planet's
         angular speed is imposed to be equal to the instantaneous
         keplerian value while the planet migrates in response to
         the nonaxisymmetric disk forces.
         \textit{Solid curves}: Simulation results for orbital 
         migration of a planet in a disk with initial surface 
         density equal to 
         $\Sigma_{p}=9\times10^{-4}\,\densu$, or about 
         $300\,\mathrm{g}\,\mathrm{cm}^{-2}$ at $a_{0}=5.2\,\AU$
         (upper migration track), and
         $\Sigma_{p}=1.5\times10^{-3}\,\densu$, or about 
         $500\,\mathrm{g}\,\mathrm{cm}^{-2}$
         (lower migration track).
         \textit{Dashed curves}: Predicted orbital migration 
         according to Type~I theory, 
         equations~(\ref{eq:tanakadotaI}) 
         (\textit{upper curve} of pair for unsaturated coorbital 
         torques) and (\ref{eq:tanakadotaIsat})  
         (\textit{lower curve} of pair for saturated coorbital
         torques).
         Analogous calculations executed for density distributions
         and disk thicknesses used in 
         section~\ref{sec:GrowingMigratingPlanets} produce 
         migration tracks that differ by $\sim 1$\%, over the 
         entire planet mass range, from those displayed in the 
         left panels of Figures~\ref{fig:a_sn}, 
         \ref{fig:a_sn_h}, and \ref{fig:a_sn_nu}.
         }
\label{fig:a_xsnk}
\end{figure*}
As discussed in section~\ref{sec:highdiskmass}, there is a possible
artificial torque that can act on a planet surrounded by a massive
disk, when the planet responds to the gravity of the disk but the
disk self-gravity is not included \citep{pierens2005}.
This torque is a consequence of the disk's axisymmetric gravitational
force in changing the planet's orbital rotation rate, but not changing
the disk's rotation rate (since the disk is not self-gravitating).
This artificial difference leads to a shift in disk resonances that
in turn leads to an artificial increase in the planet's inward
migration. It can largely be remedied by forcing the planet to rotate
at the local keplerian rate, i.e., at the same speed as the gas rotates
apart from effects of gas pressure. In this prescription, the planet
responds to the nonaxisymmetric forces of the disk that result in
migration, while undergoing orbital motion at the keplerian rate. 
This scheme is in reasonable accord with simulations that include the
full effects of disk self-gravity \citep{baruteau2008}. The full
effects of self-gravity cause a slightly faster migration rate than
this approximation suggests. We have carried out three-dimensional
simulations with such imposed keplerian planetary orbits for various
disk mass cases discussed in sections~\ref{sec:GrowingMigratingPlanets}
and \ref{sec:typeiii}. For mass distributions and disk thicknesses, as
those applied in section~\ref{sec:GrowingMigratingPlanets}, migration
tracks show negligible differences, over the entire planet mass range.
The only cases that produce changes beyond a few percent in migration
rates are those in Section~\ref{sec:typeiii}. 
In Figure~\ref{fig:a_xsnk} we plot the resulting migration for the same
disk models as in Figure~\ref{fig:a_xsn}. The migration rates are slower,
as found by \citet{baruteau2008}. But, they are still in approximate 
agreement with the predictions of migration theory.

\section{Additional Tests on Fast Migration}
\label{sec:typeiii_app}

Simulations of the orbital evolution of a fixed Saturn-mass planet 
($\Mp=0.3\,\MJup$) in a cold ($H/r=0.03$) and massive disk
($\Sigma_{p}=2\times 10^{-3}\,\densu%
\approx 670\,\mathrm{g}\,\mathrm{cm}^{-2}$ at the planet's initial
orbital radius) can lead to a buildup of gas within the planet's 
Hill sphere, which is eventually halted when a sufficiently large 
pressure gradient is established. 
The mass of material that accumulates around the planet can exceed
the planet's mass, with possible effects on migration rates. In order 
to prevent the accumulation of gas within the Hill sphere, in the 
models presented in section~\ref{sec:NongrowingPlanets}, we applied 
accreting boundary conditions near the planet, without adding the 
gas mass to the planet mass. 
In this Appendix we wish to reconsider the nonaccreting configuration
(as in MP03 and DBL05). 
The nonaccreting approach may be considered to be crudely simulating 
a case where some process prevents the planet from gaining further
mass. 

\begin{figure*}
\centering%
\resizebox{\linewidth}{!}{%
\includegraphics{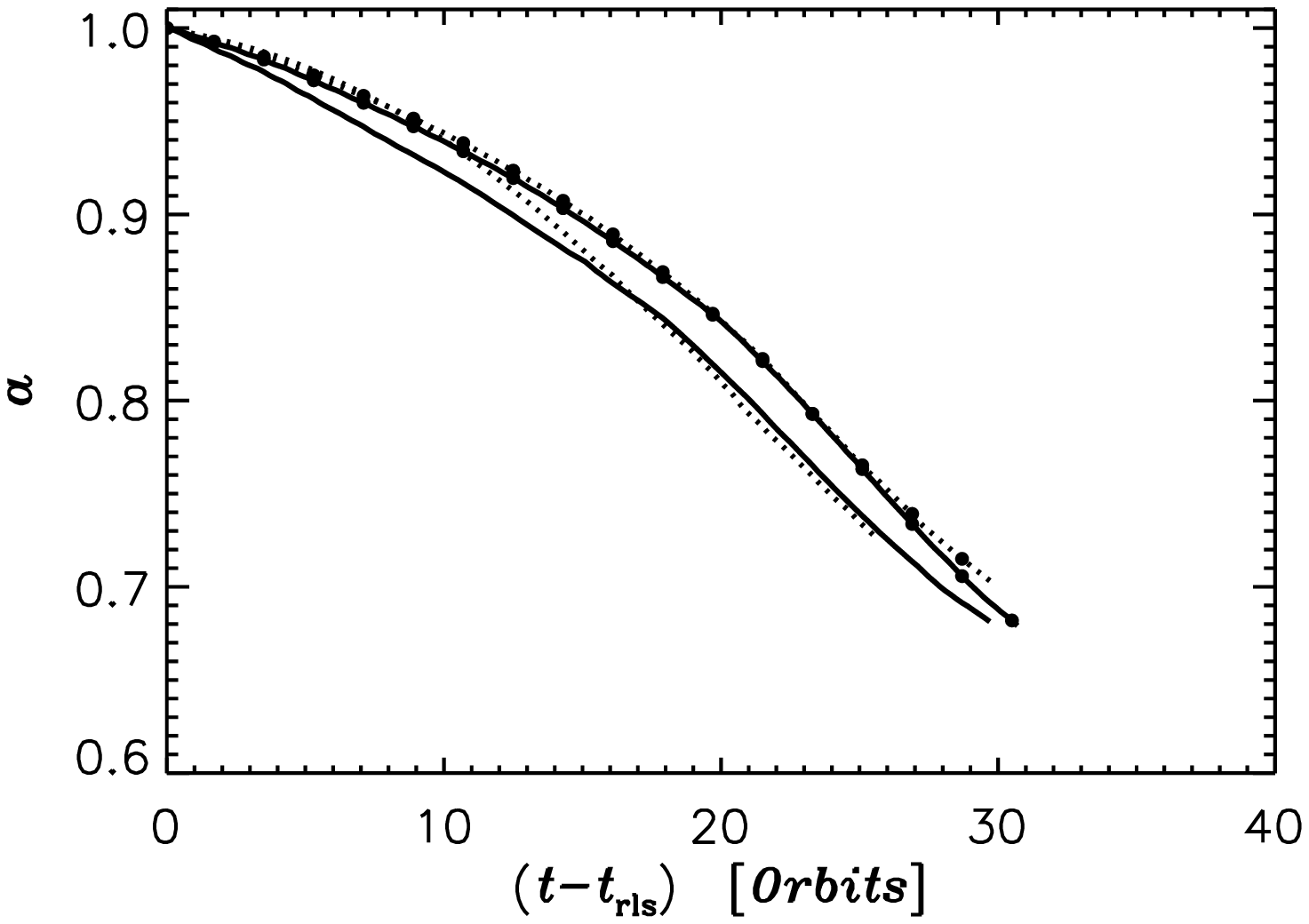}%
\includegraphics{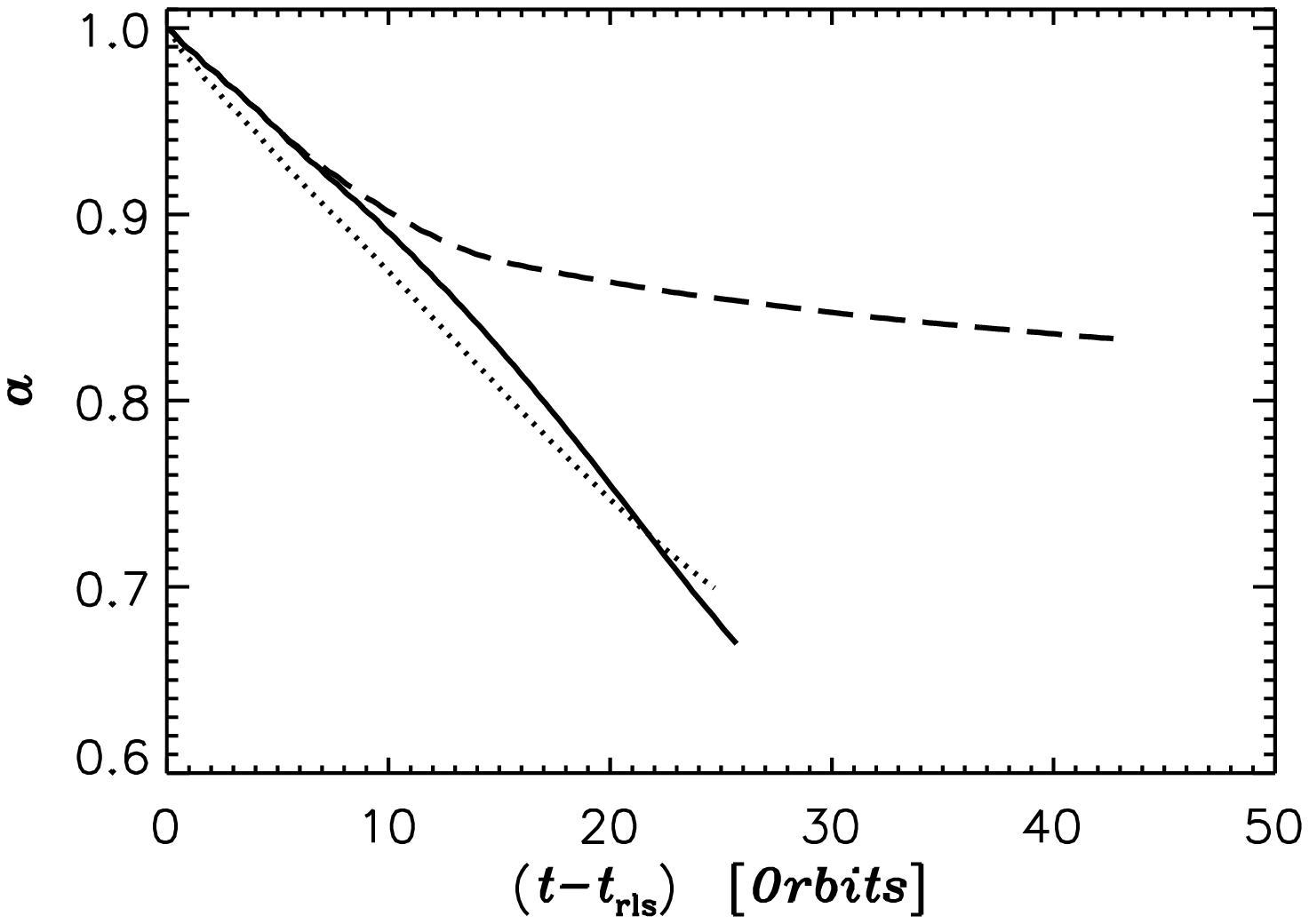}}
\caption{Orbital migration of a $0.3\,\MJup$ planet in a cold and
         massive disk (see section~\ref{sec:NongrowingPlanets} for
         details). The planet's orbit is held fixed for
         $t_{\mathrm{rls}}$ initial orbital periods.
 \textit{Left}: Comparison between two-dimensional models with 
         (\textit{solid lines}) and without (\textit{dotted lines})
         accreting boundary conditions near the fixed mass planet 
         ($t_{\mathrm{rls}}=100$). The curves marked with solid
         circles represent migration tracks obtained by excluding
         torques from within the planet's Hill sphere.
 \textit{Right}: Migration tracks from three-dimensional models
         with sudden release ($t_{\mathrm{rls}}=1$),
         and fixed (\textit{solid and dotted lines}) and variable
         (\textit{dashed line}) mass planets. The solid and
         dotted curves are for an accreting and nonaccreting planet,
         respectively. The long-dashed curve represents a case in
         which the initial planet mass ($0.3\,\MJup$) is augmented
         by the mass of the gas within $\Rhill/4$ of the planet.
         }
\label{fig:a_fast_chk}
\end{figure*}

In the left panel of Figure~\ref{fig:a_fast_chk}, migration tracks
from a two-dimensional model with an accreting planet 
(\textit{solid curves}) are compared to those obtained from a 
two-dimensional model with a nonaccreting planet 
(\textit{dotted curves}).
The gas masses within the Hill spheres are drastically different: 
$\sim 0.07\,\Mp$ and $\sim 1.6\,\Mp$ in the accreting and nonaccreting 
planet cases, respectively. 
In the nonaccreting case, the mass of the gas ($\sim 1\,\Mp$) within 
the bound region (see Fig.~\ref{fig:boundtracers}, \textit{right panel}) 
should be added to the inertial mass of the planet, thereby slowing 
migration \citep{papa2007}. This effect is indeed seen at early times 
(less than $10$ orbits after release).
The migration of the nonaccreting planet with bound gas 
(\textit{dotted curve}) is slowed by about a factor of $2$ relative 
to the accreting case (\textit{solid curve}). At later times, the
migration rates are closer, although it is not clear why. The reason
may be related to our determination that the mass of ``bound'' gas 
decreases at later times in the nonaccreting case (see also right 
panel of Fig.~\ref{fig:boundtracers}).

The left panel of Figure~\ref{fig:a_fast_chk} also plots the
orbital evolution when torques from within the Hill sphere are not
taken into account 
(\textit{solid and dotted curves with solid circles}).
The similarity of these migration tracks to the other plotted in the
figure indicates that torques from gas in this region do not
dominate the migration rates in this particular case.

In the nonaccreting case, dense gas that accumulates around the planet
could be thought of as forming an envelope, once it becomes bound to
the planet. A massive envelope would then participate in both the 
gravitational and inertial mass of the planet. We set up a 
three-dimensional model, with the same disk properties mentioned
above (see section~\ref{sec:SimulationsSetup} for details), and planet
mass $\Mp=M_{c}+M_{e}$, where $M_{c}=0.3\,\MJup$ is a ``core'' mass and 
$M_{e}=M_{e}(t)$
is the mass of the gas within $\Rhill/4$ of the planet. Given the large
initial mass in the coorbital region ($\sim 2\,\MJup$), the planet
rapidly gains mass, growing beyond $1\,\MJup$ in less than $25$ 
initial orbits. The planet is released in a smooth disk after 
a few orbits. 
The orbital radius evolution is shown as a dashed line
in the right panel of Figure~\ref{fig:a_fast_chk}, together with those
from three-dimensional models with a fixed mass planet and accreting
(\textit{solid line}) and nonaccreting (\textit{dotted line}) 
boundary conditions near the planet.
The initial migration rates are similar in all three configurations but
migration starts to rapidly slow down when the planet mass, in 
dashed-line case, grows beyond $\Mp\approx 0.8\,\MJup$.
This behavior resembles that seen in Figure~\ref{fig:a_03sr} 
(\textit{dotted line with solid circles}).

\begin{figure*}
\centering%
\resizebox{\linewidth}{!}{%
\includegraphics{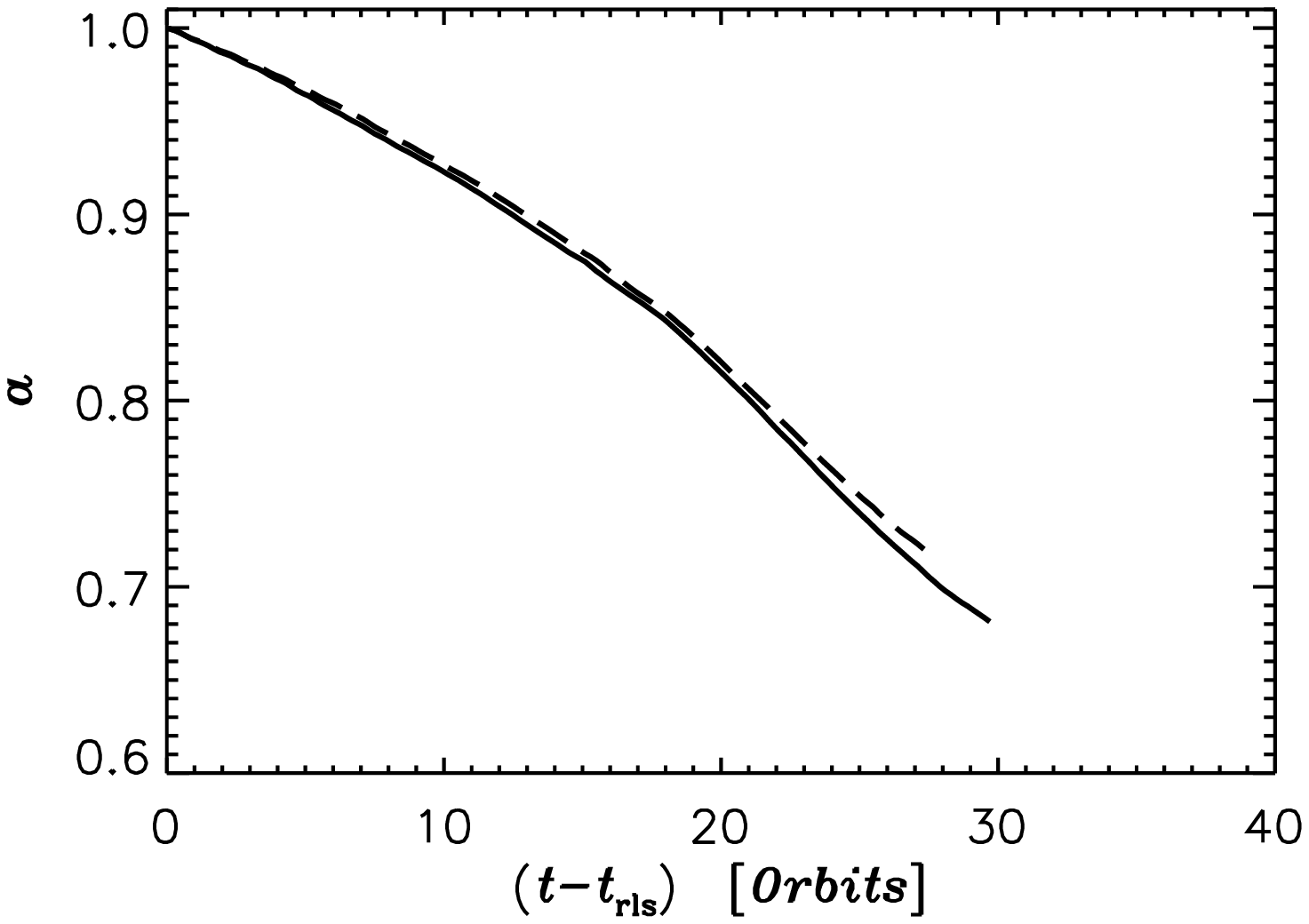}%
\includegraphics{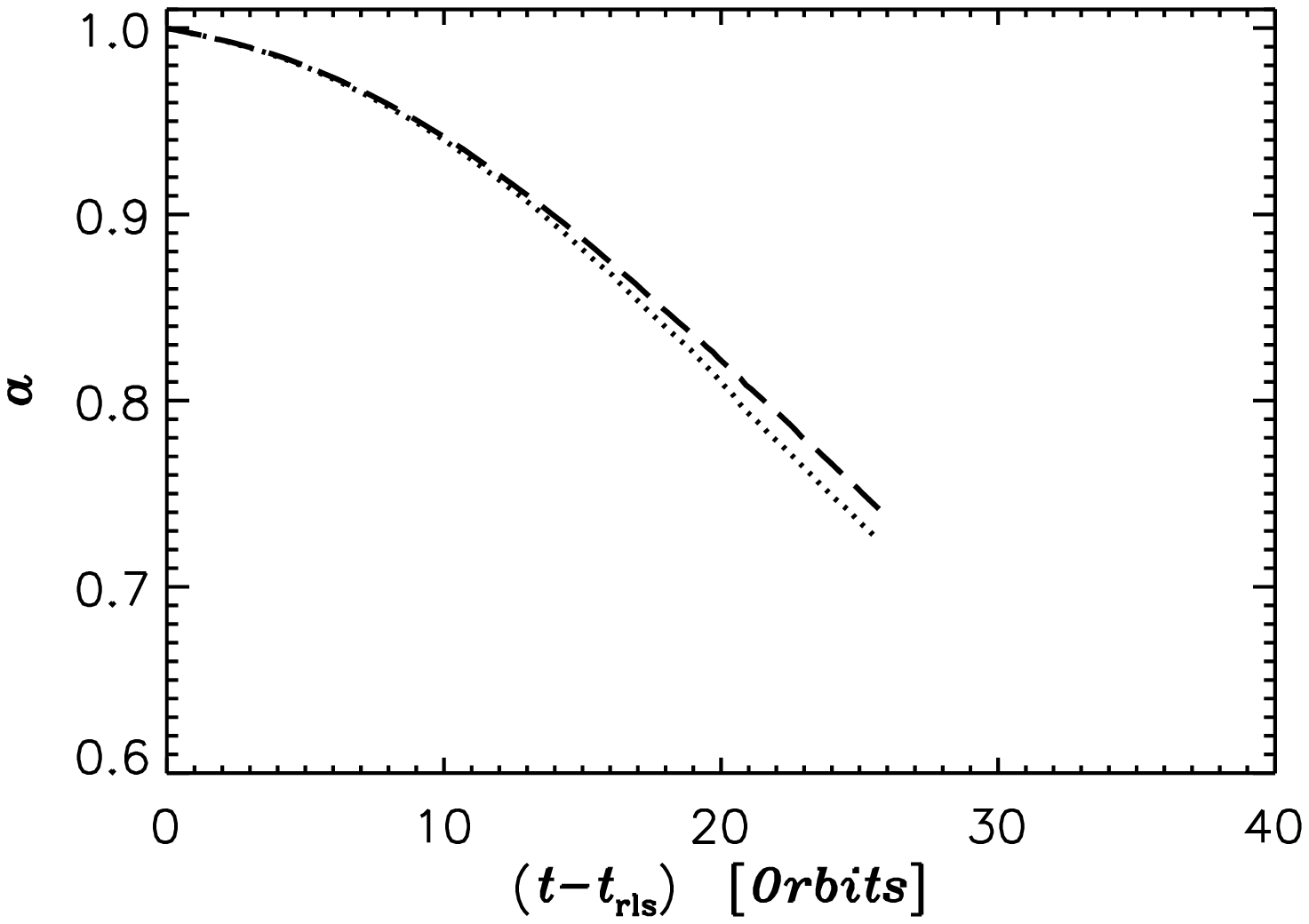}}
\caption{Numerical convergence tests for two-dimensional models
         with and without accreting boundary conditions near the
         planet (see Fig.~\ref{fig:a_fast_chk}).
         The planet's orbit remains fixed for $t_{\mathrm{rls}}=100$
         initial orbital periods.
 \textit{Left}: Comparison between simulations with an accreting planet
         and coarsest grid resolution in the coorbital region 
         $\Delta r=a_{0}\,\Delta\phi=0.014\,a_{0}$ 
         (\textit{dashed curve}) and
         $3.5\times 10^{-3}$ (\textit{solid curve}).
 \textit{Right}: Comparison between models with a nonaccreting planet
         and a linear resolution in the coorbital region around the
         planet of $0.02\,\Rhill$ (\textit{dotted curve}) and 
         $0.01\,\Rhill$ (\textit{dashed curve}).
         }
\label{fig:a_fast_res}
\end{figure*}
Figure~\ref{fig:a_fast_res} displays numerical convergence tests for
the accreting (\textit{left}) and nonaccreting (\textit{right}) planet
models presented in the left panel of Figure~\ref{fig:a_fast_chk}.
The two simulations in the left panel have coarsest (linear) resolutions
in the coorbital region that differ by a factor of $4$, in both radial 
and azimuthal directions. Calculations in the right panel have resolutions
in the coorbital region around the planet that differ by a factor of $2$
in each direction.

\subsection{Gas Bound to the Planet}
\label{sec:BoundGas}

In the calculations with an accreting planet, torque contributions
from within $\Rhill/2$ of the planet are ignored. By following fluid
paths, here we show that most of this material is captured and
eventually accreted by the planet.

In a nonstationary flow, streamlines can be used as a proxy for fluid 
trajectories only over short distances and periods of time. Therefore,
we track trajectories of fluid parcels by deploying tracer (massless) 
particles in the flow and then following their motion. This procedure
allows us to obtain a reliable determination of fluid paths regardless
of whether the flow is close or far from steady state.

The equations of motion of each particle are integrated every
hydro-dynamical time-step by interpolating the velocity field at the
particle's location and by advancing its position in time via a 
second-order Runge-Kutta method.
Both spatial and temporal interpolations are performed by using the
velocity field with the highest resolution available, i.e., that 
belonging to the most refined grid level in which the particle resides.
The spatial interpolation is based on a monotonized harmonic mean 
\citep{vanleer1977}, which is second-order accurate and capable of 
handling discontinuities and shock conditions.
Hence, trajectories are formally second-order accurate in both space
and time.

\begin{figure*}
\centering%
\resizebox{\linewidth}{!}{%
\includegraphics{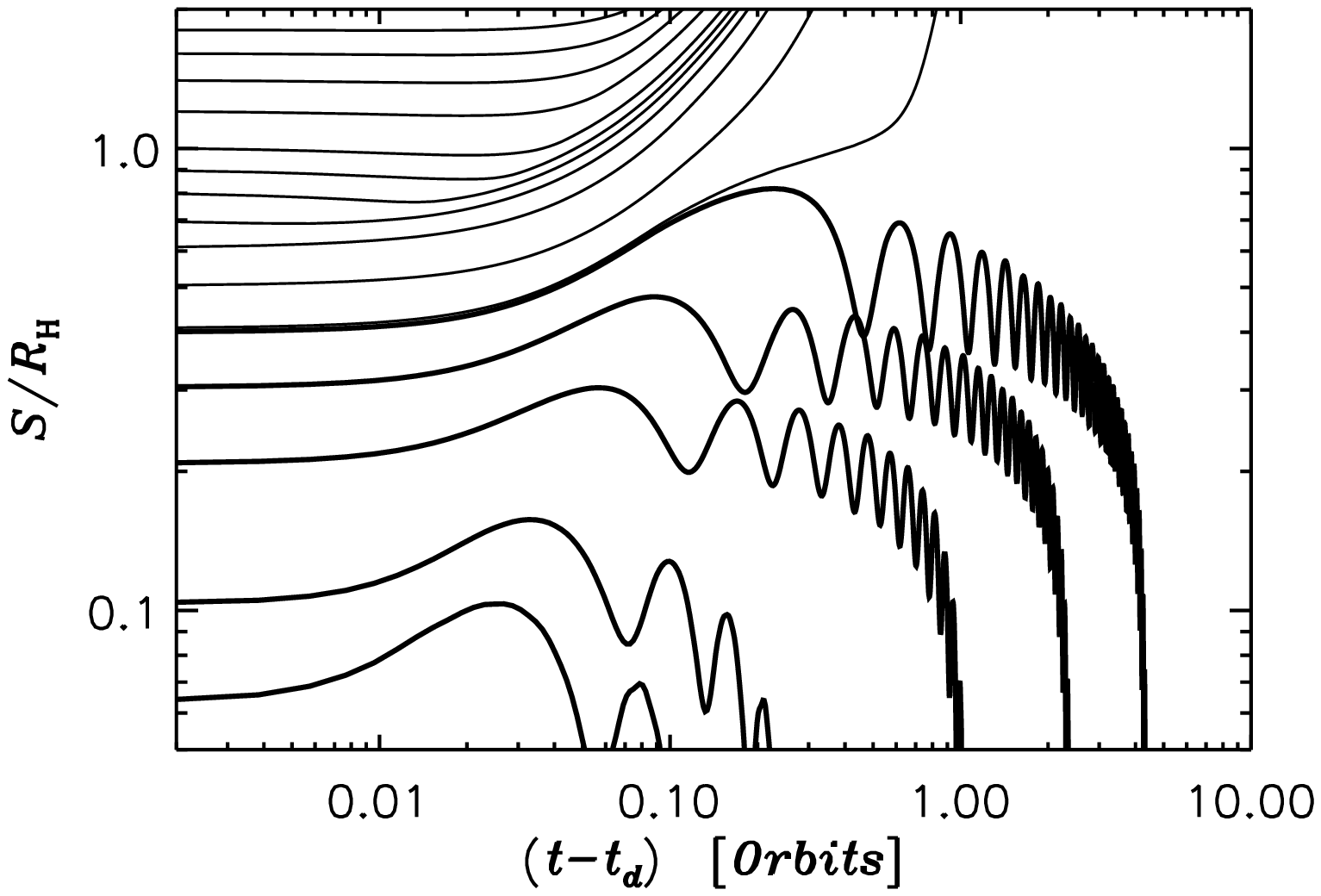}%
\includegraphics{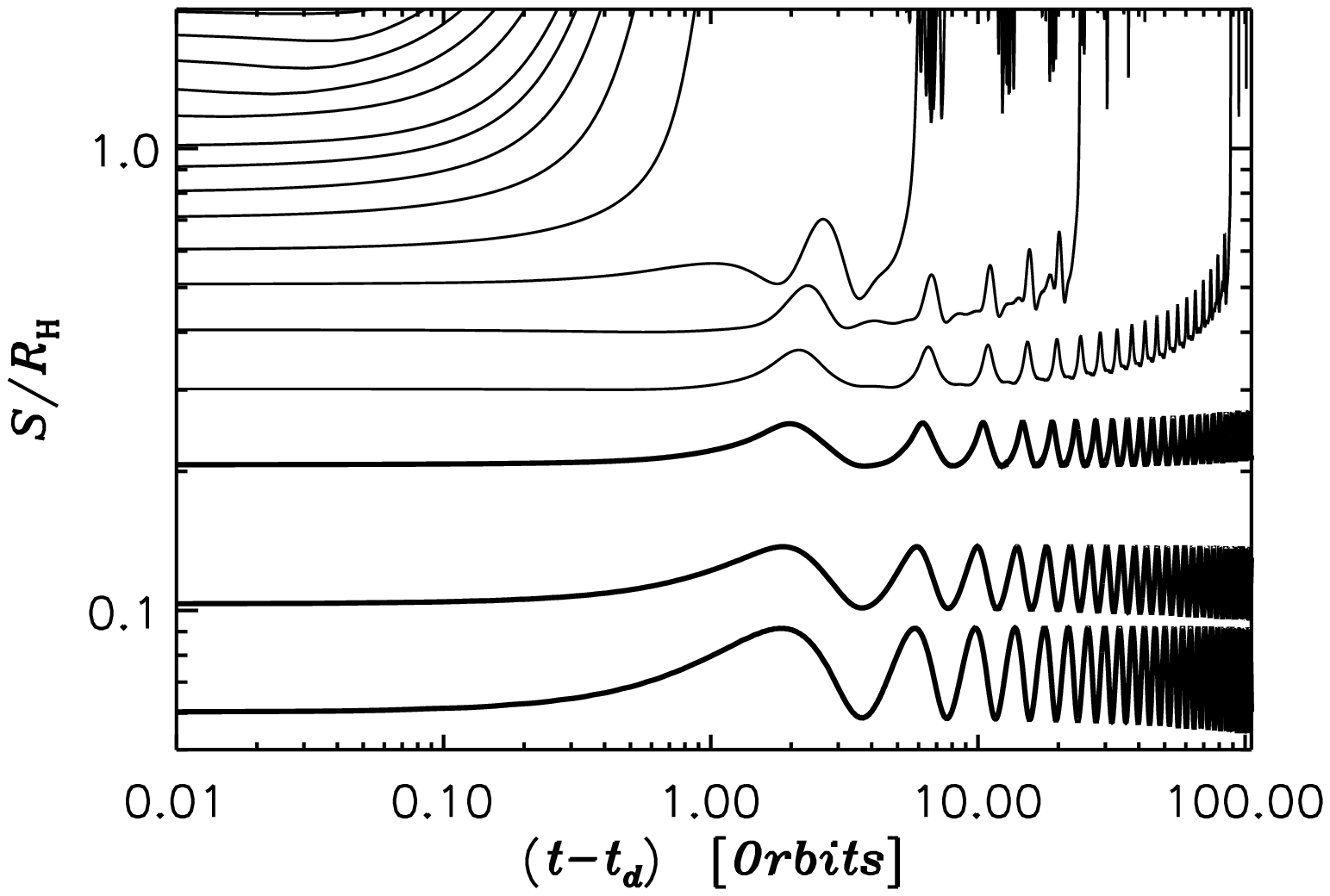}}
\caption{Tracer particles deployed within about $2\,\Rhill$ of a
         $0.3\,\MJup$ planet orbiting in the cold disk model
         ($H/r=0.03$) discussed in this Appendix and in 
         section~\ref{sec:NongrowingPlanets}. The plot shows
         the distance, in units of $\Rhill$, from an accreting
         (\textit{left}) and nonaccreting (\textit{right}) planet.
         Distances of trajectories that return to the disk are
         indicated as thin lines, otherwise they are indicated as
         a thick lines.
         }
\label{fig:boundtracers}
\end{figure*}
Here we employ tracer particles to estimate the size to
the region occupied by gas bound to nonmigrating planets. Tracers are
deployed in the disk within about $2\,\Rhill$ of a Saturn-mass planet
(\Mp=$0.3\,\MJup$). We use both models with and without accreting
boundary conditions near the planet discussed in this Appendix 
(Fig.\ref{fig:a_fast_chk}, \textit{left panel}) and in 
section~\ref{sec:NongrowingPlanets}.
Figure~\ref{fig:boundtracers} shows the distance from the planet, $S$,
of particles as a function of time. Tracers are deployed at a time 
$t_{d}$, when the mass within the Hill sphere has reached a nearly
steady value. The distance along the trajectories is normalized to the
Hill radius, $\Rhill$. 
The left panel refers to the accreting planet case, whereas the right
panel refers to the nonaccreting planet case. Thin curves mark
distances of trajectories that are not captured and thus return to
the disk. Thick curves mark distances of trajectories that are
captured in the planet's gravitational potential (\textit{left panel})
or otherwise remain within $\Rhill/2$ of the planet, over the simulated
evolution (\textit{right panel}). In the accreting case, bound
trajectories rapidly decay towards the planet and it is therefore
possible to make a clear distinction between bound and unbound
trajectories.
In the nonaccreting case, the distinction is less clear and may apply
only over a given amount of time.

% Bibliography:
%\bibliographystyle{apj}
%\bibliography{charlie}

\begin{thebibliography}{60}
\expandafter\ifx\csname natexlab\endcsname\relax\def\natexlab#1{#1}\fi

\bibitem[{{Alibert} {et~al.}(2005){Alibert}, {Mordasini}, {Benz}, \&
  {Winisdoerffer}}]{alibert2005a}
{Alibert}, Y., {Mordasini}, C., {Benz}, W., \& {Winisdoerffer}, C. 2005, A\&A,
  434, 343

\bibitem[{{Artymowicz}(1993)}]{pawel1993a}
{Artymowicz}, P. 1993, \apj, 419, 155

\bibitem[{{Artymowicz}(2004)}]{pawel2004}
{Artymowicz}, P. 2004, in KITP Conference: Planet Formation: Terrestrial and
  Extra Solar

\bibitem[{{Baruteau} \& {Masset}(2008)}]{baruteau2008}
{Baruteau}, C., \& {Masset}, F. 2008, \apj, 678, 483

\bibitem[{{Bate} {et~al.}(2003){Bate}, {Lubow}, {Ogilvie}, \&
  {Miller}}]{bate2003}
{Bate}, M.~R., {Lubow}, S.~H., {Ogilvie}, G.~I., \& {Miller}, K.~A. 2003,
  MNRAS, 341, 213

\bibitem[{{Beutler}(2005)}]{beutler2005}
{Beutler}, G. 2005, {Methods of celestial mechanics. Vol. I: Physical,
  mathematical, and numerical principles} (Methods of celestial
  mechanics.~Vol.~I / Gerhard Beutler.~In cooperation with Leos Mervart and
  Andreas Verdun.~Astronomy and Astrophysics Library.~Berlin: Springer, ISBN
  3-540-40749-9, 2005, XVI, 464 pp.~99 figures, 11 in color, 32 tables and a
  CD-ROM.)

\bibitem[{{Blandford} \& {Payne}(1982)}]{blandford1982}
{Blandford}, R.~D., \& {Payne}, D.~G. 1982, \mnras, 199, 883

\bibitem[{Bodenheimer \& Pollack(1986)}]{bodenheimer1986}
Bodenheimer, P., \& Pollack, J.~B. 1986, Icarus, 67, 391

\bibitem[{{Bryden} {et~al.}(1999){Bryden}, {Chen}, {Lin}, {Nelson}, \&
  {Papaloizou}}]{bryden1999}
{Bryden}, G., {Chen}, X., {Lin}, D.~N.~C., {Nelson}, R.~P., \& {Papaloizou},
  J.~C.~B. 1999, ApJ, 514, 344

\bibitem[{{Butler} {et~al.}(2006){Butler}, {Wright}, {Marcy}, {Fischer},
  {Vogt}, {Tinney}, {Jones}, {Carter}, {Johnson}, {McCarthy}, \&
  {Penny}}]{butler2006}
{Butler}, R.~P., {Wright}, J.~T., {Marcy}, G.~W., {Fischer}, D.~A., {Vogt},
  S.~S., {Tinney}, C.~G., {Jones}, H.~R.~A., {Carter}, B.~D., {Johnson}, J.~A.,
  {McCarthy}, C., \& {Penny}, A.~J. 2006, \apj, 646, 505

\bibitem[{{D'Angelo} {et~al.}(2005){D'Angelo}, {Bate}, \&
  {Lubow}}]{gennaro2005}
{D'Angelo}, G., {Bate}, M.~R., \& {Lubow}, S.~H. 2005, MNRAS, 358, 316 (DBL05)

\bibitem[{{D'Angelo} {et~al.}(2002){D'Angelo}, {Henning}, \&
  {Kley}}]{gennaro2002}
{D'Angelo}, G., {Henning}, T., \& {Kley}, W. 2002, A\&A, 385, 647

\bibitem[{{D'Angelo} {et~al.}(2003){D'Angelo}, {Kley}, \&
  {Henning}}]{gennaro2003b}
{D'Angelo}, G., {Kley}, W., \& {Henning}, T. 2003, ApJ, 586, 540

\bibitem[{{D'Angelo} {et~al.}(2006){D'Angelo}, {Lubow}, \&
  {Bate}}]{gennaro2006}
{D'Angelo}, G., {Lubow}, S.~H., \& {Bate}, M.~R. 2006, \apj, 652, 1698

\bibitem[{{Dobbs-Dixon} {et~al.}(2007){Dobbs-Dixon}, {Li}, \&
  {Lin}}]{dobbs-dixon2007}
{Dobbs-Dixon}, I., {Li}, S.~L., \& {Lin}, D.~N.~C. 2007, \apj, 660, 791

\bibitem[{{Eggleton}(1983)}]{eggleton1983}
{Eggleton}, P.~P. 1983, \apj, 268, 368

\bibitem[{{Fendt}(2003)}]{fendt2003}
{Fendt}, C. 2003, \aap, 411, 623

\bibitem[{{Flaherty} \& {Muzerolle}(2008)}]{flaherty2008}
{Flaherty}, K.~M., \& {Muzerolle}, J. 2008, \aj, 135, 966

\bibitem[{{Godon}(1996)}]{godon1996}
{Godon}, P. 1996, MNRAS, 282, 1107

\bibitem[{{Godon}(1997)}]{godon1997}
---. 1997, ApJ, 480, 329

\bibitem[{{Gold\-reich} \& {Tre\-maine}(1979)}]{gt1979}
{Gold\-reich}, P., \& {Tre\-maine}, S. 1979, ApJ, 233, 857

\bibitem[{{Gold\-reich} \& {Tre\-maine}(1980)}]{gt1980}
---. 1980, ApJ, 241, 425

\bibitem[{{Guillot}(2005)}]{guillot2005}
{Guillot}, T. 2005, Annual Review of Earth and Planetary Sciences, 33, 493

\bibitem[{{Haisch} {et~al.}(2001){Haisch}, {Lada}, \& {Lada}}]{haisch2001}
{Haisch}, K.~E., {Lada}, E.~A., \& {Lada}, C.~J. 2001, ApJ, 553, L153

\bibitem[{{Hartmann} {et~al.}(1998){Hartmann}, {Calvet}, {Gullbring}, \&
  {D'Alessio}}]{hartmann1998}
{Hartmann}, L., {Calvet}, N., {Gullbring}, E., \& {D'Alessio}, P. 1998, \apj,
  495, 385

\bibitem[{{Hourigan} \& {Ward}(1984)}]{hourigan1984}
{Hourigan}, K., \& {Ward}, W.~R. 1984, Icarus, 60, 29

\bibitem[{{Hubickyj} {et~al.}(2005){Hubickyj}, {Bodenheimer}, \&
  {Lissauer}}]{hubickyj2005}
{Hubickyj}, O., {Bodenheimer}, P., \& {Lissauer}, J.~J. 2005, Icarus, 179, 415

\bibitem[{{Hubickyj} {et~al.}(2007){Hubickyj}, {Lissauer}, {D'Angelo}, \&
  {Bodenheimer}}]{hubickyj2007}
{Hubickyj}, O., {Lissauer}, J.~J., {D'Angelo}, G., \& {Bodenheimer}, P. 2007,
  AGU Fall Meeting Abstracts, A6

\bibitem[{Kley(1999)}]{kley1999}
Kley, W. 1999, MNRAS, 303, 696

\bibitem[{{Li} {et~al.}(2005){Li}, {Li}, {Koller}, {Wendroff}, {Liska},
  {Orban}, {Liang}, \& {Lin}}]{li2005}
{Li}, H., {Li}, S., {Koller}, J., {Wendroff}, B.~B., {Liska}, R., {Orban},
  C.~M., {Liang}, E.~P.~T., \& {Lin}, D.~N.~C. 2005, \apj, 624, 1003

\bibitem[{{Lin} \& {Papaloizou}(1986)}]{lin1986}
{Lin}, D.~N.~C., \& {Papaloizou}, J. 1986, ApJ, 309, 846

\bibitem[{{Lin} \& {Papaloizou}(1993)}]{lin1993}
{Lin}, D.~N.~C., \& {Papaloizou}, J.~C.~B. 1993, in Protostars and Planets III,
  749--835

\bibitem[{{Lissauer} \& {Stevenson}(2007)}]{lissauer2007}
{Lissauer}, J.~J., \& {Stevenson}, D.~J. 2007, in Protostars and Planets V, ed.
  B.~{Reipurth}, D.~{Jewitt}, \& K.~{Keil}, 591--606

\bibitem[{{Lubow} \& {D'Angelo}(2006)}]{lubow2006}
{Lubow}, S.~H., \& {D'Angelo}, G. 2006, ApJ, 641, 526

\bibitem[{Lubow {et~al.}(1999)Lubow, Seibert, \& Artymowicz}]{lubow1999}
Lubow, S.~H., Seibert, M., \& Artymowicz, P. 1999, ApJ, 526, 1001

\bibitem[{{Marcy} {et~al.}(2005){Marcy}, {Butler}, {Fischer}, {Vogt}, {Wright},
  {Tinney}, \& {Jones}}]{marcy2005}
{Marcy}, G., {Butler}, R.~P., {Fischer}, D., {Vogt}, S., {Wright}, J.~T.,
  {Tinney}, C.~G., \& {Jones}, H.~R.~A. 2005, Progress of Theoretical Physics
  Supplement, 158, 24

\bibitem[{{Masset} {et~al.}(2006){Masset}, {D'Angelo}, \& {Kley}}]{masset2006}
{Masset}, F.~S., {D'Angelo}, G., \& {Kley}, W. 2006, \apj, 652, 730

\bibitem[{{Masset} \& {Papaloizou}(2003)}]{masset2003}
{Masset}, F.~S., \& {Papaloizou}, J.~C.~B. 2003, ApJ, 588, 494 (MP03)

\bibitem[{{Mihalas} \& {Weibel Mihalas}(1999)}]{M&M}
{Mihalas}, D., \& {Weibel Mihalas}, B. 1999, {Foundations of radiation
  hydrodynamics} (New York: Dover, 1999)

\bibitem[{{Nelson} \& {Benz}(2003)}]{anelson2003b}
{Nelson}, A.~F., \& {Benz}, W. 2003, ApJ, 589, 578

\bibitem[{{Nelson} {et~al.}(2000){Nelson}, {Papaloizou}, {Masset}, \&
  {Kley}}]{rnelson2000}
{Nelson}, R.~P., {Papaloizou}, J.~C.~B., {Masset}, F., \& {Kley}, W. 2000,
  MNRAS, 318, 18

\bibitem[{{Ogilvie} \& {Lubow}(2006)}]{ogilvie2006}
{Ogilvie}, G.~I., \& {Lubow}, S.~H. 2006, \mnras, 370, 784 (OL06)

\bibitem[{{Paczy{\'n}ski}(1971)}]{paczynski1971}
{Paczy{\'n}ski}, B. 1971, \araa, 9, 183

\bibitem[{{Papaloizou} {et~al.}(2007){Papaloizou}, {Nelson}, {Kley}, {Masset},
  \& {Artymowicz}}]{papa2007}
{Papaloizou}, J.~C.~B., {Nelson}, R.~P., {Kley}, W., {Masset}, F.~S., \&
  {Artymowicz}, P. 2007, in Protostars and Planets V, ed. B.~{Reipurth},
  D.~{Jewitt}, \& K.~{Keil}, 655--668

\bibitem[{{Pierens} \& {Hur{\'e}}(2005)}]{pierens2005}
{Pierens}, A., \& {Hur{\'e}}, J.-M. 2005, \aap, 433, L37

\bibitem[{{Pollack} {et~al.}(1996){Pollack}, {Hubickyj}, {Bodenheimer},
  {Lissauer}, {Podolak}, \& {Greenzweig}}]{pollack1996}
{Pollack}, J.~B., {Hubickyj}, O., {Bodenheimer}, P., {Lissauer}, J.~J.,
  {Podolak}, M., \& {Greenzweig}, Y. 1996, Icarus, 124, 62

\bibitem[{{Pudritz} \& {Norman}(1986)}]{pudritz1986}
{Pudritz}, R.~E., \& {Norman}, C.~A. 1986, \apj, 301, 571

\bibitem[{{Shu} {et~al.}(1994){Shu}, {Najita}, {Ostriker}, {Wilkin}, {Ruden},
  \& {Lizano}}]{shu1994}
{Shu}, F., {Najita}, J., {Ostriker}, E., {Wilkin}, F., {Ruden}, S., \&
  {Lizano}, S. 1994, \apj, 429, 781

\bibitem[{Tanaka {et~al.}(2002)Tanaka, Takeuchi, \& Ward}]{tanaka2002}
Tanaka, H., Takeuchi, T., \& Ward, W. 2002, ApJ, 565, 1257

\bibitem[{{Tanigawa} \& {Watanabe}(2002)}]{tanigawa2002}
{Tanigawa}, T., \& {Watanabe}, S. 2002, ApJ, 580, 506

\bibitem[{{van Leer}(1977)}]{vanleer1977}
{van Leer}, B. 1977, JCP, 23, 276

\bibitem[{Ward(1997)}]{ward1997}
Ward, W. 1997, Icarus, 126, 261

\bibitem[{{Ward}(1986)}]{ward1986}
{Ward}, W.~R. 1986, Icarus, 67, 164

\bibitem[{{Ward}(1992)}]{ward1992}
{Ward}, W.~R. 1992, in Lunar and Planetary Institute Conference Abstracts,
  Vol.~23, Lunar and Planetary Institute Conference Abstracts, 1491--1492

\bibitem[{{Ward} \& {Hourigan}(1989)}]{ward1989}
{Ward}, W.~R., \& {Hourigan}, K. 1989, \apj, 347, 490

\bibitem[{{Wuchterl}(1991)}]{wuchterl1991b}
{Wuchterl}, G. 1991, Icarus, 91, 53

\bibitem[{{Wuchterl}(1993)}]{wuchterl1993}
---. 1993, Icarus, 106, 323

\bibitem[{{Yuan} \& {Cassen}(1994)}]{yuan1994}
{Yuan}, C., \& {Cassen}, P. 1994, \apj, 437, 338

\bibitem[{{Zhou} \& {Lin}(2007)}]{zhou2007}
{Zhou}, J.-L., \& {Lin}, D.~N.~C. 2007, \apj, 666, 447

\bibitem[{Ziegler \& Yorke(1997)}]{ziegler1997}
Ziegler, U., \& Yorke, H.~W. 1997, Computer Physics Communications, 101, 54

\end{thebibliography}
% If necessary, insert .bbl file after this line and comment the two lines
% above.
%%++++++++++++++++++++++++++++++++++++++++++++++++++++++++++++++++++++++%%

%%++++++++++++++++++++++++++++++++++++++++++++++++++++++++++++++++++++++%%

\end{document}